\documentstyle[prd,aps,floats,epsfig,eqsecnum]{revtex} 
\newcommand{\be}{\begin{equation}}
\newcommand{\ee}{\end{equation}}
\newcommand{\bea}{\begin{eqnarray}}
\newcommand{\eea}{\end{eqnarray}}
\begin{document} 
\preprint{LPTHE-99-24; PITT-99}
\title{\bf NON-EQUILIBRIUM QUANTUM PLASMAS IN SCALAR QED: PHOTON
PRODUCTION, MAGNETIC  AND DEBYE MASSES AND CONDUCTIVITY.} 
\author{\bf D. Boyanovsky$^{(a,b)}$, H. J. de Vega$^{(b,a)}$,
M. Simionato$^{(b,d)}$} 
\address
{(a) Department of Physics and Astronomy, University of 
Pittsburgh, Pittsburgh  PA. 15260, U.S.A \\ 
(b) LPTHE, Universit\'e Pierre et Marie Curie (Paris VI) et Denis Diderot 
(Paris VII), Tour 16, 1er. \'etage, 4, Place Jussieu, 75252 Paris cedex 05, 
France\\
(d) Istituto Nazionale di Fisica Nucleare, Frascati, Italy} 
\maketitle 
\begin{abstract} 
We study the generation of a non-equilibrium plasma in scalar QED with
$N$-charged  scalar fields in the cases:
a) of a supercooled second order phase transition through spinodal 
instabilities  and  b) of
parametric amplification  when the order parameter oscillates with
large amplitude around the minimum of the potential. The focus is to study the
non-equilibrium electromagnetic properties of the plasma, such as
photon production, electric and magnetic screening and conductivity.  
A novel kinetic equation is introduced to compute photon production
far away from equilibrium in the large $ N $ limit and lowest order in
the electromagnetic coupling. During the early stages of the dynamics
the photon density grows exponentially and asymptotically the frequency
distribution becomes $ N_{ph}(\omega)\sim \alpha \; m^2 /
[\lambda^2 \; \omega^3] $  with  
$ \lambda $ the scalar self-coupling and $ m $ the scalar mass. In the case of
a phase transition, electric and magnetic fields are correlated on
distances $ \xi(t) \sim  \sqrt{t/m} $ during the early stages of the evolution
and the power  
spectrum is peaked at low momentum. This aspect is relevant for 
the generation of primordial magnetic fields in the Early Universe and
for photoproduction as a potential experimental signature of the
chiral phase transition. Magnetic and  Debye screening masses are
defined out of equilibrium as generalizations of the equilibrium
case. While the magnetic mass vanishes out of equilibrium in this
abelian model, we introduce an effective time and wave-number dependent 
magnetic mass  that reveals the different processes that contribute to
screening and their time scales. The Debye mass  turns out to be $
m^2_{Deb} \sim \alpha \; m^2/\lambda $ for a supercooled phase
transition  while in the case of an oscillating order parameter an
interpolating time dependent Debye mass grows  as $ \alpha
\sqrt{m\,t}/\lambda $ due to a non-linear resonance at low momentum in
the charged particle distribution. It is shown how the transverse
electric conductivity builds up during the formation of the
non-equilibrium plasma. Its long wavelength limit reaches a value $
\sigma_{k\approx 0} \sim \alpha \; m/\lambda $ at the end of the stage
of linear instabilities. It is shown that the electric conductivity
stays {\bf finite} for all $ k $ including $ k = 0 $ for {\bf finite}
time. In the asymptotic regime it attains a form analogous to the
equilibrium case  but in terms of the non-equilibrium particle
distribution functions.  
\end{abstract} 

\tableofcontents

\section{Introduction and motivation}

The study of the dynamics of  phenomena strongly out of equilibrium is
very relevant in cosmology where it  plays a fundamental role in the
consistent description of inflationary scenarios, baryogenesis and of
generation of primordial magnetic fields. Also in Relativistic Heavy Ion
collisions where it now acquires further phenomenological importance since the
Relativistic Heavy Ion Collider (RHIC) at Brookhaven begins
operation. RHIC and the forthcoming Large Hadron Collider at CERN will  
 probe the quark-gluon plasma and the chiral phase transitions in an
extreme environment of high temperature and density. These experimental
programs have inspired  intense theoretical efforts to understand the
formation, evolution and potential experimental signatures of the
quark-gluon plasma\cite{qgp,books} as well as relaxation and transport
phenomena on unprecedented short time scales. There are several
fundamental questions which define to a large extent the theoretical
aspects of this program: how does the quark-gluon plasma form and
equilibrates from the evolution of the parton distribution functions?
what are the time scales for electric and magnetic screening that 
dress the gluons and cut-off small angle scattering? how does a
hydrodynamic picture of the space-time evolution of the plasma emerge?
what are the experimental signatures?. These and other fundamental but
extremely difficult questions are being addressed 
from many different perspectives. An important approach that seeks to
describe the space-time evolution of partons is based on transport
equations that describe partonic cascades starting from a microscopic
description and incorporate semi-phenomenologically some 
screening corrections in the scattering cross
sections\cite{geiger,wang,eskola}. A correct description of electric
and magnetic screening is very important in this program since both
act as infrared cutoffs in transport cross sections and determine
energy losses in the plasma. 
Amongst the several potential experimental signatures proposed
to detect the QGP, photons and dileptons are deemed to be 
clean probes of the quark-gluon plasma because they only interact
electromagnetically\cite{qgp,books,alam} and their mean-free paths are  
much larger than the size of the fireball $\approx 20 \mbox{fm}$.
Hence, these electromagnetic probes could provide 
clean signatures of equilibration or out of equilibrium phenomena
unhindered by the strong interactions.  
Non-equilibrium phenomena associated
with a quenched chiral phase transition could have potentially
important electromagnetic signatures in the photon 
spectrum if there are strong charged pion fluctuations during the
phase transition. A preliminary study in this 
direction was pursued in\cite{photop2} where it was indicated that
departures from equilibrium in the photon 
distribution at low momentum could provide a signature of a
supercooled chiral phase transition\cite{rajagopal}. In cosmology,
post-inflationary 
phase transitions or the fast evolution of an inflaton field after
inflation could generate the hot plasma that describes the 
standard big bang scenario with a radiation dominated
Friedman-Robertson-Walker cosmology at the end of inflation\cite{inflation}.  
Furthermore, non-equilibrium effects during cosmological phase
transitions had been conjectured to generate the primordial 
magnetic fields that could act as seeds to be amplified by dynamo
mechanisms as an explanation  for the observed galactic magnetic
fields\cite{peebles,widrow}. Theoretical 
models for generation of primordial magnetic fields involve strong
fluctuations of charged fields that lead to non-equilibrium 
electromagnetic currents\cite{baym,ahonen,enqvist}, much  like the
strong fluctuations in the pion fields during a possible  
supercooled chiral phase transition and the possibility of photon
production associated with these fluctuations\cite{photop2}.   

Thus we see that physically relevant non-equilibrium physical
phenomena are common to cosmology and the quark-gluon plasma and 
chiral phase transition and it has been conjectured that indeed
primordial electromagnetic fields can be generated from strong 
electromagnetic fluctuations at the quark-hadron phase
transition\cite{spergel,olinto}. An important ingredient both in the
quark-gluon plasma as well as in the formation of astrophysical and
cosmological plasmas is a description of the transport properties, 
in particular the screening masses and the  electrical
conductivity. Screening masses are an important ingredient in charmonium 
suppression which is one of the potential probes of the QGP\cite{satz}
and regulate the infrared behavior of transport
coefficients\cite{transport,robinfra}. 

The electrical conductivity plays an important role in the formation and
correlations of primordial magnetic fields in the early universe and contributes to ohmic heating and therefore energy
losses and entropy production in the QGP. The electrical conductivity in the early universe was estimated in\cite{widrow} and (equilibrium) screening corrections were included in\cite{hosoya}. More recently the electrical conductivity of the plasma at temperatures near the electroweak scale was calculated in\cite{heiselberg} including Debye and dynamical (Landau damping) screening
of electric and magnetic interactions.

Hence, there are common relevant problems in cosmology, astrophysics and ultrarelativistic heavy ion collisions that seek a deeper
understanding of the physics of the formation of a plasma beginning from a non-equilibrium initial
state of large energy density, its evolution, the onset of electric and magnetic screening phenomena and the generation of
seeds of bulk electric and magnetic fields, i.e. photon production.  

A first principles description of the formation of a hot plasma  and its dynamical evolution from an initial
state of large energy density beginning from QCD or the Standard Model would be a desirable goal, but clearly an extremely
complicated task. 

\vspace{2mm}
{\bf The goals of this work:}

In this article we study a model that bears many of the important
aspects of QCD and the Standard Model which combined 
with a non-perturbative framework allows us to provide quantitative
and qualitative answers to many of the questions 
associated with the formation and evolution of a non-equilibrium
plasma.   

The model that we propose to study is scalar QED with 
$N$-charged scalar fields coupled to {\bf one} $ U(1) $ photon field and
one neutral scalar field that plays the role of an order parameter for a
phase transition. The model is such that the $U(1)$ local gauge
symmetry associated with the photon field is {\bf not} spontaneously 
broken much in the same manner as the usual electromagnetic field in
the Standard Model. Besides, this model being a suitable framework to
study the questions posed above, we will argue that it is potentially
relevant to the description of photon production during the chiral
phase transition of QCD. Therefore, the dynamics and mechanisms revealed in
this model could prove to be very valuable in the description of the
generation of primordial magnetic fields during one of the QCD phase 
transitions in the early universe and also in photon production during
the chiral phase transitions in heavy ion collisions.  

Furthermore scalar QED has been shown to share many properties of
spinor QED and QCD in leading order in  the hard thermal loop
approximation\cite{rebhan,thoma}, hence the model studied in this
article can serve as a useful and relevant testing ground to study
similar questions in QED and QCD.   

Since the non-equilibrium processes that lead to the formation of the
 plasma are non-perturbative, we resort to the large $ N $ limit as a
 consistent framework to study the non-perturbative dynamics. We take the
electromagnetic coupling to be perturbative and compute 
various quantities, such as the rate of photon production,  magnetic
and Debye masses and the transverse conductivity to leading order in
the large $ N $ limit and to lowest order in the electromagnetic
coupling, discussing the validity of weak coupling in each case.  

The focus of this work centers
on the following aspects: i) The description of the formation of a
non-equilibrium plasma of charged particles during  
a stage of strong non-equilibrium evolution beginning from an initial
state of large energy density. ii) The production 
of photons and therefore of electric and magnetic fields from the
strong fluctuations of the charged fields. This aspect 
is relevant for the formation of primordial magnetic fields in the
early universe and also for photon production during 
non-equilibrium stages for example of the chiral phase transition,
where the charged fields would be the pions.   
iii) The dynamical aspects of electric and magnetic screening. 
We study in detail the magnetic and Debye  
masses and the time scale of the different processes that contribute
to screening. iv) 
The {\em non-equilibrium} transverse electrical conductivity. We
analyze in detail the build-up of conductivity as 
the plasma is forming and its asymptotic limit, comparing to the
equilibrium case.

 In particular two
important situations are studied: {\bf a):}
 a `quenched' (or supercooled) second order phase transition in which
the initial 
state of large energy density is the false vacuum (the quantum state
is localized at the top of the potential). 
The dynamics in this case is described by the process of spinodal
decomposition and phase separation, characterized by 
the exponential growth of long-wavelength unstable fluctuations. These
instabilities and the ensuing large fluctuations of the charged fields
and particle production result in the formation of a non-equilibrium
plasma and the non-perturbative production of photons and therefore of
electric and magnetic fields. The spinodal instabilities are shut-off
by the non-linearities and the resulting  plasma possesses a
non-equilibrium distribution function of  charged scalars
peaked at low momenta.  {\bf b):} The stage of large 
amplitude oscillations of the order parameter around the minimum of
the potential. This stage arises for example {\bf after} a phase transition 
in which the order parameter has rolled down the potential hill and is
oscillating around one of the minima of the 
potential. Such would be the case in the case of the chiral phase
transition where a small explicit symmetry breaking 
term (that gives mass to the pions) will force the isoscalar order
parameter to evolve towards the minimum. This stage 
is characterized by parametric amplification of quantum fluctuations
of the charged fields and again results in non-perturbative production
of charged scalars {\bf and of photons}\cite{photop2}. This stage is
also relevant in cosmology and describes 
the reheating process {\bf after} an inflationary phase transition or
in chaotic inflationary models\cite{inflation}. 
The phenomenon of parametric amplification of quantum fluctuations
during the oscillatory phase of the order parameter, 
the inflaton in the cosmological setting, has been recognized as a
very efficient mechanism of particle  production and reheating in the early
universe\cite{parametric,nuestros}. Parametric amplification of pion
fluctuations  after a supercooled chiral phase transition has also
been recognized to be an important possibility in heavy ion
collisions\cite{mrowmuller}. Both non-equilibrium phenomena are
non-perturbative in the scalar quartic self-coupling. Therefore, the
dynamics in the scalar sector is studied consistently in leading order
in the large $ N $ expansion, while electromagnetic phenomena are
studied to lowest order in $\alpha$. 

Spinodal instabilities  or parametric amplification of quantum
fluctuations of {\bf charged fields} result in the  
formation of a non-equilibrium plasma. In both cases strong
fluctuations in the electromagnetic currents result in 
the production of photons i.e. electric and magnetic fields as well as
screening currents generating screening masses and an
electrical conductivity in the medium.   

Thus, our main objectives are to study the {\bf dynamics} of formation
 of the non-equilibrium plasma, photon production and the power
 spectrum in the generated electric and magnetic fields, the onset of
 electric and magnetic screening phenomena described in real time and
 the build up of conductivity in the medium. Equilibrium aspects of
 hot scalar QED had been previously studied\cite{rebhan,htl} and  we
 will compare the non-equilibrium aspects to the equilibrium case to
 highlight the differences and  similarities. 

\vspace{2mm}

{\bf Results:}  
{\bf a) Photon production:} 

We have derived a consistent kinetic
equation to describe photon production in situations 
strongly out of equilibrium and used this equation to lowest order in
$ \alpha $ (the electromagnetic coupling) and 
leading order in the large $ N $ limit for the charged fields, to
obtain the spectrum of photons produced via 
spinodal and parametric instabilities. In the case of spinodal
instabilities which correspond to the case of a supercooled 
(second order) phase transition we have obtained the power spectrum
and correlation function of the electric and 
magnetic fields generated during the non-equilibrium stage. We find
that there is a {\em dynamical} correlation  
length that grows as $ \xi(t) \sim \sqrt{t} $ at short times. It
determines the spatial correlations of the electromagnetic fields. The
 power spectrum is peaked at long-wavelength with an amplitude
$ \sim \alpha/\lambda^2 $ with $ \lambda $ the quartic self-coupling
of the charged scalar fields. In the case of parametric amplification
the power spectrum peaks near the center of parametric resonance
bands; the amplitude being also $ \sim \alpha/\lambda^2 $ but the electric and 
magnetic fields have small correlation lengths. In the asymptotic
regime  the distribution of produced photons as function of 
frequency $\omega$ behaves as $ \sim \alpha \; m^2 /[\omega^3
\; \lambda^2 ] $. This entails a logarithmically infrared  divergent 
 number of photons but a finite total energy. In the case when the
plasma is generated by spinodal instabilities, the asymptotic photon
distribution continues to grow proportional to $ \log m\tau $  due to
collinear singularities. These behaviors points to the necessity of a
resummation perhaps via the dynamical renormalization group introduced in
reference\cite{rgir}.

{\bf b) Magnetic and Debye screening masses:} We introduce a
definition of the magnetic and Debye screening masses 
out of equilibrium which are the natural extension of that in
equilibrium\cite{kapusta,lebellac,thoma}. We find that the magnetic
mass out of equilibrium {\bf vanishes} at order $ \alpha $ through
cancellations akin to those that take place in equilibrium.
Furthermore, we introduce  an effective  magnetic mass that describes
non-equilibrium screening phenomena for long-wavelength fluctuations
as a function of time and which reveals the different time scales 
of the processes that contribute to the cancellation of the magnetic
mass. Asymptotically for long times and in the long-wavelength limit
we find that processes which are the non-equilibrium counterpart of
Landau damping contribute on time scales which are much longer than
typical production and annihilation processes. 

The extrapolation of this time dependent effective magnetic mass to
the zero momentum limit at finite time 
reveals an unexpected instability in the time evolution of transverse
electromagnetic mean fields during the 
time scales  studied  in this article. This is  a rather
weak instability presumably  
related to photon production although the precise relation is not
clear and deserves further study.  

In the case of spinodal instabilities we find that the (electric
screening) Debye mass at leading order in $ e^2 $ and $ 1/N $ is finite
and given by $ m^2_{Deb} =  8\; |m_R|^2\;  e^2/\lambda + {\cal O}(1) $.
In the case of parametric amplification the Debye mass
grows monotonically with time as $ \sqrt{t} $ times a coefficient of
order $ {\cal O}(e^2/\lambda) $. This result is a consequence of 
non-linear resonances  \cite{destri} which make the charged
particle distribution strongly peaked at small momentum, 
$ {\cal N}_k(t=\infty)\sim  |m_R|^2 /[\lambda \; k^2] $. Since 
the Debye mass is determined by the {\em derivative} of the
distribution function, the singularity at small momentum 
results in a divergent Debye mass for asymptotically long time. 
This  result, valid to first order in $ \alpha $, strongly suggests that 
a resummation of electromagnetic corrections will be required 
in the case of parametric resonance. Such a program lies outside the
scope of this work and will be the subject of a forthcoming study. 

{\bf c) Transverse electric conductivity:} As the plasma of charged
particles forms the medium becomes conducting. We study the 
transverse electrical conductivity from linear response out of
equilibrium (Kubo's conductivity) as a function of time to lowest 
order in the electromagnetic coupling. The
early time behavior during the stage of spinodal or parametric
instabilities results in a rapid build up of the conductivity which 
attains a non-perturbative value $ {\cal O} (\alpha \; m/\lambda) $ 
at the end of this stage. We find that the conductivity is {\bf
finite} for all $ k $ (including $ k = 0 $) at {\bf finite} time. 
Asymptotically at long times, the conductivity attains a form similar to the 
equilibrium case (to lowest order in $\alpha$) but in terms of the
non-equilibrium distribution functions. \\ 

This feature of the asymptotic conductivity must apply to other
physical magnitudes for asymptotic times. Namely, one can compute
their $ t \to \infty $ limit just replacing the thermal occupation
numbers in their equilibrium expression by the out-of-equilibrium
distribution functions. \\

The article is organized as follows: in section II the model is
introduced and the large $ N $ limit is described. In section III we
review the main features of spinodal decomposition and parametric
amplification and introduce the relevant non-equilibrium 
 Green's functions necessary for the calculations. In section IV we
study photon production both during the early stages of the 
instabilities as well as at asymptotically long times, In section V 
we study photon production {\em in equilibrium} to contrast and compare 
to the non-equilibrium results. In section VI 
we study magnetic screening and the magnetic 
mass out of equilibrium. Just as in the equilibrium case in this
abelian theory, we show that the magnetic mass vanishes, but  
point out the different time scales for the processes involved. 
A suitably defined effective magnetic mass describes non-equilibrium
aspects of magnetic screening on intermediate time scales.   Section VII
studies the Debye (electric) screening mass, and it is argued 
that in the case of parametric amplification the Debye mass diverges
because as a result of a singular distribution function for 
the charged scalars at low momentum.
In section VIII we study Kubo's (linear response) transverse electrical
conductivity to lowest order in $\alpha$. In particular we 
focus on the build-up of conductivity during the early stages of
formation of the plasma. We compare the conductivity in the 
asymptotic time regime to the result in equilibrium. 
Our conclusions are summarized in section IX. Here we also
discuss the limit of validity of our studies and the potential 
phenomenological implications of the results from this model. An
Appendix is devoted to a novel kinetic equation that describes 
photon production away from equilibrium. 

\section{The model: SQED with $N$ charged scalars in the large $N$ limit}

We focus on the non-equilibrium dynamics of the formation of  relativistic
quantum plasma at high density after a phase transition, either via
long-wavelength spinodal instabilities in the early stages  
of a rapid (quenched) second order phase transition or by parametric
amplification of quantum fluctuations as the order parameter 
 oscillates
around the equilibrium minimum. Previous work\cite{nuestros,eri97,destri}
 revealed that both types of phenomena are non-perturbative in the
scalar self-coupling, 
hence we propose to use the large $ N $ limit as a consistent tool to
study non-equilibrium phenomena non-perturbatively. Our 
main goals are to provide a quantitative understanding of several
important processes that are of interest both in cosmology as well as in 
the formation of a quark-gluon plasma:  i) non-equilibrium production 
of photons, i.e. the
non-equilibrium generation of electromagnetic fields,
ii) the dynamics of screening
and generation of electric and magnetic masses strongly out of
equilibrium, iii) the
build-up of conductivity in the non-equilibrium plasma.  

We consider a version of scalar quantum electrodynamics with $N$
charged scalar fields  $ \Phi_r $ to
be collectively referred to as pions 
coupled to a neutral field
$ \sigma $ is such a way that the scalar sector of the theory has an 
$ O(2N+1)$ isospin symmetry. The coupling to the electromagnetic field 
reduce this symmetry to an $ SU(N)_{global}\times U(1)_{local} $. 
When we consider the breaking of the isospin symmetry, the neutral 
scalar field $ \sigma $ will acquire an expectation value, but {\bf not}
the charged fields $ \Phi_r $. There are two main reasons 
for this choice a) this  allows  to separate the Higgs
phenomenon and generation of mass for the vector field from truly
non-equilibrium effects and b) we seek to describe a
phenomenologically relevant model, in particular the role of
non-equilibrium 
pion fluctuations during the chiral phase transition wherein
electromagnetism is not spontaneously broken by  chiral symmetry  
breaking.

 The same methods can be used to study the Higgs phenomenon out of
equilibrium and we expect to report on such 
study in the near future. Furthermore, as we seek to describe some
relevant phenomenology for low energy QCD, this model 
describes the large $ N $ limit of the $O(4)$ gauged linear sigma
model that describes the three pions. Electromagnetism 
is unbroken but isospin is broken by the coupling of the charged pions
to electromagnetism and this is captured by the 
model under consideration.

In this abelian theory it is  straightforward to provide a {\em gauge
invariant} description by requiring that the 
set of first class constraints, $ \Pi_0 = 0 \; ; \; \vec{\nabla}\cdot
\vec{E}-\rho=0$  annihilate the  physical states\cite{gauginv,boltzshang}
with $ \Pi_0 $ being the canonical momentum conjugate to the temporal
component of the vector field and $ \vec{\nabla}\cdot \vec{E}-\rho=0 $
is Gauss' law and $ \rho $ is the charge density. This procedure is
described in detail in\cite{gauginv,boltzshang} where it is shown to
be  
equivalent to  a gauge-fixed formulation in Coulomb's gauge. The
instantaneous Coulomb interaction is traded by a  
Lagrange multiplier $ A_0(\vec x,t) $ {\em not} to be confused with the
original temporal component of the gauge field. The issue 
of gauge invariance is an important one because we will study the
distribution function of charged scalar fields and by  
providing a gauge invariant description from the beginning we avoid
potential ambiguities.  

In this formulation we introduce the physical fields 

$$
(\sigma,\Phi_r,\Phi_r^\dagger,\vec{A}_T^i,A_0),\quad r=1,\dots,N
$$
The electromagnetic potential is a physical field which
satisfies the transversality condition
$$
\nabla\cdot \vec A_T=0,
$$
whereas  $A_0$ is the Lagrange multiplier associated with the Gauss' law 
constraint 
$$\nabla\cdot E=-\nabla^2 A_0=\rho.$$
Thus $A_0$ is a non-propagating  field completely specified by the
charge density evolution.

To simplify expressions, we now use the following notations,
$$
\Phi^\dagger\Phi=\sum_{r=1}^N\Phi^\dagger_r\; \Phi_r 
\;,\quad\partial_\mu\Phi^\dagger 
\partial^\mu\Phi=\sum_{r=1}^N\partial_\mu\Phi_r^\dagger\;
\partial^\mu\Phi_r\;,\quad 
\Phi^\dagger\nabla\Phi=\sum_{r=1}^N\Phi_r^\dagger\;\nabla\Phi_r \; ,\quad
\Phi^\dagger\dot\Phi=\sum_{r=1}^N\Phi_r^\dagger\;\dot\Phi_r \;.
$$
With these notations the Lagrangian density is written
\begin{equation}
{\cal L}={\cal L}_1+{\cal L}_2+{\cal L}_3 \label{totallagra}
\end{equation}
with
\be
{\cal L}_1=\frac12\partial_\mu\sigma\; 
\partial^\mu\sigma+\partial_\mu\Phi^\dagger   \;
\partial^\mu\Phi-m^2\left(\frac12\sigma^2+\Phi^\dagger\Phi\right)
-\frac\lambda{2N} 
\left(\frac12\sigma^2+\Phi^\dagger\Phi\right)^2,\label{L1}
\ee
\be
{\cal L}_2=\frac12\partial_\mu\vec A_T\cdot\partial^\mu \vec A_T+\frac12
(\nabla A_0)^2 \label{L2}
\ee
and
\be
{\cal L}_3=-\frac{ie}{\sqrt N}\vec A_T\cdot\left(\Phi^\dagger\nabla\Phi-
\nabla\Phi^\dagger\Phi\right)-
\frac{e^2}N(\vec A_T^2-A_0^2)\;\Phi^\dagger\Phi-\frac{ie}{\sqrt N}\; A_0\left(
\Phi\dot\Phi^\dagger-\Phi^\dagger\dot\Phi\right) \label{L3}
\ee
We have rescaled the couplings with the proper powers of $ N $ so that
$e\; , \lambda$ are fixed in the 
large $ N $ limit. This rescaling allows a consistent identification of
terms as powers of $ 1/N $.  

We focus  on the evolution of initial states with a nonperturbatively
large energy density ( of order $ m^4/\lambda\gg m^4  $ ) in two
different situations: 
{\bf i) $ m^2<0 $}: this case corresponds to a symmetry breaking
potential. We will choose the neutral 
field $\sigma$ to describe the direction of global symmetry breaking, hence
the local gauge symmetry describing electromagnetism 
 {\bf is not} spontaneously broken, i.e. 
$ \langle \Phi \rangle =0 $. A rapid (quenched or 
supercooled) symmetry breaking phase transition can be described by
assuming that $ m^2 $ changes sign suddenly from positive 
describing a symmetric potential to negative describing a symmetry
breaking potential\cite{eri97,destri,boysinglee}. The long-wavelength 
modes become unstable and grow exponentially, this is the process of
spinodal decomposition and the hallmark of phase separation.  

We  emphasize that in the case under consideration the
choice of negative sign {\bf does not} result in the spontaneous
breakdown of the gauge symmetry, since in this model 
 the gauge field {\bf does not} couple to the order parameter
$ \sigma $. Indeed, the global gauge symmetry is 
always unbroken as the charged fields do {\bf not} acquire an
expectation value.  

{\bf ii) $m^2>0 ~, \langle \sigma \rangle(t=0) \approx
m/\sqrt{\lambda}$:} in this case the 
expectation value of the sigma field will oscillate inducing large
parametric amplification of the 
$\Phi$ field. In both cases the quantum fluctuations of the fields
will become non-perturbatively large in the scalar self-coupling and these
will be treated in the leading order in the large $ N $ limit (mean
field)\cite{eri97,destri}.  
The electromagnetic interaction instead, being of order $\alpha$ will only 
give small corrections to the scalar field evolution: thus 
the backreaction of the
gauge field on the evolution of the scalar field will  be
neglected. Therefore to leading order in $ N $ the equations of motion
for the scalar sector are the same as those obtained in refs. \cite{destri} in 
absence of electromagnetic coupling.

Assuming in general that the sigma field acquires a non-equilibrium
expectation value we shift 
\be
\sigma(\vec x,t) = \sqrt{N} \, \varphi(t) + \chi(\vec x,t) \quad ;
\quad \langle  \chi(\vec x,t) \rangle = 0 \label{expecval} 
\ee 
where the expectation value is taken in the time evolved density matrix or
initial state.  
The large $ N $ limit in the
scalar sector can be obtained either by introducing an auxiliary
field\cite{largen} or equivalently in leading order by the 
Hartree-like factorizations\cite{eri97}
\begin{eqnarray}
&&(\Phi^{\dagger}\Phi)^2 \rightarrow 2 \langle
\Phi^{\dagger}\Phi\rangle \Phi^{\dagger}\Phi \label{hartreefac}\\ 
&&\chi \Phi^{\dagger}\Phi \rightarrow \chi \langle
\Phi^{\dagger}\Phi\rangle \label{chifac} 
\end{eqnarray}
The non-linear terms of the $\sigma$ field lead to subleading
contributions in the large $ N $ limit, and to leading 
order the dynamics is completely determined by the $ N $ complex scalars
$\Phi$. The factorization that leads to the 
leading contribution in the large $ N $ limit makes the Lagrangian for
these fields quadratic (in the absence of the gauge 
coupling) at the expense of a self-consistent condition: thus
charged fields $\Phi$ acquire a self-consistent time 
dependent mass. The dynamics is determined
by the equation of motion of $\varphi(t)$ and by 
the Heisenberg equations of the charged fields.  

\section{Spinodal and Parametric Instabilities: summary of main features}

Before we begin our study of non-equilibrium photon production and the
emergence 
of dynamical masses, we review the main features associated with the
non-equilibrium dynamics of 
the scalar fields to provide the physical picture and the basic ideas
upon which we will elaborate  
with the inclusion of the gauge fields. For more details the reader is
referred to\cite{eri97,destri}.  
 As mentioned above the leading order in the large $ N $ limit can be 
obtained by a Hartree-like factorization that turns the Lagrangian
into a quadratic form. The equation 
of motion for the expectation value
$\varphi(t)$ [see eq.(\ref{expecval})] is given by 
\be
\ddot{\varphi}(t)+ m^2\,
\varphi(t)+\frac{\lambda}{2}\,\varphi^3(t)+\frac{\lambda}{N}\langle
\Phi^{\dagger} \Phi\rangle \; \varphi(t)=0\; .
\label{unscaledeqn}
\ee
Introducing the usual decomposition 
\be
\Phi_r(t,\vec x)= \int {d^3 k \over \sqrt{2 \, (2\pi)^3}}
 \left[ a_r(\vec k) \; f_k(t)\;
e^{i\vec k\cdot \vec x}+ b_r^\dagger(\vec k)\; f^*_k(t)\;
e^{-i\vec k\cdot \vec x} \right]\; , \label{phidecompo}
\ee
\be
\Phi_r^\dagger(t,\vec x)=\int {d^3 k \over \sqrt{2 \, (2\pi)^3}}
\left[ b_r(\vec k)\; f_k(t)\;
e^{i\vec k\cdot \vec x}+ a_r^\dagger(\vec k)\; f^*_k(t)\;
e^{-i\vec k\cdot \vec x} \right] \; , \label{phidaggerdecompo}
\ee
\noindent we find that the charged fields obey the Heisenberg
equations if the mode functions $ f_k(t) $ obey the 
following equations of motion\cite{eri97,destri}
\be
\left[\frac{d^2}{dt^2}+k^2+m^2+\frac{\lambda}{2}\varphi^2(t)+
\frac{\lambda}{N}\langle\Phi^{\dagger}\Phi\rangle\right] \;
f_k(t)=0 \label{unsaledeqnsofmot}
\ee
We will choose the initial state to be the  state annihilated by
the $a_r(\vec k),~ b_r(\vec k)$ operators and determined by the 
following initial conditions on the mode functions,
\be 
f_k(0)= \frac{1}{\sqrt{W_k}} \quad ; \quad\dot{f}_k(0)= -iW_k~f_k(0)\; .
\label{iniconds} 
\ee
The frequencies $ W_k $ will be chosen in the particular cases to be
analyzed below. This choice of initial state with the 
initial conditions given by (\ref{iniconds}) corresponds to the vacuum
of the Fock quanta of oscillators of frequencies $ W_k $.  
This initial state can be generalized straightforwardly to a thermal
density matrix, but the main physical mechanisms can be 
highlighted in a simpler manner by the choice of this state. With
this choice one finds 
\be
\frac{\lambda}{N}\langle\Phi^{\dagger} \Phi\rangle =
\frac{\lambda}{4}\int \frac{d^3k}{(2\pi)^3}\;|f_k(t)|^2 
\ee
This expectation value is ultraviolet divergent, therefore the renormalization
must be carried out consistently in terms of mass and coupling 
constant and the reader is referred to \cite{eri97,destri} for details. 

It proves convenient to introduce  dimensionless variables in terms of
the renormalized mass and coupling 
\be
\tau=|m_R| \; t\quad,\quad  q={ k \over |m_R|}\quad,\quad
g=\frac{\lambda_R}{8\pi^2}\quad, \quad \Omega_q =
\frac{W_k}{|m_R|}\label{dimensionless1} 
\ee
\be
\eta^2(\tau)=
\frac{\lambda_R}{2|m_R|^2}\; \varphi^2(t)\quad,\quad
\varphi_q(\tau)=|m_R|^\frac12 f_k(t)\quad . \label{dimensionless2}
\ee
and the subtracted self-consistent self-energy\cite{eri97,destri}
\be
g\Sigma(\tau)=g\int_0^{\infty} q^2\; dq
        \Biggr\{|\varphi_q(\tau)|^2-|\varphi_q(0)|^2
+\frac{\Theta(q-1)}{2q^3}\biggl[-\eta^2(0)+\eta(\tau)^2+
g\Sigma(\tau)\biggr]
        \Biggr\}. \label{gsigma}
\ee

From now on we set the only dimensional variable in the problem
$ |m_R|\equiv 1 $ and all dimensionful quantities will be in units of
$ |m_R| $. 

To leading order in the large $ N $ limit the dynamics is
completely determined by the  following equations of
motion\cite{eri97,destri} 
\be
\ddot\eta(\tau)\pm \eta(\tau)+\eta^3(\tau)+g\Sigma(\tau)\;\eta(\tau)=0\;,
\label{zeromodeeqn} 
\ee

\begin{equation}
\left[\frac{d^2}{d\tau^2}\pm 1+q^2+\eta^2(\tau)+g\Sigma(\tau)
\right]\varphi_q(\tau)=0\;.
\label{modeeq}
\end{equation}

Two different cases correspond to the different signs in the evolution
equations above.  

The negative sign is associated with tree level potentials that allow
global $ O(2N+1) $ broken symmetric  ground states, whereas the positive
sign determines a potential with a symmetric minimum. 
 As it will be discussed in detail below, the
non-equilibrium dynamics in the {\em broken symmetry} 
case is described for early times by the process of spinodal
decomposition and phase 
ordering and triggered by long-wavelength instabilities 
just as in a typical second order phase transition during a rapid
quench through the critical temperature\cite{boysinglee}.  

For positive sign, the physical situation that we want to
describe is the case when  the order parameter has an initial 
value corresponding to a large amplitude $ \varphi (t=0) $ of order
$ {\cal O} (m_R/\sqrt{\lambda_R}) $ i.e. $ \eta(0) = {\cal O} (1) $ [see
eq. (\ref{dimensionless2})]. 
The subsequent  non-equilibrium evolution of the order parameter is described
in terms of large amplitude oscillations around the minimum of the
potential. This situation would describe the dynamics  
{\em after} the phase transition when the order parameter has rolled
down the potential hill and undergoes large amplitude 
oscillations near the minimum. In cosmology this situation also
describes the period of reheating in  
chaotic scenarios\cite{parametric,nuestros}. As can be seen from the
equation of motion (\ref{modeeq}) the effective mass 
for the charged field modes oscillates in time leading to parametric
amplification\cite{parametric,nuestros,eri97}.  

In this case the phenomenon is that of energy transfer from the `zero
mode' i.e. from the expectation value of the order parameter to the
modes with non-zero wavevectors as a consequence of parametric
amplification of quantum fluctuations.  

Thus, the physics is very different between the two cases and the only
feature in common is that either through the 
growth of long-wavelength fluctuations via spinodal instabilities or
the growth of fluctuations via parametric 
amplification the ensuing non-equilibrium dynamics results in the
production of a dense plasma of charged particles strongly 
out of equilibrium. 

The initial conditions on the  order parameter (condensate) are chosen to be
\be
\eta(0)=\eta_0\quad,\quad \dot\eta(0)=0\;, \label{inicondzeromode}
\ee
and the initial conditions  on the mode functions are [see
eqs.(\ref{iniconds}) and (\ref{dimensionless2})] 
\begin{equation} \label{bc}
\varphi_q(0)=\frac1{\sqrt{\Omega_q}}\quad,\quad
\dot\varphi_q(0)=-i\sqrt{\Omega_q} \; ,
\end{equation}
where the dimensionless frequencies $\Omega_q$ will be determined in
each particular case below.

\subsection{Broken symmetry: spinodal instabilities} 

Consider the case in which the system is undergoing a sudden phase
transition out of equilibrium  
from an initial disordered state at large temperature
very rapidly to almost zero temperature, i.e. a
quenched phase transition with a vanishing order
parameter\cite{eri97,boysinglee}. For $ \tau>0 $ the equations of motion
are those for a broken symmetry 
case with the $ (-) $ sign in eq.(\ref{modeeq}) with $
\eta(0) \ll 1 ~,\dot{\eta}(0)=0 $. For simplicity, we shall consider the case
$ \eta(0) = 0 $ which entails $ \eta(\tau)\equiv 0 $. 

Furthermore, we see that for very weak coupling and early times i.e. when
the back reaction from the term $ g\Sigma(\tau) $ in  (\ref{modeeq}) 
can be neglected, there is a band of {\em spinodally unstable}
wave-vectors $0\leq q\leq 1$. The modes in this unstable band 
will grow exponentially initially. Because we are describing an
initial condition corresponding to a sudden quench, we impose 
the initial condition that at the initial time the mode functions
describe particles of the stable phase, i.e., we choose the 
initial frequencies for the modes in the unstable band to be given
by\cite{boysinglee,eri97,destri}  
\be
\Omega_q = \sqrt{q^2+1} \quad \mbox{for} \quad q^2<1 \label{unstfreqs}
\ee

the short wavelength modes are not affected by the sudden quench and we choose
\be
\Omega_q = \sqrt{q^2-1} \quad \mbox{for} \quad q^2>1 \label{stablefreqs}
\ee
However, we emphasize that detailed numerical studies reveal that the
dynamics is not very sensitive to the choice 
of the initial frequencies for weak coupling\cite{eri97,destri}. 

The important feature is that this initial state has
non-perturbatively large energy density, of 
order $ |m_R|^4/\lambda_R $ as compared to the broken symmetry vacuum
state, for which $ |\eta| =1 $.  

As discussed in refs.\cite{eri97,destri} the ensuing dynamics is
strongly out of equilibrium. The modes with 
wavevectors in the unstable band begin growing exponentially and their
contribution to the self-consistent expectation value  
$ g\Sigma(\tau) $ causes it to grow exponentially. This instability is
the hallmark of the process of phase separation and determines the
emergence of correlated regions\cite{boysinglee,eri97,destri}:  
these are the familiar spinodal instabilities associated with the 
process of phase separation and phase ordering. The contribution of
these unstable modes to $ g\Sigma(\tau) $ dominates 
the early time dynamics and when $ g\Sigma(\tau) $ becomes of $ {\cal
O}(1) $ and  competes with the tree level term  
$ (-1) $ in the evolution equations for the mode functions
(\ref{modeeq}) these instabilities shut-off through the backreaction. 
This defines a new {\bf dynamical} time scale that determines the
onset of full non-linear evolution and is estimated to
be\cite{eri97,destri} 
\be \label{tnolin}
\tau_{NL} = \frac{1}{2}\ln\left[\frac{1}{g}\sqrt{\frac{8}{\pi}}\right]
+ {\cal O}(\ln |\ln g| )
\ee
Thus, two different regimes emerge: 

i) the early time regime for
$\tau \leq \tau_{NL}$ in which the back-reaction can be neglected and
the evolution of the mode functions is essentially linear and
dominated by the spinodally unstable wave-vectors for which the mode functions
grow exponentially (linear instabilities). 

ii) the late time regime for $\tau \geq \tau_{NL}$ for which the
effective mass squared $ {\cal M}^2(\tau) = -1+g\Sigma(\tau) +
\eta^2(\tau) $ tends to zero and the mode functions become effectively
massless\cite{destri}.    

\subsubsection{Early time regime} 

For $\tau \leq \tau_{NL}$ and weak coupling, the effects of the
back-reaction can be neglected and the mode functions obey 
a linear equation of motion. Whereas the modes outside the spinodally
unstable band oscillate and their amplitudes remain bound in time, 
those in the unstable band grow exponentially. For the case $ \eta_0
\ll 1 $ we can neglect 
at early times both the quantum fluctuations $ g \Sigma(\tau) $ and $
\eta^2(\tau)  $  in the mode equations (\ref{modeeq}). 
The explicit solution is thus\cite{destri} 

\begin{equation}
\varphi_q(\tau)=\alpha_q\ \exp\left(\tau\sqrt{1-q^2}\right)+\alpha_q^*\ 
\exp\left(-\tau\sqrt{1-q^2}\right)
\label{early.broken}
\end{equation}
where the coefficient $ \alpha_q $ is determined by the initial
conditions (\ref{bc}), i.e. 
$$ \varphi_q(0)=(1+q^2)^{-1/4}\quad ,\quad \dot\varphi_q(0)=-i(1+q^2)^{1/4},$$
we find
\be
\alpha_q=\frac{\sqrt{1-q^2}-i\sqrt{1+q^2}}{2\sqrt{1-q^2}\;(1+q^2)^{1/4}}. \label{alfas}
\ee

A feature of the solution (\ref{early.broken}) with (\ref{alfas}) that
will become important  is 
that when the exponentially damped solution becomes negligible as
compared to the exponentially growing one, the phase of the mode
functions $\varphi_q(\tau)$ {\em freezes}, i.e. 
becomes constant in time and is a slowly varying function of $q$ for
long wavelengths.  

\subsubsection{Late time regime}

For times $ \tau > \tau_{NL} $ the effective mass term ${\cal M}^2(\tau) =-1+ g\Sigma(\tau)$ vanishes leading to the
sum rule\cite{eri97,destri}
\be
g\Sigma(\infty) =1 \label{brokesymsumrule}
\ee
and the mode functions obey a massless wave equation. The asymptotic solutions are given by\cite{destri}
\begin{equation}\label{asimodoR}
\varphi_q(\tau)=A_q\; e^{i q \tau}+B_q\; e^{-i q\tau} 
\end{equation}
where the coefficients $A_q~, B_q$ are both non-vanishing because  the
Wronskian is constant and determined by the initial conditions
\be
-2i =W[\varphi_q,\varphi^*_q] =
\dot{\varphi}_q(\tau)\varphi^*_q(\tau)-\varphi_q(\tau)\dot{\varphi}^*_q(\tau)=
-2iq\left[|B_q|^2-|A_q|^2\right] \label{wronsk} 
\ee
leading to the important result
\be
|B_q|^2-|A_q|^2 = \frac{1}{q}\label{wronskbro}
\ee
Furthermore, the  sum rule (\ref{brokesymsumrule})
is asymptotically dominated by the modes in the unstable band 
\be
g\Sigma(\tau) \buildrel{\tau \to \infty}\over= g \int_0^1 q^2 dq
\left[|A_q|^2+|B_q|^2\right] + 
\mbox{oscillating~terms}\label{sumru} 
\ee
where the oscillating terms vanish as $ 1/\tau $. We
conclude\cite{destri} that for the modes in the unstable band
\be
|A_q|^2 = {\cal O}(1/g)= |B_q|^2\;,\quad 0<q<1 \label{nonpertbrok}
\ee
determining that $ A_q~, B_q $ are both  of $ {\cal O}(1/\sqrt{g}) $
whereas $ A_q~, B_q $ are of order one elsewhere.

The following sum rules arise from a) the vanishing 
of the effective mass and b) conservation of energy\cite{destri}

\begin{eqnarray}
\int^1_0 q^2\; dq \; |A_q|^2 = \frac{1}{2g}+{\cal O}(1) \label{sumrule1} \\
\int^1_0 q^4 \; dq\; |A_q|^2 = \frac{1}{8g}+{\cal O}(1)\label{sumrule2}
\end{eqnarray}
Furthermore,  the small $q$ behavior of 
$A_q$ and $B_q$ is given by\cite{destri},
\begin{equation}\label{smallq}
A_q  \buildrel{q \to 0}\over= -\frac{i}{2q}\left[ K + i q L + {\cal
O}(q^2) \right] \; , \; B_q  \buildrel{q \to 0}\over=
\frac{i}{2q}\left[ K - i q L + {\cal O}(q^2) \right]
\end{equation} 
where Im$L {\bar K} =1 $ according to the Wronskian condition\cite{destri}.

The non-zero coefficient $ K $ is determined by  the linear growth in
time of the mode $ \varphi_{q=0}(\tau) $ in this case with broken
symmetry\cite{destri}. For small coupling it is found numerically to be given by\cite{destri},
\begin{equation}\label{faseK}
K = K_+ - i K_- \quad \mbox{where} \quad K_{\pm} = \frac{1}{\sqrt{g}}\left\{1.1\ldots 
 \mp 0.003\ldots\; {g} + {\cal O}(g^{2})\right\}~ .
\end{equation} 
This asymptotic behavior for small momentum will prove to be important
for a quantitative analysis of the magnetic mass.

\subsection{Unbroken symmetry: parametric amplification}
In the unbroken symmetry case, corresponding to the choice of the plus
 sign in the equations of motion 
 (\ref{zeromodeeqn})-(\ref{modeeq}) the frequencies $\Omega_q$ are chosen
 to be\cite{destri} 
\be
\Omega_q = \sqrt{q^2+1+\eta^2(0)}\label{inifrequnbro}
\ee
and the  initial condition for the dimensionless order parameter is
chosen to be 
\be 
\eta(0) \equiv\eta_0 = {\cal O}(1)\quad ; \quad \dot{\eta}(0)=0
\label{unbrokeninicondzeromode} 
\ee
In this case the `zero mode' (expectation value) $ \eta(\tau) $
oscillates around the minimum of the potential resulting in an
oscillatory time dependent mass term for the modes $ \varphi_q(\tau) $.  

\subsubsection{Early time regime}

Neglecting the backreaction of the fluctuations, an oscillatory 
time dependent mass leads to parametric amplification of the mode
functions which are Floquet solutions. 
These solutions are characterized by parametric instability bands. 

For weak coupling the early time behavior of $ \eta(\tau) $ and the mode
functions $ \varphi_q(\tau) $ can be found by neglecting 
the backreaction terms in the equations of motion
(\ref{zeromodeeqn})-(\ref{modeeq}) for the unbroken symmetry case. The  
equation for the zero mode with the initial conditions
(\ref{unbrokeninicondzeromode}) has as solution a simple elliptic  
function\cite{destri}. Inserting this elliptic function,  the evolution
equation  for each mode $ \varphi_q(\tau) $ becomes a  Lam\'e equation that 
can be analytically solved in terms of Jacobi theta functions, the details
are given in \cite{destri}. The important  
feature is that this Lam\'e equation has {\em only} one band of
parametric instability for real $ q $. The unstable band  
corresponds to  wavevectors\cite{destri}
\be
0< q < \frac{\eta_0}{\sqrt{2}}. \label{unstablebandunbroken}
\ee
The modes in the unstable band grow exponentially in time, whereas
those in the stable region $\eta_0/\sqrt{2}<q<\infty$ oscillate 
in time with constant amplitude.

The explicit solution with boundary conditions (\ref{bc}) for the mode
functions in the unstable band is given by
\be
\varphi_q(\tau)=\alpha_q\ U_q(-\tau)+\alpha_q^*\ U_q(\tau) \label{earlyunbro}
\ee
with the Floquet solution $U_q(\tau)$ given in \cite{destri}. 

With the choice of frequencies (\ref{inifrequnbro}) the coefficient $\alpha_q$ is
found to be given by  
\be
\alpha_q=\frac1{2\sqrt{\Omega_q}}\left(1-\frac{2i\Omega_q}{W_q}\right) ~~; ~~
W_q=-2q\ \sqrt\frac{\eta_0^2/2+1+q^2}{\eta_0^2/2-q^2}.\label{alfaqunbro}
\ee
The Floquet solutions $ U_q(\tau) $ are derived in detail in\cite{destri}
and depend on the initial condition $ \eta_0 $ through the nome 
 $ \hat q(\eta_0) $. Since in this case $ \hat q(\eta_0) <
e^{-\pi}=0.0432139...$ for any initial condition $\eta_0$\cite{destri}, we can
express  $ \hat q(\eta_0) $ by the excellent approximation
\be
\hat q(\eta_0)=\frac12\frac{(1+\eta_0^2)^{1/4}-(1+\eta_0^2/2)^{1/4}}
{(1+\eta_0^2)^{1/4}+(1+\eta_0^2/2)^{1/4}}\label{nome}
\ee 
with an error smaller than $\sim 10^{-7}$\cite{destri}. In addition
we can use the approximation $ \hat q(\eta_0)\ll1 $ and the Floquet solutions 
simplify in this limit to
\be
U_q(-\tau)=e^{B_q \tau}\frac{\sin\left(\pi v_q-
\sqrt{1+\eta_0^2}\ \tau\right)}{\sin{\pi v_q}} +{\cal O}({\hat q})
\label{growfloquet}
\ee
with Floquet index 
\be
B_q=4\sqrt{1+\eta_0^2}\ \hat q(\eta_0)\ \sin{2\pi v_q}+{\cal O}({\hat
q}^2)\quad,\quad \sin{\pi v_q}=\sqrt{1-\frac2{\eta_0^2}q^2}+{\cal O}({\hat
q})\quad,\quad \cos{\pi v_q}= \frac{\sqrt2}{\eta_0}q+{\cal O}({\hat
q})\quad . \label{floquetindex} 
\ee 
Therefore the backreation $ g\Sigma(\tau) $ grows exponentially at
early times because of the parametric instabilities. The exponential 
envelope of the backreation term is given by\cite{destri} 
\be
g\Sigma(\tau) = \frac{g}{\tilde{N}\sqrt{\tau}} \; e^{\tilde{B}\;\tau}
\ee
where $ \tilde{N} $ and $ \tilde{B} $ can be found in\cite{destri}. When the
backreaction competes with the tree level term, i.e. 
$ g\Sigma(\tau) \approx 1+\eta^2_0/2 $ the full nonlinearities must be
taken into account, this equality determines the 
{\em non-linear} time scale $ \tau_{NL} $ given by\cite{destri}
\be \label{tnolinp}
\tau_{NL} \approx
\frac{1}{\tilde{B}}\ln\left[\frac{\tilde{N}(1+\eta^2_0/2)}{g\sqrt{\tilde{B}}}
\right] 
\ee 
Detailed analytic and numerical
studies in\cite{destri} reveal that most of the particle production
occurs during the time interval $\tau \leq \tau_{NL}$. 

For $ \tau_{NL}>\tau > 1 $ the modulus squared of the mode functions
$ |\varphi_q(\tau)|^2 $ is peaked at the value of 
$ q $ at which the Floquet index is maximum, this value is given
by\cite{destri} 
\be
q^*= \frac{1}{2}\; \eta_0\left[1-\hat{q}(\eta_0)\right]+{\cal O}({\hat
q}^2) \label{maxfloquet}
\ee

\subsubsection{Late time regime}

The parametrically resonant band, $ 0 \leq q \leq \eta_0/\sqrt2 $, is
shut-off by the non-linearities 
[the term $  g\Sigma(\tau) $] for times $ \tau \gtrsim \tau_{NL} $. 
Two non-linear resonant bands appear in this regime. 
One near $ q = 0 $ and the other just below $ q = \eta_0/\sqrt2 $.
The width of these nonlinear resonances diminishes in time. We have
for the  non-linear resonant bands\cite{destri} 
\be
0<q^2<\frac{K_1}{\tau} \quad \mbox{and} \quad
\frac{\eta^2_0}{2}-\frac{K_2}{\tau}< q^2 <\frac{\eta^2_0}{2}
\label{bandas} 
\ee
(with $K_1,K_2$ determined in ref.\cite{destri})
and the phase space for these  small resonant regions becomes increasingly
smaller at late times. 

Asymptotically, the effective mass oscillates around the  constant
value\cite{destri}  
\be
{\cal M}^2(\infty) = 1+g\Sigma(\infty) =
1+\frac{\eta^2_0}{2}\label{asymassunbro} 
\ee
and the mode functions can be written as
\be
\varphi_q(\tau)=A_q(\tau)\; e^{i \omega_q \tau}+B_q(\tau)\; e^{-i
\omega_q\tau} ~~;~~ \omega_q = \sqrt{q^2+{\cal M}^2(\infty)}  
\label{asymodounbro}
\end{equation}
where the amplitudes $ A_q(\tau) $ and $ B_q(\tau) $ depend  on the
slow time scale $ \tau/\tau_{NL} $ for $ \tau > \tau_{NL} $ and are
defined by\cite{destri} 
\begin{eqnarray}
&&A_q(\tau)  =  \frac{1}{2} e^{-i\omega_q\tau}\left[\varphi_q(\tau)-i\;
\frac{\dot{\varphi}_q(\tau)}{\omega_q} \right] \label{aofq} \\  
&&B_q(\tau)  =  \frac{1}{2} e^{+i\omega_q\tau}\left[\varphi_q(\tau)+i\;
\frac{\dot{\varphi}_q(\tau)}{\omega_q} \right] \label{bofq}  
\end{eqnarray} 
These amplitudes vary slowly in time and in particular
$\omega_q|A_q(\tau)|^2$ is identified with the number of asymptotic
particles of mass $ {\cal M}^2(\infty) $\cite{destri}.  

For wavevectors inside the small bands of {\em non-linear resonances}
(\ref{bandas}), these amplitudes grow with a power law\cite{destri}, whereas   
 the modes outside from these  resonant regions oscillate
with constant amplitude. The fact that the width 
of these resonances diminishes at longer times is a consequence of the
non-linearities. A very important 
consequence is that asymptotically for all modes with $ q \neq
0, \; \eta_0/\sqrt{2} $ 
\be
\lim_{\tau \rightarrow \infty}~ A_q(\tau) = A_q \quad , \quad
\lim_{\tau \rightarrow \infty}~ B_q(\tau) = B_q 
\ee
where $ A_q $ and $ B_q $ are constants.
 
Hence, the mode functions with $q\neq 0~,~ \eta_0/\sqrt{2}$ 
asymptotically behave as
\be\label{asimodo}
\varphi_q(\tau)\buildrel{\tau\gg 1}\over=A_q\; e^{i \omega_q
\tau}+B_q\; e^{-i \omega_q\tau} 
\ee
Asymptotically the constancy of the Wronskian leads to
\be
|B_q|^2 - |A_q|^2= \frac{1}{\omega_q} \label{wronskunbro}
\ee
Furthermore, for the modes with wavectors in the resonant band
$0<q<\eta_0/\sqrt{2}$,
\be
|A_q| = {\cal O}\left( \frac{1}{\sqrt{g}} \right) = |B_q| 
\ee
Just as in the broken symmetry case, there are two 
important sum rules 
as a result of the asymptotic value of the effective mass and of
conservation of energy. In this case these read\cite{destri}
\begin{equation}\label{reglaS}
\int_0^{\eta_0/\sqrt2} dq\; q^2\;
|A_q|^2=\frac1{4g}\;\eta_0^2+O\left(g^0\right), 
\quad \int_0^{\eta_0/\sqrt2} dq\; q^4\;  |A_q|^2=\frac1{32 
g}\;\eta_0^4+O\left(g^0\right) \; ,
\end{equation}

\subsection{Formation of the plasma}

\begin{figure}
\centerline{ \epsfig{file=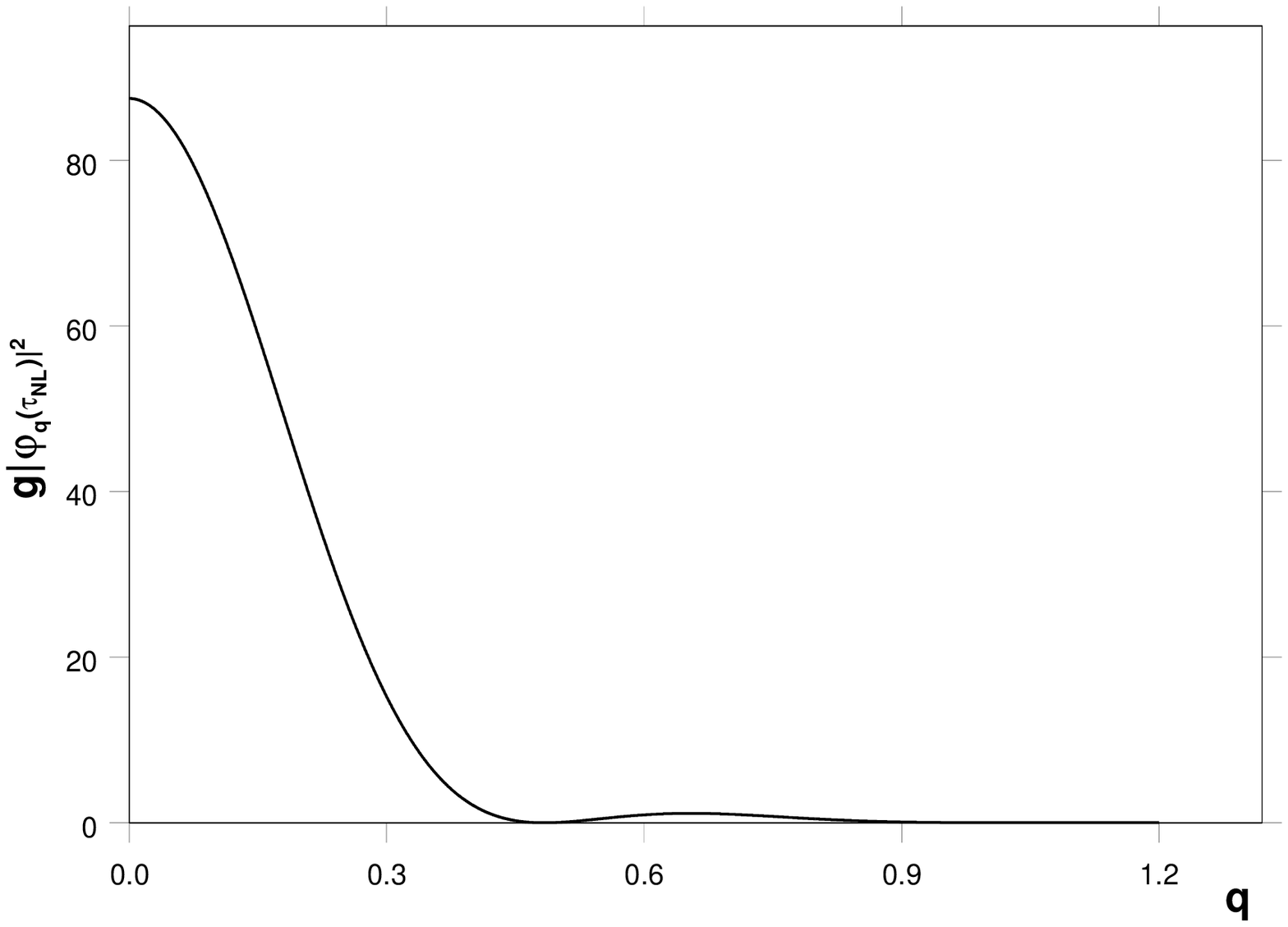,width=7in,height=8in}}
\caption{$g|\varphi_q(\tau=\tau_{NL})|^2$ for broken symmetry for
$ g=10^{-4} $ and $ \eta_0 = 0 $.  \label{modebroken}} 
\end{figure}

\begin{figure}
\centerline{ \epsfig{file=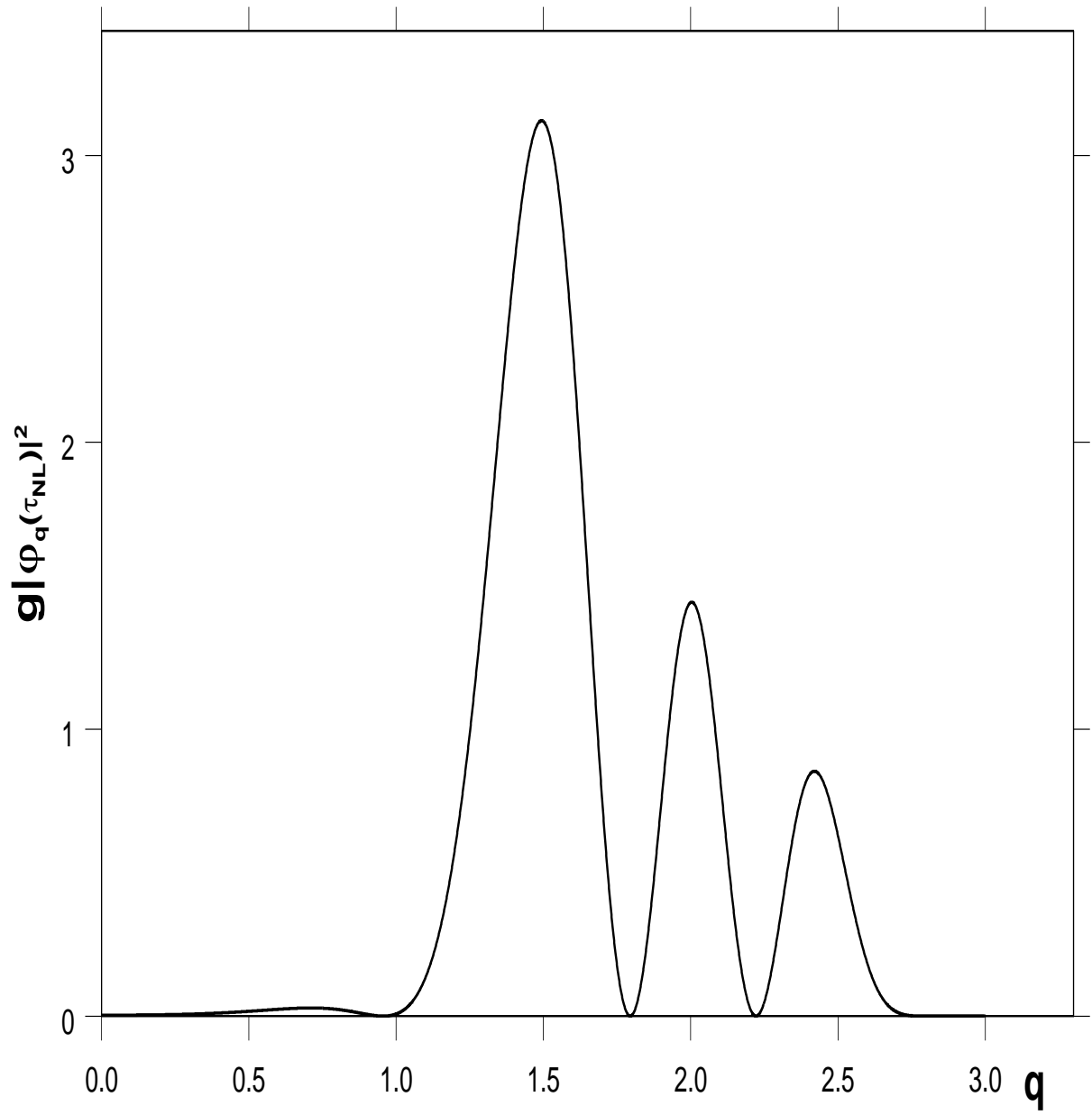,width=7in,height=8in}}
\caption{$g|\varphi_q(\tau=\tau_{NL})|^2$ for unbroken symmetry for
$g=10^{-4}~~;~~\eta_0=4.0$.  \label{modeunbroken}} 
\end{figure}

The main conceptual feature that emerges from the summary above is
that in both situations, broken or unbroken symmetry,  
spinodal or parametric instabilities lead to profuse particle
production. The particles that are produced are charged 
scalars, these are produced in pairs of total zero momentum, and the
distribution of produced particles is localized in 
the region of instabilities. In the case of broken symmetry the
distribution is peaked in the region $0\leq q \leq 1$ and 
in the case of unbroken symmetry in the region
$ 0\leq q \leq \eta_0/\sqrt{2} $. In
both cases the amplitude of the mode functions in these 
regions become $ |\varphi_q(\tau)|^2 = {\cal O}(1/g) $ i.e. non-perturbatively
large. This amplitude is associated with the 
number of particles created\cite{eri97,destri} (see below) and
therefore we conclude that during the period of spinodal 
or parametric instabilities $ 0<\tau\leq \tau_{NL} $ a {\bf dense plasma
of charged particles is formed} as a result of these 
instabilities. This plasma is neutral and is described by the
distribution functions of the particles, which is proportional to 
$ |\varphi_q(\tau)|^2 $\cite{eri97,destri} (see below) and is clearly a
non-equilibrium distribution in the sense that it cannot 
be described by a thermal distribution at some temperature. These
distribution functions had been obtained numerically
in\cite{eri97,destri}. Figs. \ref{modebroken} and
\ref{modeunbroken} display  $ g|\varphi_q(\tau_{NL})|^2 $ for the
broken and unbroken symmetry   
cases respectively, it is clear that the square of the mode
functions become of order $ {\cal O}(1/g) $ at $ \tau \approx
\tau_{NL} $ for wavevectors in  the unstable bands.   

Furthermore the distribution of particles continues to evolve for
$\tau > \tau_{NL}$ and this evolution is more marked in the unbroken
symmetry case.  

We now turn to the description of the electromagnetic properties of
this non-equilibrium plasma of charged particles  for 
which we need the non-equilibrium Green's functions of the charged
scalar fields.

\subsection{Basic ingredients: real time non-equilibrium Green's functions}
The proper description of real time non-equilibrium evolution is in
terms of the time evolution of 
an initial density matrix. A formulation in terms of a path integral
along a complex contour in time 
allows to use the familiar tools of quantum field theory to study
non-equilibrium phenomena. In this, the  Schwinger-Keldysh or CTP (closed time
path) formulation\cite{ctp,eri97}, the essential ingredients 
are the non-equilibrium Green's functions. In particular there are four 
possibile two-point functions, denoted by indices  $ (a,b)\;\in\{+,-\} $ 
which  correspond to the evolution along the forward 
and backward time branches. 
\begin{itemize}

\item{Transverse photon propagators} 

Since photons will be treated perturbatively, we need the bare photon 
Green's functions. Furthermore we will consider that the initial state
is the photon vacuum. Therefore 
the relevant real time Green's functions for transverse photons are given by
$$
{\langle}{A}^{(a)}_{Ti}(\vec{x},t){A}^{(b)}_{Tj}(\vec{x^\prime}, 
t^{\prime}){\rangle}=-i\int {d^3k\over{(2\pi)^3}}\;{\cal G}_{ij}^{ab} 
(k;t,t^\prime)\;e^{-i\vec{k}\cdot(\vec{x}-\vec{x^\prime})}\;, 
$$
where the explicit form of ${\cal G}_{ij}^{ab} 
(k;t,t^\prime)$ is
\begin{eqnarray} 
&&{\cal G}_{ij}^{++}(k;t,t^{\prime})={\cal P}_{ij}(\vec{k}) \;
\left[{\cal G}_k^{>}(t,t^{\prime})\Theta(t-t^{\prime}) 
+{\cal G}_k^{<}(t,t^{\prime})\Theta(t^{\prime}-t) \right]\;, \label{phot++}\\ 
&&{\cal G}_{ij}^{--}(k;t,t^\prime)= {\cal P}_{ij}(\vec{k}) \;
\left[{\cal G}_k^{>}(t,t^{\prime})\Theta(t^{\prime}-t) 
+{\cal G}_k^{<}(t,t^{\prime})\Theta(t-t^{\prime}) \right]\;, 
\label{phot--}\\ 
&&{\cal G}_{ij}^{\pm}(k;t,t^\prime)={\cal P}_{ij}(\vec{k}) \;
{\cal G}_k^{<}(t,t^{\prime})\;; \; {\cal G}_{ij}^{\mp}(k;t,t^\prime)={\cal P}_{ij}(\vec{k}) \;
{\cal G}_k^{>}(t,t^{\prime}) \label{photpm}
\end{eqnarray} 
and ${\cal P}_{ij}(\vec{k})$ is the transverse projection operator,
\begin{equation} 
{\cal P}_{ij}(\vec{k})=\delta_{ij}-\frac{k_ik_j}{k^2}\;.\label{projector} 
\end{equation}  
At tree level
\begin{eqnarray} 
&&{\cal G}_k^{>}(t,t^{\prime})=\frac{i}{2k}\;
e^{-ik(t-t^\prime)}\;,\label{phot>} \\ 
&&{\cal G}_k^{<}(t,t^{\prime})=\frac{i}{2k}\;e^{ik(t-t^{\prime})}\; .
\label{phot<}
\end{eqnarray}

\item{Scalar propagators} 

The scalar propagators are truly non-equilibrium and can be written in
the general form
$$
{\langle}{\Phi_r}^{(a)\dagger}(\vec{x},t){\Phi_s}^{(b)}(\vec{x^\prime}, 
t^{\prime}){\rangle}=-i\delta_{rs}\int {d^3k\over{(2\pi)^3}}\; G_k^{ab}
(t,t^\prime) \; e^{-i\vec{k}\cdot(\vec{x}-\vec{x^\prime})}\;, 
$$
where $ (a,b)\;\in\{+,-\},\ r,s=1,\dots, N $.  With the field
expansion given by (\ref{phidecompo})-(\ref{phidaggerdecompo}) 
in terms of the (dimensionful)
mode functions $ f_k(t) $ obeying the equations of motion
(\ref{unsaledeqnsofmot}) we obtain in the $ N = \infty $
limit,

\begin{eqnarray} 
&&G_k^{++}(t,t^\prime)=G_k^{>}(t,t^{\prime})\Theta(t-t^{\prime}) 
+G_k^{<}(t,t^{\prime})\Theta(t^{\prime}-t)\; , \label{gplpl} 
\\ 
&&G_k^{--}(t,t^\prime)= G_k^{>}(t,t^{\prime})\Theta(t^{\prime}-t)+ 
G_k^{<}(t,t^{\prime})\Theta(t-t^{\prime})\;, \nonumber
\\ 
&&G_k^{+-}(t,t^\prime)=G_k^{<}(t,t^{\prime})~~; ~~
G_k^{-+}(t,t^\prime)=G_k^{>}(t,t^{\prime}), 
\label{gplmin}\\ 
&&G_k^{>}(t,t^{\prime})=\frac i2f_k(t)f^*_k(t^\prime)\; 
,\label{greater} 
\\
&&G_k^{<}(t,t^{\prime})=\frac i2f_k(t^\prime)f^*_k(t)\; .\label{lesser} 
\end{eqnarray} 

\end{itemize}

An important property of the mode functions in the asymptotic region
will allow us to establish a correspondence between 
the non-equilibrium results to be obtained below and the more familiar
equilibrium results. In both cases, broken or unbroken 
symmetry, after the non-linear time scale the mode functions become
those of a free field theory [see eqs.(\ref{asimodoR}) and (\ref{asimodo})]. 
The Wronskian conditions (\ref{wronskbro}) and (\ref{wronskunbro}) allows
to write the modulus of the coefficients $ A_q~, B_q $ in  the form
\be\label{nfuera}
|B_q|^2 = \frac{1}{\omega_q}\;[1+{\cal N}_q] ~~; ~~ |A_q|^2 =
\frac{1}{\omega_q}\; {\cal N}_q \label{occupation} 
\ee 
with $\omega_q= q$ for broken symmetry or $\omega_q=
\sqrt{q^2+{\cal M}^2(\infty)}$ for unbroken symmetry. 

$ {\cal N}_q $  describes the distribution of asymptotic  charged scalar
particles  created during the rapid non-equilibrium stages of spinodal
decomposition or parametric instabilities\cite{eri97,destri}. 
These are {\bf non-equilibrium} distribution functions, a result of profuse 
particle production during the stage of spinodal instabilities or
parametric amplification. The number of charged scalars produced 
during these stages is non-perturbatively large, since for the wave vectors
in the unstable bands $ {\cal N}_q $ is of order $ 1/g $.

The asymptotic behavior of the functions $ G_k(t,t') $ when both time
arguments are in the asymptotic region, much larger than the
non-linear time scale can be written in the illuminating form 
\bea
&&G_k^>(t,t') \buildrel{\tau,\tau' \gg 1}\over= \frac{i}{2\omega_k}\left[
\left( 1+{\cal N}_k \right) \; e^{-i\omega_k(t-t')}+ 
{\cal N}_k  \;e^{i\omega_k(t-t')}\right] + i\, Re\left[A_k \; B^*_k \;
e^{i\omega_k(t+t')} \right] ~~, \cr \cr
&&G_k^<(t,t')= G_k^>(t',t) \label{asygreen} 
\eea
\noindent where the mixing terms proportional to
$e^{\pm i\omega_k(t+t')}$ are a signal that the non-equilibrium
behavior remains in the asymptotic region.  
In most circumstances these rapidly varying oscillatory terms lead to
contributions that vanish very rapidly by dephasing.  

The first two terms of the  function (\ref{asygreen}), that depend on
the difference of the time arguments can be compared 
to that of a free field theory {\em in equilibrium}

\be
G^>_{k,equil}(t,t') = \frac{i}{2\omega_k}\left\{ \left[1+n_k\right] \; 
e^{-i\omega_k(t-t')} + n_k \; e^{i\omega_k(t-t')}\right\}  ~~; ~~
G_{k,equil}^<(t,t')= G_k^>(t',t) \label{equilgreen} 
\ee 

\noindent where $ n_k $ is the thermal distribution function. Thus
we see that the part of the asymptotic non-equilibrium Green's
functions that depend on the difference of the time arguments has the
form  of the free field {\em equilibrium} Green's functions 
but in terms of the {\em non-equilibrium} distribution functions
${\cal N}_k$. This formal similarity will allow us to compare the 
non-equilibrium results in the asymptotic regime to those more
familiar in equilibrium field theory and to interpret the different
processes in the medium. 

It is convenient to summarize the main features that will be
responsible for the phenomena studied below.
\begin{itemize}
\item{In either case, spinodal instabilities in the case of the
potential allowing broken symmetry states or 
parametric instabilities associated with the oscillatory evolution of
the order parameter in the case of 
unbroken symmetry, there are strong fluctuations that lead to
non-perturbative particle production of the order $ {\cal O}(1/g) $. At times 
larger than the non-linear time, when the effects of the backreaction
become of the same order as the tree  
level terms, the occupation number of modes in the unstable band is
non-perturbatively large [$ {\cal O}(1/g) $]. The state of the 
system can be best characterized as a {\em non-equilibrium} dense
plasma. The distribution function of the 
created particles is not an equilibrium one and has a finite limit for
infinite time.} 
\item{As a result of non-perturbative particle production, the Green's
functions of the scalar fields, determined 
by the mode functions $ f_k(t) $, are those of a {\em plasma} strongly
out of equilibrium and will provide a  
{\em non-perturbative} contribution to the photon polarization. }
\item{To leading order in the large $ N $ limit and to lowest order in
$ \alpha=e^2/4\pi $, the photon polarization is 
given by the diagrams shown in fig. \ref{loops}a,b.

The loop is in
terms of the full scalar propagator in the leading 
order in the large $ N $ limit, which receives contributions from the
mean-field and background as depicted in fig. \ref{loops}c.} 
\end{itemize}

We now have all of the ingredients to study the electromagnetic
signatures of these  non-perturbative phenomena to leading order in
the large $ N $ limit and to lowest order in $ \alpha=e^2/4\pi $.

We emphasize again that these phenomena have nothing to do with the
ordinary Higgs mechanism. In this  model  
the global gauge symmetry is {\bf not} spontaneously broken by the
initial state even when the potential for 
the scalar fields allows for broken symmetry and remains unbroken
throughout the dynamical evolution.  

\section{Photon production via spinodal and parametric instabilities}

We study the production of photons both via the spinodal
instabilities associated with the process of 
phase ordering (in the case of broken symmetry potentials) and via the 
parametric instabilities associated with the non-equilibrium evolution of the
order parameter around the symmetric minimum (in the case of unbroken 
symmetry potentials). As described in the
previous section, we carry out this study to leading 
order in the large $ N $ limit and to lowest order in
the electromagnetic coupling. This is similar 
to the formulation in\cite{toimela,rusk} for the rate of photon
production in the QGP to all orders in $\alpha_s$ and to 
lowest order in $\alpha_{em}$. 

However, our  approach differs fundamentally from the usual approach in the 
literature\cite{toimela,rusk,kapustagale,lichard,bdrs}, which relies
on the computation of the photoproduction {\em rate} from processes
that satisfy energy conservation, i.e. on shell. This results from the
use of Fermi's Golden Rule in the computation of a transition
probability from a state in the far past to another state in the far future. 

Instead our computation relies on obtaining the integrated photon number
at {\bf finite time} $t$ from the time evolution of an  {\em initial
state} at $t_0$. Clearly this approach is 
more appropriate in out-of-equilibrium situations where transient,
time-dependent phenomena are relevant. 

Non-equilibrium time dependent transient phenomena cannot be captured
by the usual rate calculation based on Fermi's Golden 
Rule, since such calculation will obtain  the  number of  produced
photons divided by the total time 
$ t $ in the limit when $ t \to \infty $. This definition is insensitive
to the non-energy conserving processes, which are subleading in the long time
limit but could dominate at finite time, and could potentially lead to
grossly disparate estimates of the total number of 
photons produced in a situation in which a plasma has a finite
lifetime as is the case in heavy ion collisions.  

To lowest order in $ \alpha $ and leading order in the large $ N $ 
the leading process giving rise to photoproduction is the {\em off-shell}
production of a pair of charged pions and one-photon from the initial
vacuum strongly out equilibrium. Thus, we consider the transition
amplitude for the process $ |\tilde{\bf 0}>\to|\pi^+\pi^-\gamma> $ to
order $ e $, more precisely the amplitude  
for the interaction to create a pair of scalars with momentum $ \vec
q+\vec k $ and $ \vec q $  respectively and a photon of momentum $
\vec k $ and polarization $ \lambda $. The initial state $ |\tilde{\bf
0}> $  at time $t_0$ is the Fock vacuum for pions and photons but its
evolution is non-trivial because it is {\em not} an eigenstate of 
the Hamiltonian, nor is it perturbatively close to an eigenstate. In
the case of spinodal instabilities this state is unstable, and it
decays via the production of pions and photons. In the case of
parametric instabilities, this state  involves a dynamical expectation value
for the $\sigma$ field with non-perturbatively large amplitude (i.e, $
\eta_0 \sim 1 $). 

The lowest order contribution to this amplitude in the electromagnetic
coupling is given by   
\be
{\cal A}_{q,k,\lambda}(t)=
<\pi^+\pi^-\gamma| \; i\int_{t_0}^t dt_1\; d^3 x\;  \vec J(t_1,\vec x)\cdot 
\vec A_T(t_1,\vec x)|\tilde{\bf 0}> \label{amplitude}
\ee
where $ \vec J(t_1,\vec x) $ is the electromagnetic current 
\be
\vec J(t_1,\vec x)=\frac{ie}{\sqrt N}\sum_{r=1}^N
\left(\Phi_r^\dagger\nabla\Phi_r-\nabla\Phi_r^\dagger\Phi_r\right)
\label{current} 
\ee 
If the (transverse) photon field is expanded in terms of creation and
annihilation operators of Fock quanta associated with 
the vacuum at the initial time
\be
\vec A_T(t,\vec x)=\sum_{\lambda=1,2}\int {d^3 k \over \sqrt{2 k\, (2\pi)^3}}
\left[\vec\epsilon_\lambda(\vec k) \;
e^{-i k t+i\vec k\cdot \vec x} \;a_\lambda(\vec k)+
\left(\vec\epsilon_\lambda(\vec k) \; e^{-ikt+i\vec k\cdot \vec x}\right)^* \;
a_\lambda^\dagger(k)\right]\label{photonfield}  
\ee
and the scalar fields are expanded as in
eqs.(\ref{phidecompo})-(\ref{phidaggerdecompo}),  
we find the amplitude to be given by 
\be
{\cal A}_{q,k,\lambda}(t)=\frac{e}{\sqrt{2 N k}}\int_{t_0}^t
dt_1\ \vec\epsilon_\lambda(\vec k)\cdot\vec q_T\;
e^{ik t_1}\ f_{q}^*(t_1)\;f_{|\vec q+\vec k|}^*(t_1).
\ee

Squaring the amplitude, summing over $q$ and $r$ and $\lambda$ and using
$$\sum_{\lambda=1}^2\epsilon_\lambda^i(\vec k)\;\epsilon_\lambda^j(\vec k)=
{\cal P}^{ij}(\vec k)
$$ 
we finally obtain that the total number of photons of momentum $ k $ produced
at time $ t $ per unit volume from the initial vacuum state at time
$ t_0 $ is given by  
\begin{equation}\label{ph.spectrum}
N_{ph}(k,t)=(2\pi)^3\frac{d^6 N(t)}{d^3x \, d^3 k}=\frac{e^2}{2k}\int
\frac{d^3q}{(2\pi)^3} \; q^2 \; (1-\cos^2\theta)\left|\int_{t_0}^t
f_q(t_1) \; f_{|\vec q+\vec k|}(t_1)\; e^{-i k
t_1}\;dt_1\right|^2  
\end{equation}
where $\theta$ is the angle between $\vec q$ and $\vec k$,
$\vec q\cdot\vec k=q\; k\cos\theta$. 
The same formula can be obtained as a particular case  of the
generalized kinetic equation for the photon distribution function
obtained in Appendix A. We refer the reader to this Appendix for a
more detailed discussion of the kinetic equation and its regime of validity. 

We point out that if the mode
functions $ f_q(t) $ are replaced with the usual exponentials $
\exp(-i\omega_q\; t)/\sqrt{\omega_q}$, and the limits $ t_0\to-\infty $
and $t\to\infty$  are taken, the familiar energy-conservation Dirac 
delta function is recovered and therefore 
the process is kinematically forbidden {\em in the vacuum}. 
Furthermore, the discussion in 
the previous section highlighted that during the stage of spinodal
instabilities or parametric amplification, 
the mode functions in the unstable bands grow exponentially. Hence 
the modes in the unstable band
will lead to an explosive production of 
photons during these early stages. Clearly the
maximum production of photons will occur in 
the region of soft momenta, with the wavevector $k$ in the unstable
bands. In this manner the scalar mode functions 
 with wavevectors $\vec q$ and $\vec{q}+\vec k$ will be in the
unstable bands leading to four powers of the exponential 
growth factor. Thus we will focus on the production of soft photons
studying the case of broken symmetry (spinodal  
instabilities) and unbroken symmetry (parametric instabilities)
separately. Having recognized the emergence of a dynamical time scale
$$
  \tau_{NL}\sim\ln\left(\frac{1}{g}\right)
$$
[see eqs.(\ref{tnolin}) and (\ref{tnolinp}) for more detailed expressions]
that separates the linear from the non-linear behavior, we analyze
both regions $ \tau< \tau_{NL} $ and $ \tau > \tau_{NL} $ separately.
 
\subsection{Photon production via spinodal instabilities:}

\subsubsection{ $\tau < \tau_{NL}$}

The  number of produced photons of wavelength $ k $ per unit volume at
time $ t $ is given by eq.(\ref{ph.spectrum}). Obviously the integrals
in this expression can be computed 
numerically\cite{photop2} since the mode functions are known
numerically with high precision \cite{destri}. However, the summary of
properties of mode functions for $ \tau\leq \tau_{NL} $ and 
$ \tau > \tau_{NL} $ provided in the previous section allows us to furnish an
{\em analytical} reliable estimate for the photon 
production. During the early, linear stages, we can insert the
expression for the mode functions given by
eqs.(\ref{early.broken})-(\ref{alfas}). Furthermore, we focus on 
small $ k $ so that $ \vec q $ and $ \vec q + \vec k $ are 
in the spinodally unstable bands and keep only the exponentially
increasing terms which dominate the integral at intermediate times.  
The time integral can now be performed and we find (using dimensionless units)
\be
 N_{ph}(k,\tau)=\frac{e^2}{2k}\int \frac{d^3q}{(2\pi)^3}\; q^2\;
(1-\cos^2\theta)\;  |\alpha_q \; \alpha_{q+k}|^2\ 
\frac{\exp\left[2\; \tau \left(\sqrt{1-q^2}+\sqrt{1-(\vec q+\vec k)^2}
\right)\right]} {\left[\sqrt{1-q^2}+\sqrt{1-(\vec q+\vec
k)^2}\right]^2+k^2}. \label{photonumberspino} 
\ee

Furthermore,  the dominant contribution to the $q-$integral arises
 from the small $q$ region justifying  the non-relativistic
approximation $q\ll1, q+k\ll1$. Hence $ N_{ph}(k,\tau) $ becomes
\begin{equation}\label{interm}
\frac1{16}\frac{e^2}{2k} e^{\tau\;(4-k^2)}\int_0^\infty\frac{dq}{(2\pi)^2}
\int_{-1}^1 dx \;  q^4 \;(1-x^2)\;e^{-2\tau(q^2+q k x)}
\end{equation}

\begin{figure}[t]
  \begin{center}
  \mbox{\epsfig{file=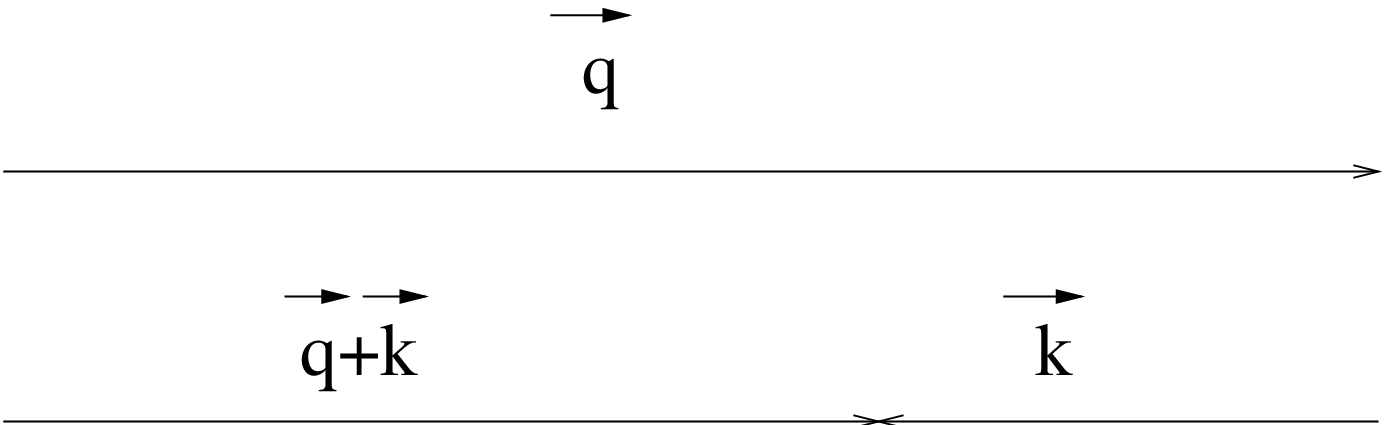,width=7in,height=3in}}
  \end{center}
  \caption{{{\small Kinematical configuration corresponding to saddle
  point for photoproduction in the broken symmetry case}}\label{collinear}}
\end{figure}

For
$ k^2 \tau \gg 1 $ we can use the approximation
$$
\int_{-1}^1 dx\ (1-x^2)\ e^{-A x}= 2e^A\left[\frac1{A^2}+{\cal O}(1/A^3)\right]
$$
to perform the angular integration. Notice that the dominant 
region corresponds to $ x=-1 $. That is $ \vec q $ and $ \vec k $ in
opposite directions. Physically, 
this corresponds to two charged scalars with parallel momenta $ \vec q $
and $ \vec q - \vec k $ emitting a collinear photon with momentum $ -\vec k $
(see fig. \ref{collinear}).

In  this regime  the photon spectrum becomes
\be
 N_{ph}(k,t)=\frac1{64\pi^2}\frac{e^2}{2k} \; e^{\tau(4-k^2)}\int_0^\infty dq\
\frac1{(2q k \tau)^2}\ 2 q^4 \ e^{2\tau q k-2\tau q^2}
\ee
where the factor $1/(k\tau)^2$ arises from the angular integration.

Now it is possible to compute the momentum integral via a saddle point
approximation. Using the saddle point $ q=\bar q \equiv k/2 $ we
obtain  
\begin{equation} \label{broken.spectrum}
N_{ph}(k,\tau)=\frac{N_1 \; e^2}{\tau^{5/2}k}\ e^{\tau[4-k^2/2+{\cal O}(k^4)]}+
{\cal O}(1/\tau)
\end{equation}
where the proportionality factor $N_1$ is given by
$$
N_1=\frac{\sqrt{2}}{2048\, \pi^{3/2}}.
$$
We see that for  $\tau\leq \tau_{NL}$ the
number of produced photons grows exponentially with time. The
production is mostly abundant for soft photons $ k \ll 1$.
However, the derivation of (\ref{broken.spectrum}) only holds in the region 
in which the saddle point expansion is reliable, i.e. for $ k^2 \tau
\gg 1 $. 

The $ k\to 0 $ limit can be studied directly. In such case
the angular integration in eq.(\ref{interm}) is straightforward and the momentum
integration can be done using the result 
$$
\int_0^\infty dq\; q^4 \exp(-2\tau
q^2)=\frac3{64}\frac{\sqrt{2\pi}}{\tau^{5/2}}\; . 
$$ 
leading to the same result as eq.(\ref{broken.spectrum}), 
$$
N_{ph}(k,\tau) \buildrel{k \to 0}\over=\frac{N_1 \;
e^2}{\tau^{5/2}\; k}\ e^{4\tau}+ {\cal O}(1/\tau)
$$

We thus find an exponentially growing number of emitted photons (as $
\sim e^{4\,\tau} $) for $ \tau < \tau_{NL} $. Since $  e^{4\tau_{NL}}
\sim g^{-2} $, we see that the total number of emitted photons at
$ \tau\approx \tau_{NL} $  is of the order
\be
N_{ph}(k,t_{NL}) \sim \frac{ e^2}{k\; g^2}\label{nphotonNL}
\ee
and is predominantly peaked at very low momentum as a consequence of
the fact that the long-wavelength fluctuations are  growing exponentially as a consequence of the spinodal instability.
 The power spectrum for the electric and magnetic
fields produced during the stage of spinodal 
growth of fluctuations is
\be
\langle |E(k,\tau)|^2 \rangle  \approx \langle |B(k,\tau)|^2 \rangle
\approx k \; N_{ph}(k,\tau)\; .  \label{energy.density}
\ee
Two important results can be inferred for the generation of electric
and magnetic fields 
\begin{itemize}
\item{ At the spinodal time scale $\tau \approx \tau_{NL}$ the power
spectrum is localized at small momenta and with amplitude $ \sim \alpha/g^2$.} 
\item{ Taking the spatial Fourier transform at a fixed given time we
can obtain the correlation length of the generated 
electric and magnetic fields. A straightforward calculation for $\tau
\leq \tau_{NL}$ using eq. (\ref{broken.spectrum}) reveals that 
\be
\langle \vec E(\vec r,\tau) \cdot \vec E(\vec 0,\tau)\rangle \sim \langle \vec
B(\vec r,\tau) \cdot \vec B(\vec 0,\tau) \rangle\sim
 e^{-\frac{r^2}{\xi(\tau)^2}} ~~; ~~ \xi(\tau) \sim \sqrt{\tau} \label{EMcorre}
\ee
The {\em dynamical} (dimensionful) correlation length $\xi(\tau) \sim
\sqrt{\tau}$ is the same as that for the scalar fields before 
the onset of the full non-linear
regime\cite{eri97,destri}. Therefore, at early and intermediate
times the generated electric and 
magnetic fields track the domain formation process of the scalar
fields and reach an amplitude $\sim \alpha/g^2$ at time 
scales $\tau \approx \tau_{NL}$ over length scales $ \approx |m_R|^{-1}
\left[ \ln(1/g)\right]^{1/2}$.}  

\end{itemize}

\subsubsection{$ \tau >  \tau_{NL} $}
We now split the time integral in eq.(\ref{ph.spectrum}) into two
pieces, one from $ 0 $ up  
to $ \tau_{NL} $ and a second one from $ \tau_{NL} $ up to $ \tau $. In the
first region we  use the exponentially growing modes as in the
evaluation above, and in the second region we use the asymptotic form
of the mode functions given by eq.(\ref{asimodo}). The time integral in this
second region can now be performed explicitly and we find  
\begin{eqnarray}\label{gordan}
 N_{ph}(k,\tau) &=& \frac{e^2}{8 \, \pi^2 \; k}\int_0^1 q^4\, dq \int_{-1}^{+1}
 dx \; (1-x^2) \cr \cr 
&&\left| \int_0^{\tau_{NL}} d\tau_1 \; \varphi_q(\tau_1)\;
 \varphi_{|\vec q+\vec k|}(\tau_1) \; e^{-ik \tau_1} +
A_q \, A_{|\vec q+\vec k|} \, { e^{i\left( q +
 |\vec q+\vec k| - k \right)\tau} - e^{i\left( q +
 |\vec q+\vec k| - k \right)\tau_{NL}} \over q + |\vec q+\vec k| - k }
\right. \cr \cr 
&+& A_q \, B_{|\vec q+\vec k|} \, { e^{i\left( q -
 |\vec q+\vec k| - k \right)\tau} - e^{i\left( q -
 |\vec q+\vec k| - k \right)\tau_{NL}} \over q - |\vec q+\vec k| - k }
 - B_q \, A_{|\vec q+\vec k|} \, { e^{-i\left( q -
 |\vec q+\vec k| + k \right)\tau} -  e^{-i\left( q -
 |\vec q+\vec k| + k \right)\tau_{NL}} \over q - |\vec q+\vec k| + k }
\cr \cr 
&-&\left.  B_q \, B_{|\vec q+\vec k|} \, { e^{-i\left( q +
 |\vec q+\vec k| + k \right)\tau} - e^{-i\left( q +
 |\vec q+\vec k| + k \right)\tau_{NL}}\over q + |\vec q+\vec k| + k }
\right|^2\left[ 1 + {\cal O}\left( g \right)\right] \; .
\end{eqnarray}
where $|\vec q+\vec k| = \sqrt{ q^2 + k^2 + 2\,k\,q\,x } $. 
The momentum integration is restricted to the region of the spinodally
unstable band since only in this 
region the modes acquire non-perturbatively large amplitudes. The integration over $ q > 1 $ only provides perturbative
 $ {\cal O}( g ) $ corrections. 

The contribution of the asymptotic region $\tau \gg \tau_{NL}$ in
eq.(\ref{gordan}) displays potentially resonant 
denominators. As long as the time argument  $ \tau $ remains finite
the integral is finite, but in the limit of large $ \tau \gg \tau_{NL} $ 
the resonant denominators can lead to secular divergences. In the long
time limit we can separate the terms that lead to potential 
secular divergences from those that remain finite at all times. Close
inspection of eq.(\ref{gordan}) shows that  
asymptotically for large time  the square modulus
of the second, third and fourth terms yield potential secular
divergences. The    
square modulus of  the last term is always bound in time and
oscillates since the denominator never vanishes. In addition, the
cross 
terms either have finite limits or are subdominant for $ \tau \to
\infty $. The square modulus of the first term is 
 $ N_{ph}(k,t_{NL}) $  given by eq. (\ref{nphotonNL}). In order to
recognize the different contributions and to 
establish a relationship with the equilibrium case it proves useful to
use the definitions given in eq. (\ref{occupation}). We find 
the following explicit expression for the
dominant contributions asymptotically at late times, 
\begin{eqnarray}\label{gorda2}
&& N_{ph}(k,\tau) \buildrel{\tau\gg  1}\over= N_{ph}(k,\tau_{NL})+
 \frac{e^2}{4 \, \pi^2 \; 
 k}\int_0^1 \frac{q^4\, dq}{q~|\vec q +\vec k|} \int_{-1}^{+1} 
 dx \; (1-x^2) \; \times   \cr \cr
&& \left\{ {\cal N}_q \, {\cal N}_{|\vec q+\vec k|} ~~{1  - 
 \cos{\left[\left( q + |\vec q+\vec k| - k \right)
\left(\tau-\tau_{NL}\right)\right]}  
 \over \left( q + |\vec q+\vec k| - k \right)^2} \right. \cr\cr
&&+ {\cal N}_q \, \left[1+{\cal N}_{|\vec q+\vec k|}\right]~~
 {1 - \cos{\left[\left( q
 - |\vec q+\vec k| - k \right)\left(\tau-\tau_{NL}\right)\right]}   
\over \left( q - |\vec q+\vec
 k| - k\right)^2 } \cr \cr 
&& \left. +  \left[1+{\cal N}_q\right] \, {\cal N}_{|\vec q+\vec k|}~~
 {1 - \cos{\left[\left( q - |\vec q+\vec
 k| + k \right)\left(\tau-\tau_{NL}\right)\right]} 
  \over \left( q - |\vec q+\vec k| + k
 \right)^2}\right\} \left[ 1 + {\cal O}\left( g
 \right)\right] + {\cal O}\left(\tau^0\right) \!  \! \! \!
\end{eqnarray}

The first term, containing the factor $ {\cal N}_q \, {\cal
N}_{|\vec q+\vec k|}$, corresponds to the 
process $\pi^+\pi^- \rightarrow \gamma$, i.e. massless charged scalar
annihilation into a photon, the second and third 
terms (which are equivalent upon  re-labelling $\vec q \rightarrow
-\vec q-\vec k$) correspond to bremsstrahlung contributions 
in the medium, $\pi^\pm \rightarrow \pi^\pm + \gamma$. 

The following relation 
$$
1-x^2 = - {1 \over 4 \, q^2  \,k^2}
\left( q + |\vec q+\vec k| - k \right)\left( q - |\vec q+\vec
 k| + k\right) \left( q - |\vec q+\vec  k| - k\right)
\left( q + |\vec q+\vec k| + k\right) 
$$
ensures that there are only simple poles in the integrand of eq.(\ref{gorda2}).
 
Asymptotically for long time the integrals in
eq.(\ref{gordan})-(\ref{gorda2}) have the typical structure\cite{rgir} 
\begin{equation}\label{lnsecu}
\int_0^{\infty} { dy \over y} \left( 1 - \cos yt \right) \; p(y)
\buildrel{t \to 
\infty}\over= p(0) \log \left[\mu e^{\gamma}\, t \right] + 
\int_0^{\infty} { dy \over y} \left[  p(y) -  p(0) \; \theta(\mu - y )
\right] + {\cal O}\left( {1 \over t } \right)  \; ,
\end{equation}
where $ p(y) $ is a continuous function, $ \mu $ an arbitrary scale and
and $ \gamma = 0.5772157 \ldots $ is the Euler-Mascheroni constant.
Notice that the expression (\ref{lnsecu}) does not depend on the scale $\mu$,
as can be easily seen by computing its derivative with respect to $\mu$.

 Therefore the simple poles arising  from the collinear singularities
 translate in logarithmic secular terms appearing for late times
 according  to eq.(\ref{lnsecu}). 

The  denominators in eqs.(\ref{gordan})-(\ref{gorda2}) vanish leading to 
collinear singularities,  i.e. kinematical configurations where the photon
and a charged particle have parallel or antiparallel momentum. More
precisely, the denominators in eq.(\ref{gorda2}) vanish  at the
following points:
$$
|\vec q+\vec k| = k-q \; , \; q-k\;  \mbox{and} \; q+k\; ,
$$
 corresponding  to $ \cos\theta =x = -1,  \; -1 $ and $
 +1  $, respectively.  

It is convenient to perform the angular integration using the variable
$ \xi \equiv |\vec q+\vec k|$ with $dx = { \xi\; d\xi \over q \, k} $.
Since the most relevant contribution 
arises from the region of momenta inside the spinodally unstable band
with $ {\cal N}_q = {\cal O}(1/g) \gg 1 $ 
the angular integration simplifies and we  find 
\begin{equation}\label{nftinf}
 N_{ph}(k,\tau) \buildrel{\tau \gg 1}\over=\frac{e^2}{2 \, \pi \; 
 k^3}\; \log \mu \, \tau \; \int_0^1 q\, dq  \; {\cal N}_q  
\left[|q-k| \; {\cal N}_{|q - k|} + (q+k)\; {\cal N}_{q+k}  \right]
\left[ 1 + {\cal O}\left(g \right)\right] + {\cal O}\left( \tau^0 \right)
\end{equation}
Here $ | q- k| $ stands for the absolute value of the difference
between the numbers $ q $ and $ k $. If we restore dimensions and
we recall that $ {\cal N}_q $ is of order $ {\cal O}(1/g) $ for $ 0<q<1 $,
we find  that the logarithmic term has a coefficient $ \sim e^2
|m_R|^3/[g^2\, k^3] $. This remark will become useful when we compare later 
to a similar logarithmic behavior in the case where the
scalars are in thermal equilibrium (sec. VIII).  

These  logarithmic infrared divergences lead to
logarithmic secular terms much in the same manner as in ref.\cite{rgir} and
indicate an obvious breakdown of the perturbative expansion. They must
be resummed to obtain consistently the real time evolution of the
photon distribution function. The dynamical renormalization group 
program introduced in ref.\cite{rgir} provides a consistent
framework to study this resummation. 

A similar logarithmic behavior of the occupation number has been found
in a kinetic description near equilibrium in 
the hard thermal loop approximation\cite{htl}. 

Furthermore, we note that the evolution equation for the photon
distribution function under consideration has neglected 
the build-up of population of photons, and therefore has neglected the
inverse processes, such as charged-scalar production 
from photons and inverse bremsstrahlung. These processes can be
incorporated by considering the full kinetic equation described in 
Appendix A. Hence a consistent program to establish the production of
photons beyond the linear regime must i) include the 
inverse processes in the kinetic description and ii) provide a
consistent resummation of the secular terms. We postpone the study 
of photon production in the asymptotic regime including these
non-linear effects to a forthcoming article.  

\subsection{Photon production via parametric amplification}
We now study the process of photon production during the stage of
oscillation of the order parameter 
around the minimum of the tree level potential in the unbroken
symmetry case. This case corresponds to the  
 evolution equations (\ref{zeromodeeqn})-(\ref{modeeq}) with the plus
sign and  with the initial conditions
(\ref{inifrequnbro})-(\ref{unbrokeninicondzeromode}). We begin by
studying the early time regime.  

\subsubsection{$\tau < \tau_{NL}$}

The dominant contribution to the production of photons again arises
from the exponentially growing terms in 
the parametrically unstable band. Hence we keep only the exponentially
growing Floquet solution (\ref{growfloquet})  
with Floquet index given by (\ref{floquetindex}). 

In order to perform the time integration we focus  on the
exponentially increasing terms and  neglect the oscillatory contributions  in the product
$$
\sin\left(\pi v_q-
\sqrt{1+\eta_0^2}\ \tau\right)\sin\left(\pi v_{q+k}-
\sqrt{1+\eta_0^2}\ \tau\right)=\sin^2\left(\pi v_q-
\sqrt{1+\eta_0^2}\ \tau\right)+{\cal O}(k)$$
$$=\frac12+{\cal O}(k)+\mbox{oscillatory terms}.
$$
In keeping only the exponentially growing contribution and neglecting
 the oscillatory parts we  evaluate the envelope 
 of the number of photons averaging over the fast oscillations. 

\begin{figure}[t]
  \begin{center}
  \centerline{\epsfig{file=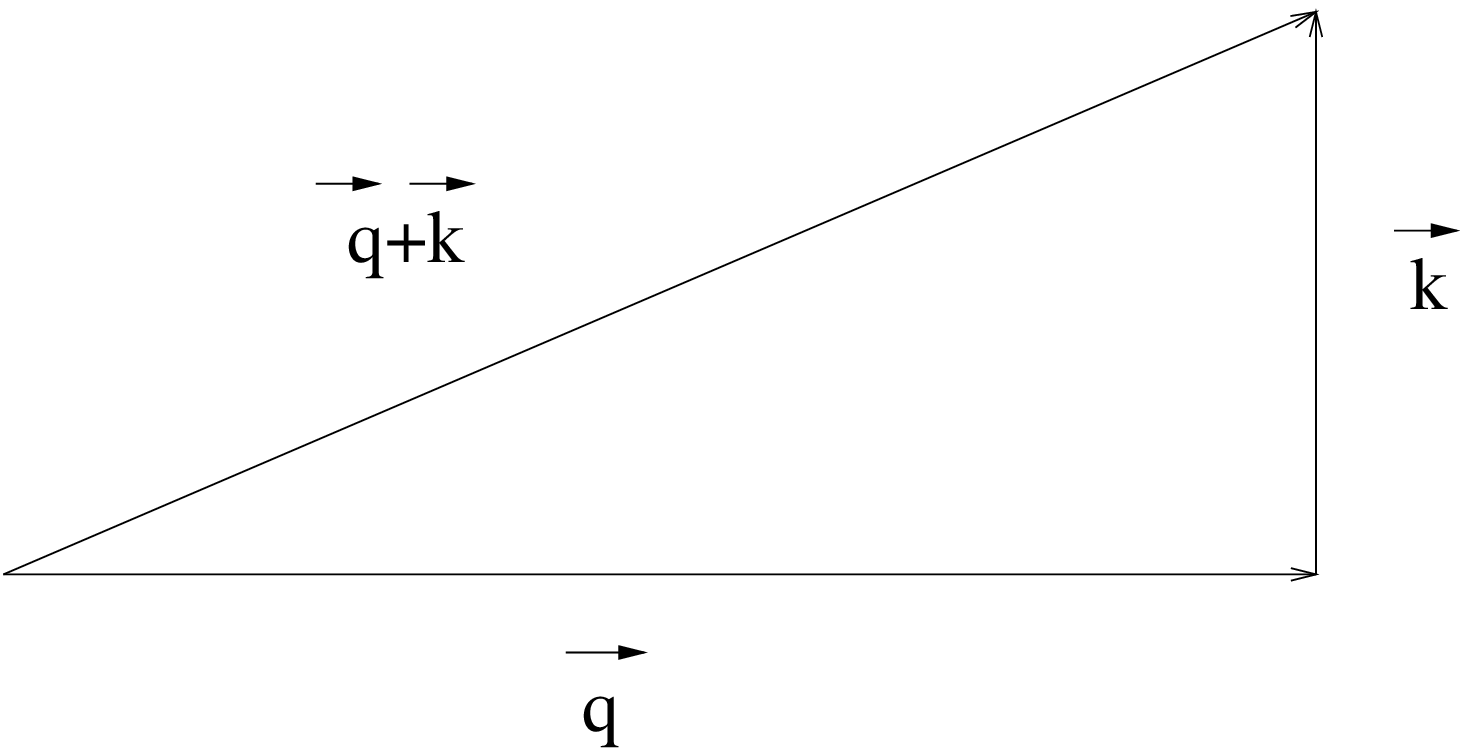,width=7in,height=3in}}
  \end{center}
  \caption{{{\small Kinematical configuration corresponding to saddle
  point for photoproduction in the unbroken symmetry case}}\label{perpendicular}}
\end{figure}

With these considerations we now have to evaluate the following
integral according to eq.(\ref{ph.spectrum}), for  $ \tau \gg 1 $, but $
\tau\leq\tau_{NL} $ 
\begin{equation} \label{ph.spectrum.II}
N_{ph}(k,t)=\frac{e^2}{8k}\frac1{(2\pi)^2}
\int_0^\infty dq\int_{-1}^1dx\; (1-x^2)\ q^4
\left|\frac{\alpha_q\alpha_{q+k}}{\sin\pi v_q\sin\pi v_{q+k}}\right|^2
\frac{\exp[(B_q+B_{q+k})\tau]}{(B_q+B_{q+k})^2+k^2}.
\end{equation}
As noted in\cite{destri} the Floquet index is maximum at $q =
\eta_0/2$ and this is the dominant region  
in the $q-$integral. The fact that during the stage of parametric
resonance the integral is dominated by a region of non-vanishing 
 $q$  is a striking contrast with the broken symmetry case and a
consequence of the structure of the parametric resonance.  
As before, the strategy is to evaluate the integral for large times by
the saddle point method. For $q$ near $\eta_0/2$ and for small $k$
the  saddle point is given by  
$$
\bar q=\eta_0/2+x^2 k^2 + {\cal O}(k^3),
$$
therefore the $q-$integral in the saddle-point approximation yields  the result
\begin{equation} \label{ensilla}
I= \frac{e^2}{256\pi^2}\frac{\eta^4_0}{k\ \tau^{1/2}}\exp\left
[\tau\ 16\sqrt{1+\eta_0^2}\ \hat q\left(1-
\frac{k^2 x^2}{2\eta_0^2}\right)\right]+O\left({1\over\tau}\right)\;.
\end{equation}
For $k^2\tau \gg1$ the angular integral (over $x$) is dominated by the
region near $x=0$ and can be evaluated by using  
another saddle point expansion. In this limit the photon production
process is dominated by the emission of photons at 
right angles with the direction of the scalar with momentum $q$.  Physically
this corresponds to two charged scalars with momenta $ \vec q $
and ${\vec q} + {\vec k} $ emitting a  photon with momentum $ \vec k $
with ${\vec q}\cdot{\vec k}=0 $ (see fig. \ref{perpendicular}). This is another
difference with the broken symmetry case wherein the 
production of low momentum photons was dominated by collinear emission.

In this limit $ k^2\tau \gg 1 $ the saddle point approximation to the
angular integral yields the final 
result for the photon distribution function 
\be
N_{ph}(k,t)= e^2 \; \frac{N_2(\eta_0)}{k^2\tau}\ e^{4\tau\left[\hat B(\eta_0) 
+{\cal O}(k^3)\right]}+O\left({1\over\tau}\right)\; ,\quad k^2\tau \gg 1
\label{distfununbro1} 
\ee
where the coefficient $\hat B(\eta_0)$ in the exponential is given by
$$
\hat B(\eta_0)=4\ \sqrt{1+\eta_0^2}\ \hat q(\eta_0)
$$
with $\hat q(\eta_0)$ the nome given by eq. (\ref{nome}), and the  factor $
N_2(\eta_0) $ is given by 
$$
N_2(\eta_0)={\frac1{32768\pi}\frac{\eta_0^6\sqrt{1+\eta_0^2}}
{(5\ \eta_0^2+4)(3\ \eta_0^2+4)^2\ \hat q^3(\eta_0)}}.
$$
and we note that an additional power $ \tau^{-1/2} $ in
eq.(\ref{ensilla}) arose from the angular saddle point integration. 

We find that there is a strong dependence on the initial condition of
the order parameter $ \eta(\tau) $, i.e. on $ \eta_0 $, which
determines completely the energy density in the initial state. This is
consistent with the strong dependence on the initial conditions of the mode 
functions that determine the evolution of the scalar fields\cite{destri}. 

In particular, we obtain for large $ \eta_0 $ 
\be
N_{ph}(\tau) \buildrel{\eta_0 \gg 1}\over= \frac{e^2\; \eta_0}{C_1 \;
k^2 \, \tau }\; e^{C_2\,\eta_0\,\tau}  ~~; ~~
C_1=373.83\ldots  ~~, ~~ C_2= 0.69142\ldots\; ,\quad k^2\tau\gg1 \; .  
\ee
Furthermore we also point out that in the region $k^2\tau\gg 1$  there
is an enhancement in the photon spectra at small 
momenta as compared to the broken symmetry case. This is a consequence
of the photon emission at right angles ($x=0$) in contrast 
with the collinear emission ($x=\pm1$) for the broken symmetry case. 
   
In the range $k^2\tau \leq 1$ the saddle point evaluation of the 
angular integral is not reliable, however in 
the very small $k$ limit the angular integration can be done directly.
We find
\begin{equation} \label{kchico}
N_{ph}(k,t)\buildrel{k \to 0}\over=\frac{ e^2 \; N_3(\eta_0)}{k\;\tau^{1/2}}\; 
e^{4\; \hat B(\eta_0)\; \tau}+O\left({1\over\tau}\right)\; ,\quad
k^2\tau\ll1 \; ,
\end{equation}
where the proportionality factor $ N_3(\eta_0) $ takes the value
$$
N_3(\eta_0)={\frac1{6144\pi^{3/2}}\frac{\sqrt2 \; \eta_0^5 \;(1+\eta_0^2)^{3/4}}
{(5\ \eta_0^2+4)(3\ \eta_0^2+4)^2\  \;\hat q^{5/2}(\eta_0)}}\; .
$$
In particular,  for large $ \eta_0 $ we obtain
\be
N_{ph}(\tau) \buildrel{\eta_0 \gg 1}\over= \frac{e^2\; \eta_0^{1/2}}{C_1' \;
k \, \tau^{1/2} }\; e^{C_2\,\eta_0\,\tau}  ~~; ~~
C_1'=422.60\ldots  ~~, ~~ C_2= 0.69142\ldots\; ,\quad k^2\tau\ll1 \;. 
\ee
This analysis reveals that the soft photon spectrum diverges as
$ 1/k$  and not as $ 1/k^2 $ at $ k\to0$ . That guarantees the electromagnetic
energy density (\ref{energy.density}) is infrared finite.

\subsubsection{$ \tau > \tau_{NL} $}

For times $ \tau > \tau_{NL} $ we use  the asymptotic form of
the  mode functions given by eq. (\ref{asymodounbro}), we
insert eq.(\ref{asymodounbro}) in the expression
(\ref{ph.spectrum}) and we split the 
time integral into two domains $ 0<\tau_1<\tau_{NL} $ and
$ \tau_{NL}<\tau_1< \tau$. The integral 
from $\tau_{NL}<\tau_1<\tau$ is performed explicitly with these
asymptotic mode functions thus obtaining an expression 
analogous to eq.(\ref{gordan}). In this case, however, the upper limit of
the momentum integration is $ q_{max}= \eta_0/\sqrt2 $ i.e. the upper limit
of the resonant band  which gives the dominant contribution 
$ {\cal O}(1/g^2) $.
The integration over momenta $ q>q_{max} $ gives a
correction perturbative in $ g $. We obtain ,
\begin{eqnarray}\label{gordan.unbroken}
 N_{ph}(k,\tau) &=& \frac{e^2}{8 \, \pi^2 \; k}\int_0^{\eta_0/\sqrt2} 
q^4\, dq \int_{-1}^{+1}
 dx \; (1-x^2) \cr \cr 
&&\left| \int_0^{\tau_{NL}} d\tau_1 \; \varphi_q(\tau_1)\;
 \varphi_{|\vec q+\vec k|}(\tau_1) \; e^{-ik \tau_1} +
A_q \, A_{|\vec q+\vec k|} \, 
{ e^{i\left( \omega_q + \omega_{|\vec q+\vec k|} - k \right)\tau} 
- e^{i\left( \omega_q + \omega_{|\vec q+\vec k|} - k \right)\tau_{NL}} 
\over \omega_q  + \omega_{|\vec q+\vec k|}  - k }
\right. \cr \cr 
&+& A_q \, B_{|\vec q+\vec k|} \, { e^{i\left( \omega_q -
 \omega_{|\vec q+\vec k|} - k \right)\tau} - 
e^{i\left( \omega_q - \omega_{|\vec q+\vec k|} - k \right)\tau_{NL}} 
\over \omega_q - \omega_{|\vec q+\vec k|}  - k }\cr\cr
& -& B_q \, A_{|\vec q+\vec k|} \, { e^{-i\left( \omega_q -
 \omega_{|\vec q+\vec k|} + k \right)\tau} -  e^{-i\left( \omega_q -
 \omega_{|\vec q+\vec k|} + k \right)\tau_{NL}} \over \omega_q
 - \omega_{|\vec q+\vec k|} + k } 
\cr \cr 
&-&\left.  B_q \, B_{|\vec q+\vec k|} \, { e^{-i\left( \omega_q +
 \omega_{|\vec q+\vec k|} + k \right)\tau} - e^{-i\left( \omega_q +
 \omega_{|\vec q+\vec k|} + k \right)\tau_{NL}}\over \omega_q +
 \omega_{|\vec q+\vec k|} + k } 
\right|^2\left[ 1 + {\cal O}\left( g \right)\right] \; . \!\!\!\!
\end{eqnarray}
We focus on studying the  small $ k $ behaviour $ 0 < k \ll 1 $ which
can be obtained with the approximation 
$$
\omega_{|\vec q+\vec k|} =  \omega_q +  {k\; q \; x \over
\omega_q} +  {\cal O}\left( k^2 \right) \; .
$$
With this approximation  the denominators in eq.(\ref{gordan.unbroken}) become,
$$
\omega_q + \omega_{|\vec q+\vec k|} - k \simeq 2 \, \omega_q
\quad , \quad \omega_q  -\omega_{|\vec q+\vec k|} - k \simeq -k 
 \left( 1 + {q\, x \over \omega_q } \right)
$$
$$
\omega_q - \omega_{|\vec q+\vec k|} + k \simeq k  \left( 1 - {q\, x
\over \omega_q } \right) \quad , \quad
\omega_q +  \omega_{|\vec q+\vec k|} + k \simeq 2 \, \omega_q
$$
We remark that since $ {\cal M}^2(\infty) $ is non-zero, these denominators
{\em never vanish}. Therefore
the integrals in eq.(\ref{gordan.unbroken}) do not generate secular
terms and they have a finite limit for $ \tau \to \infty $. For asymptotically long time  and small $ k $, 
the two denominators linear in $ k $ and their
cross-product dominate eq.(\ref{gordan.unbroken}). Isolating these dominant 
contributions we find
\begin{eqnarray}\label{N.late.unbroken}
 N_{ph}(k,\infty) &\buildrel{k \to 0}\over=& 
\frac{e^2}{4 \, \pi^2 \; k^3}\int_0^{\eta_0/2} {q^4\ dq\over
\omega_q^2}\,  {\cal N}_q \, \left( 1 +  {\cal N}_q \right)\, 
F(q^2)\left[ 1 + {\cal O}\left( g \right)\right] \; .
\end{eqnarray}
where $ F(q^2) $ is the regular function
\be
F(q^2)=\int_{-1}^{+1} dx \; (1-x^2) \; \frac{3+(q\; x/\omega_q)^2}
{[1-(q\; x/\omega_q)^2]^2}=\frac{2 \;\omega_q}q\left[\frac{4 q^2+
3 \; {\cal M}^2(\infty)}{q^2}\;\mbox{ArgTh}\frac q{\omega_q}-\frac{3\omega_q}
q\right]\; .\label{regularfunction}
\ee
With the identifications given by eq. (\ref{occupation}) we recognize
that the dominant contribution in the 
asymptotic regime to soft photon production arises from bremsstrahlung
of {\em massive} charged scalars in the 
medium. 

A noteworthy feature is  that the soft photon spectrum {\bf is strongly} enhanced for
small $ k $ since $ N_{ph}(k,\infty) $ grows as $ k^{-3} $ for $ k \to
0 $, this behavior must be compared to the distribution at early time $\tau \leq \tau_{NL}$ where we had previously 
found that $ N_{ph}(k,\tau) \propto k^{-1} $
for $ k \to 0 $ [eq.(\ref{kchico})]. 

Thus in both cases, broken and unbroken symmetry, we find that the asymptotic 
non-equilibrium photon spectrum behaves for long wavelengths as  $ 1/k^3
$ for $ k\to 0 $.  This behaviour signals an IR divergence which may
require a resummation of higher order terms in $\alpha$. This is  
 beyond the scope of this study. 

It will be found in section VIII that for charged particles {\em in 
equilibrium}, the photon spectrum has very similar features.  Therefore,
the total photon number $ N_{ph,TOT}(\tau)=\int d^3k \; N_{ph}(k,\tau) $
is logarithmically divergent at small $ k $. Nevertheless the total
energy dissipated in photons, 
$$ 
E_{ph,TOT}(\tau)=\int d^3k \;  k \; N_{ph}(k,\tau) 
$$ 
is finite at finite times. As mentioned above, for late time in the
broken phase, a  resummation in $ \alpha $ is needed to assess more
reliably the photon distribution.  

\section{Photoproduction from charged scalars in thermal equilibrium}

We compute here the photoproduction process to leading order in $ e^2
$ from charged scalars {\em in thermal equilibrium} to compare it with
the non-equilibrium case studied in sec. IV. However just as in the
non-equilibrium case, we study the production of photons as an initial
value problem, i.e, an initial state is evolved in time and the number
of photons produced during a {\em finite time} scale is computed.  
We emphasize again  that this calculation is fundamentally different
from the usual formulation of the {\em rate} obtained by assuming the
validity of Fermi's Golden Rule and energy conservation.   

We shall find that there are some striking similarities between the two
cases by identifying  the high temperature limit $ T/|m_R|\gg1 $ of
the equilibrium case with the small coupling limit $ g \ll 1 $  of the
non-equilibrium situation. In both cases the plasma has a very large  
particle density.

We highlight the most relevant aspects of the result before we engage
in the technical details so that the reader will recognize the
relevant points of the calculation.  

\begin{itemize} 

\item Both from  charged scalars {\bf in and out} of equilibrium
the photon production is {\em strongly enhanced} in the infrared 
since $ N_{ph}(k,t) $ increases as $ 1/k^3 $ whereas for early times $
N_{ph}(k,t) $ grows as  $ 1/k $.
 
\item In the {\em broken symmetry case} both {\bf in 
 and out} of equilibrium the number of produced photons
increases at late times logarithmically  in time due to collinear divergences.
The physical processes that lead to  photon production can be identified with 
collinear pair-annihilation and bremsstrahlung of pions in the medium. 

\item The distribution of produced photons
approaches a stationary value as  $ t\to\infty $ in the {\em unbroken case} 
both in and out of
equilibrium with a distribution $ N_{ph}(k,\infty)\sim 1/k^3 $. 
The relevant physical process is {\em off-shell}
bremsstrahlung $ \pi \rightarrow \pi + \gamma $. 

\end{itemize}

Consider that at the initial time $ t_0 $ there is some given
distribution of photons $ N_k(t_0)$  and charged scalars $ n_p $.  
The kinetic description provided in Appendix A leads to the following
expression for the change in 
the photon distribution when the Green's functions of all fields are
the form of the equilibrium ones given by eq. (\ref{equilgreen}) 
but in terms of $ n_p $ and $ N_k(t_0) $\cite{htl}

\begin{eqnarray} 
\dot N_k(t) 
&& =\frac{e^2}{16\pi^3k} \int \frac{d^3q}{\omega_q\omega_{|\vec q+\vec k|}} 
\;q^2  \;
(1-x^2)  \int^t_{t_0} dt' \left\{ \right. \nonumber \\ 
&&\left. \cos[(\omega_q+\omega_{|\vec q+\vec k|}+k)(t-t')] 
 \left[[1+N_k(t_0)](1+n_q)(1+n_{|\vec q+\vec k|})-N_k(t_0) \;
 n_q \; n_{|\vec q+\vec k|}\right]+ \right. 
\nonumber \\
&&\left. \cos[(\omega_q+\omega_{|\vec q+\vec k|}-k)(t-t')]
\left[[1+N_k(t_0)] \;n_q \; n_{|\vec q+\vec k|}-
N_k(t_0) \; (1+n_q)(1+n_{|\vec q+\vec k|})\right]+ 
\right. \nonumber \\ 
&&\left. \cos[(\omega_q-\omega_{|\vec q+\vec k|}+k)(t-t')]
\left[[1+N_k(t_0)](1+n_q) \; n_{|\vec q+\vec k|}-
N_k(t_0) \; n_q
\;(1+n_{|\vec q+\vec k|})\right]+ 
\right. \nonumber \\ 
&&\left. \cos[(\omega_q-\omega_{|\vec q+\vec k|}-k)(t-t')]
\left[[1+N_k(t_0)](1+n_{|\vec q+\vec k|}) \; n_q-N_k(t_0) \; 
n_{|\vec q+\vec k|}(1+n_q)\right] \right\} 
\label{kineticeqn} 
\end{eqnarray}

The different contributions in the above expression have a simple and
obvious interpretation in terms of  
gain-loss processes\cite{htl}. 

\subsection{Photoproduction at first order in $\alpha $}

In order to compare to the non-equilibrium situation described above,
we will set the initial photon  
distribution to zero, i.e. $N_k(t_0)=0$, and we will also neglect the change
in the photon population (this is also the case for the rate equation
obtained by\cite{toimela,rusk}). Integrating in time we obtain the 
expression 
\begin{equation}\label{eq.decomposition}
N_{ph}(k,t)=\frac{e^2}{8\pi^2 k}\int q^4 \; dq\int_{-1}^1 dx \;\frac{(1-x^2)}
{\omega_q\;\omega_{|\vec q+\vec k|}}\; [A_1+A_2+A_3+A_4](q,k,x)
\end{equation}
where
$$
A_1(q,k,x)=n_q \; n_{|\vec q+\vec k|}\;
\frac{1-\cos[\alpha_1(t-t_0)]}
{\alpha_1^2}\quad,\quad \alpha_1=\omega_q+\omega_{|\vec q+\vec k|}-k\;.
$$ 
$$
A_2(q,k,x)=n_q \; [1+n_{|\vec q+\vec k|}]\;
\frac{1-\cos[\alpha_2(t-t_0)]}
{\alpha_2^2}\quad,\quad \alpha_2=\omega_q-\omega_{|\vec q+\vec k|}-k\;.
$$ 
$$
A_3(q,k,x)=[1+n_q] \; n_{|\vec q+\vec k|}\;
\frac{1-\cos[\alpha_3(t-t_0)]}
{\alpha_3^2}\quad,\quad \alpha_3=\omega_q-\omega_{|\vec q+\vec k|}+k\;.
$$ 
$$
A_4(q,k,x)=[1+n_q] \; [1+n_{|\vec q+\vec k|}]\;
\frac{1-\cos[\alpha_4(t-t_0)]}
{\alpha_4^2}\quad,\quad \alpha_4=\omega_q+\omega_{|\vec q+\vec k|}+k\;.
$$ 
From this explicit expression one can easily see that 
in the zero temperature limit
there is no photoproduction up to order $e^2$ .
In fact, in the vacuum, only the term
proportional to $A_4(q,k,x)$ and corresponding to the virtual process 
$|0>\to|\pi^+\pi^-\gamma>$ remains but its contribution vanishes 
as $1/t$ in the long time limit, 
since the energy conservation condition
$$\alpha_4(q,k,x)=\omega_q+\omega_{|\vec q+\vec k|}+k=0$$ 
cannot be satisfied for positive non-zero 
$\omega_q,\omega_{|\vec q+\vec k|},k$. This observation highlights
that photon production will be  
completely determined by the plasma of charged scalars  both in and
out of equilibrium. We study in detail both  cases separately.

\subsubsection{Broken symmetry phase}

In this case we study the spectrum of photons escaping from a thermal
bath of massless scalars (Goldstone's bosons) with energy $\omega_q=q$.
The analysis is very similar to that performed in the non-equilibrium
case and hinges upon extracting the secular terms in the asymptotic limit 
$ \mu \, t\gg 1 $. These arise from  different kind of {\em on-shell}
processes: 
\begin{enumerate}
\item the term
$A_1(q,k,x)$ corresponds to the annihilation $\pi^+\pi^-\to\gamma$
in which a hard photon ($k>q$) 
is emitted in the opposite direction of the initial pion
($x=-1$);
\item  the term $A_2(q,k,x)$ corresponds to the bremsstrahlung
$\pi\to\pi+\gamma$ in which a soft photon ($k<q$) is emitted in the 
opposite direction ($x=-1$); 
\item the term $A_3(q,k,x)$ corresponds 
to the bremsstrahlung $\pi\to\pi+\gamma$ in which the photon in emitted
in the same direction of the pion ($x=1$).
\end{enumerate}
Using eq.(\ref{eq.decomposition}) with $ \omega_q=q $ we recognize that the
secular terms are of the same type as those of eq.(\ref{lnsecu}) and lead to a 
logarithmic divergence $ \log\mu\, t $ with $\mu$ an infrared
cutoff. After a detailed  
analysis similar to that carried out in the non-equilibrium case, we
obtain 
\begin{equation}\label{nfeq.broken}
 N_{ph}(k,t) \buildrel{\mu\,t \gg 1}\over= \frac{e^2}{2 \, \pi^2 \;
 k^3}\; \log \mu\,t \; \int_0^\infty\, dq\,q\, n_q  \,
\left[ n_{| q- k|} \; |q-k| + 
n_{ q+ k} \; (q+k) \right]
\end{equation}
\noindent which is remarkably {\em similar} to eq.(\ref{nftinf})
   upon the replacement for the occupation numbers. 
For a thermal distribution of charged scalars the
momentum integral is finite and  for $ T\gg k $ we find 
\begin{equation}\label{nf.eq.broken}
N_{ph}(k,t) \buildrel{T\gg k}\over= \frac{e^2}6 \; 
\frac{T^3}{k^3} \log \mu\,t \; \; . 
\end{equation}

The high temperature limit of eq.(\ref{nfeq.broken}) can be compared to
the result out of equilibrium  
[eq.(\ref{nftinf})] by identifying $ (T/m)^3 $ in the thermal case with 
$1/g^2$ in the non-equilibrium case. In other
words, $ m \; g^{-2/3} $ sets the scale of an `effective temperature'
to allow a qualitative comparison between the asymptotic 
description of photon production from charged particles with a thermal
distribution and from a non-equilibrium  plasma. However we emphasize that the 
non-equilibrium distribution is {\bf far from thermal} and such a comparison
only reflects a qualitative description. Furthermore, it becomes clear that the
logarithmic secular term signals a breakdown of the perturbative kinetic 
equation and a resummation and inclusion of inverse processes will be
required to study the long time limit.   

\subsubsection{Unbroken phase}

Also in this case the analysis is similar to the out of equilibrium 
computation: the final result is finite as $ t\to\infty $ since there are no
secular terms and we can simply neglect the oscillatory pieces. 
This is due to  the presence of a non-zero mass for the scalars: 
as a consequence $ \omega_q
=\sqrt{q^2+m^2} > q $ and the denominators
in eq.(\ref{eq.decomposition}) never vanish (there are no collinear
divergences).  However for small $ k $ the two denominators 
linear in $ k $ 
$$
\alpha_2(q,k,x)\simeq-k(1+q\; x/\omega_q),\quad
\alpha_3(q,k,x)\simeq k(1-q\; x/\omega_q),
$$ 
dominate and the formula simplifies as follows as $ t\to\infty $ :
\begin{eqnarray}
 N_{ph}(k,\infty) &\buildrel{k \to 0}\over=& 
\frac{e^2}{4 \, \pi^2 \; k^3}\int_0^{\infty} q^4\ dq \int_{-1}^{+1} dx
\; (1-x^2) \; \frac{1}{\omega_q\omega_{\vec q+\vec k}}\left\{\frac{n_q \; [1+
n_{|\vec q+\vec k|}]}{(1+q\; x/\omega_q)^2}+\frac{[1+n_q]
 \; n_{|\vec q+\vec k|}}{(1-q\; x/\omega_q)^2}
\right\} \; .
\end{eqnarray}
From this expression one extracts a clear physical interpretation of the 
photoproduction process as generated by the {\em off-shell} 
bremsstrahlung of charged
scalars in the medium. To give an estimation of $ N_{ph}(k,\infty)$ in
the small $k$ and high density limits we rewrite the previous formula as
\begin{eqnarray}\label{nfeq.unbroken}
 N_{ph}(k,\infty) & = & \frac{e^2}{4 \, \pi^2 \; k^3}\int_0^{\infty} q^4\ dq\
\frac{F_{eq}(q^2)}{\omega_q^2}\ n_q(1+n_q)\left[ 1 + {\cal O}(k)\right]\; .
\end{eqnarray}
where $ F_{eq}(q^2) $ is the regular function
$$
F_{eq}(q^2)=\int_{-1}^{+1} dx \; (1-x^2) \frac{2+2(q\; x/\omega_q)^2}
{[1-(q\; x/\omega_q)^2]^2}=\frac{8 \; \omega_q^2}{q^2}\left\{\frac{\omega_q}q
 \; \mbox{ArgTh}\frac q{\omega_q}-1\right\}\; 
$$
which is similar to the function $ F(q^2) $ found in the non-equilibrium
case given by eq. (\ref{regularfunction}).
We can  estimate the temperature dependence of the photon density in the
high temperature limit $ T\gg m $: in this limit the integral
(\ref{nfeq.unbroken}) is dominated by momenta $q\sim T$ and 
we can replace  $\omega_q$ and $F_{eq}(q^2)$
with their asymptotic expressions
$$
\omega_q\to q,\quad F_{eq}(q^2)\to8\ln\frac{2q}m,\quad \frac mT\to0
$$
leading to the result
\begin{equation}\label{nf.eq.unbroken}
N_{ph}(k,\infty)\buildrel{T\gg m,T\gg k}\over=
\frac{2e^2}3\frac{T^3}{k^3}\left[\ln\frac Tm+{\cal O}(1) 
\right]\;.
\end{equation}

\subsection{Discussion}

Here we highlight a fundamental difference between our analysis of the
photoproduction process and the typical analysis offered in the 
literature\cite{toimela,rusk,kapustagale,lichard,bdrs}.

Our approach hinges upon computing the expectation value of the number
operator of transverse photons in a state that has been evolved from
an initial time $t_0$ to the {\em finite} time $t$ at which the number
of photons is measured. By contrast, the usual approach computes the
transition probability from a state prepared in the infinite past to a
state in the infinite future. In such calculation there appears the
familiar product of delta functions which are interpreted as the
on-shell condition (energy momentum conservation) multiplied by the
volume of space-time. Dividing by this volume one obtains the
transition probability per unit volume and time which is
interpreted as the production rate: this is basically the content of
Fermi's Golden Rule.   

In our approach  we directly compute the expectation value  $\langle
\dot{N}_k\rangle(t) = R^{(+)}(k,t)$ in a time evolved state and 
obtain the photon distribution at a time $t$ by integrating this
quantity, i.e.,  $ N_{ph}(k,t) = \int_{t_0}^t dt' R^{(+)}(k,t')$. This
requires the knowledge of the dynamical photoproduction rate
$R^{(+)}(k,t')$ for all times $t_0\leq t'\leq t$. 

The usual computation via Fermi's Golden Rule takes the long time
limit and isolates the secular term that is linear in time 
by replacing $ R^{(+)}(k,t')$ by its  asymptotic limit 
\be
R^{(+)}_{as}(k)=\lim_{t\to\infty}R^{(+)}(k,t)\label{as.rate}
\ee
The condition (\ref{as.rate}) is tantamount to considering only  
{\em on-shell} processes, i.e, those that satisfy energy (and
momentum) conservation. 

Keeping {\em only} on-shell processes,  the  large time limit of the
photon number becomes 
$$
N_{ph}(k,t)=R^{(+)}_{as}(k) \cdot (t-t_0)\quad,\quad t-t_0\to\infty.
$$
However our approach includes also {\em off-shell} processes that
contribute to $ N_{ph}(k,t) $ in a {\bf finite} time interval. These
processes do not contribute to $ \dot{N}_{ph}(k,t) $ asymptotically
since they are subleading at very large time, i.e,  
$$
\lim_{t\to\infty}R^{(+)}_{off-shell}(k,t)=0
$$
however they could be {\bf dominant} at finite time. Actually, as we have seen
in the previous section, the off-shell processes are of lower order in the 
electromagnetic coupling and strongly enhanced at soft momenta. Asymptotically 
we can write the  the photon number in the form
$$
N_{ph}(k,t)=N_{off-shell}(k,t)+R^{(+)}_{as}(k)\cdot (t-t_0)\; 
$$
where $ R^{(+)}_{on-shell}(k) $ is the usual rate calculated in
equilibrium from on-shell processes whose expansion in $ \alpha $ begins 
at order $ \alpha^2 $ (or $ \alpha \, \alpha_s $ in the case of the
quark-gluon plasma). In the case of broken symmetry studied in the
previous sections  
$$
N_{off-shell}(k,t) \propto \alpha \ln\left[\mu(t-t_0)\right]
$$
Therefore  off-shell processes dominate during a  time scale  $t<t^*$ with
$$
t^*\sim \frac1{\mu \alpha} \; \ln{1 \over \alpha}
$$
This is an important point in the  application of our
novel approach to the physics of heavy ion collisions. In this case,
the lifetime of the  quark-gluon plasma is relatively  short and the
standard approach could miss important physics associated with
transient off-shell effects. 

This analysis is essential in order to understand the possible
phenomenological relevance of the transient effects. A quantitative
assessment of it requires to compare the magnitude of the
contributions to photon production from off-shell and  on-shell
processes at the {\em finite} time scale $ t $ of survival of the
quark-gluon plasma.  We intend to report
the details of our studies on these issues within the context of photon
production in the quark-gluon plasma in a forthcoming article.  

\section{The magnetic mass out of equilibrium}

The magnetic mass {\bf in thermal equilibrium} is defined as \cite{lebellac}
\be\label{masequil}
m^2_{equil,mag} = \lim_{k\rightarrow 0} \lim_{\omega\rightarrow
0}\tilde{\Sigma}^{equil}_{k,bub}(\omega) +\Sigma_{tad}^{equil}\label{mequil}
\ee 
where $ \tilde{\Sigma}_{k,bub}^{equil}(\omega) $ is the Fourier
transform of the retarded transverse polarization kernel $
\Sigma_k^{equil}(t-t') $ of the non-local part of the self-energy, 
and $ \Sigma_{tad}^{equil} $ is the tadpole contribution in thermal
equilibrium. When the evolution equation for the transverse mean-field
is studied as an initial value problem,  the
relevant kernel to study is the Laplace transform of the retarded 
self-energy\cite{htl}, i.e.  
\begin{equation}
\tilde{\Sigma}_{k,bub}^{equil}(s) =\int_0^\infty dt \; e^{-st} \;
\Sigma_{k,bub}^{equil}(t)  \; .
\end{equation}
It is important to remark that the limits must be taken in
eq.(\ref{masequil}) in the precise order displayed above because the
limits {\em do not commute}.

It is a known result  that {\em in equilibrium} the magnetic mass
vanishes in an abelian gauge theory. The general argument relies on the
structure of the Schwinger-Dyson equations, the Ward identities and
translational invariance in space and time\cite{fradkin}. 
More specifically to the scalar theory 
under consideration the vanishing of the magnetic mass to leading
order in $ \alpha $ (or alternatively to leading order in the hard
thermal loop resummation) relies on the {\em exact} cancellation
between the tadpole diagram and the zero frequency limit of the  
bubble diagram contributing to $ \Sigma_k^{equil}(\omega) $. A detailed
analysis of this cancellation reveals the 
role of the Ward identity as highlighted by 
the general result in equilibrium. 

\bigskip 

The equilibrium aspects of magnetic screening phenomena are fairly
well established in abelian theories \cite{fradkin,lebellac}, however to our
knowledge the situation  {\em out of equilibrium} has not 
received much attention. In this section we will study the dynamical
aspects  of the magnetic screening with an explicit computation at
leading order in $ 1/N $ and first order in $ \alpha $. 

The initial stage in this program is to obtain an expression for the
magnetic mass.  

This is achieved by considering the linearized evolution equation for
the transverse photon condensate or mean field which is generated as a linear 
response to an externally prescribed transverse current $\vec{\cal
J}_{T}(\vec x,t)$. Such an equation has already been obtained in
\cite{htl} and we refer the reader to that article for details. In terms of 
spatial Fourier transforms, it is given by
\begin{equation}
\left(\frac{d^2}{dt^2}+k^2\right){\cal A}_{Ti}(\vec{k},t)+\int_0^t dt'\;
\Sigma_{k,ij}(t,t')\;{\cal A}_{Tj}(\vec{k},t')={\cal J}_{Ti}(\vec
k,t)\; , \label{eq_photop}
\end{equation}
where $ \Sigma_{k,ij}(t,t') $ is the transverse retarded photon
polarization out of thermal equilibrium. It contains two contributions,
one  local in time and determined by tadpole diagrams displayed in
fig. \ref{loops}b and the other is non-local and retarded in time
and given to lowest order in $ \alpha $ by the  bubble diagram displayed
in  fig. \ref{loops}a. We have,
\begin{equation}
\label{Sigma}\Sigma_{k,ij}(t,t')=\Sigma_k^{tad}(t)\;\delta(t-t')\;\delta_{ij}+
\Sigma_k^{bub}(t,t')\;{\cal P}_{ij}(\vec k),
\end{equation}
where $ \Sigma_k^{tad}(t) $ is  the tadpole diagram
(fig. \ref{loops} b) 
\begin{equation}
\Sigma_k^{tad}(t)=
2e^2\langle\Phi^\dagger\Phi\rangle=2e^2\int_0^\infty \frac{dq}
{(2\pi)^2}\; q^2\; |\varphi_q(t)|^2
\end{equation}
and $ \Sigma_k^{bub}(t,t') $ the bubble diagram in real
time (fig. \ref{loops} a), given by\cite{htl}
\begin{equation}
\Sigma_k^{bub}(t,t')=-4e^2\int_0^\infty \frac{dq}{(2\pi)^2}\;
q^4\int_{-1}^1  dx\; (1-x^2)\;  \mbox{Im}[G^>_q(t,t') 
G^>_{|\vec q+\vec k|}(t,t')]\; . \label{sigmabub} 
\end{equation}
with $ G^<~,~G^> $ the scalar Green's functions given by
eqs.(\ref{greater})-(\ref{lesser}) and $ x $ being the
cosine of the angle between $ \vec q $ and $ \vec k $.

 The linear response to an external current is a {\em different}
problem from that of photon production studied in the 
previous section, and although the
polarization diagram shown in fig. \ref{loops} describes  both
processes, here we are interested in extracting a 
{\em different} information, which in equilibrium corresponds to the
real part of the polarization in the limit of 
zero  frequency. 

Clearly, out of equilibrium, the very concept of mass is a delicate
one, but we can make contact with the equilibrium definition
[eq.(\ref{masequil})] by a derivative expansion in
time. Writing 
\be
\Sigma_k^{bub}(t,t') = \frac{d\Gamma_k(t,t')}{dt'} ~~; ~~ \Gamma_k(t,t')=
\int_0^{t'} \Sigma_k^{bub}(t,t'') \; dt'' \label{gamma} 
\ee
and integrating by parts in  eq.(\ref{eq_photop}) we find
$$
\left(\frac{d^2}{dt^2}+k^2+\Sigma_k^{tad}(t)+\Gamma_k(t,t)
\right){\cal A}_{Ti}(\vec{k},t)-\int_0^t dt'\;
\Gamma_k(t,t')\;{ d{\cal A}_{Ti} \over dt'}(\vec{k},t')={\cal J}_{Ti}(\vec 
k,t)\; .
$$
Collecting the local terms in this equation of motion leads to the
identification  
\be
m^2_{mag}= \lim_{k\rightarrow 0} \lim_{t\rightarrow
\infty}\left[\Sigma_k^{tad}(t)+\Gamma_k(t,t)\right] \label{magmass} 
\ee 
We see from eqs.(\ref{masequil}) that this definition
reduces to the one at equilibrium in the case of time translational
invariance. The definition (\ref{magmass}) is the description  of magnetic
screening that is consistent with known equilibrium results in abelian
theories.  

With the purpose of understanding the time and wavelength
dependence of the several contributions, we now introduce a time and
$k$-dependent effective magnetic mass 
\be
m^2_{mag}(k,\tau) \equiv  \int^{\tau}_0 d\tau'\; \Sigma_k^{bub}(\tau,\tau') 
 +\Sigma_k^{tad}(\tau) = m^2_{bub}(k,\tau)+m^2_{tad}(\tau)
\label{magmassdef}
\ee 
The non-equilibrium
definition of the magnetic mass, which coincides with the equilibrium
definition in the case of time translational invariance is then
$$
m^2_{mag}=\lim_{k\rightarrow 0} \lim_{\tau \rightarrow
\infty}m^2_{mag}(k,\tau)
$$
and we  remark again that the limits must be taken in this
precise order. We now analyze generally the magnetic mass for both
cases, broken and unbroken symmetry.

We point out, however, that the effective magnetic mass
(\ref{magmassdef}) is introduced to highlight the time scale of the different 
processes that contribute to the magnetic mass (\ref{magmass}) and its
sole purpose is to provide a qualitative understanding of the 
different dynamical scales for the processes that contribute to
magnetic screening.  

Using the asymptotic form of the mode functions
(\ref{asimodo})-(\ref{asymodounbro}), the definition of the asymptotic 
occupation numbers $ {\cal N}_q $ given by eq.(\ref{occupation}) and
neglecting oscillatory terms that vanish in the asymptotic time regime
due to dephasing, the tadpole contribution to the magnetic mass becomes 
\be
m^2_{tad}(\infty)= \frac{e^2}{2\pi^2} \int_0^{\infty} \frac{q^2
dq}{\omega_q}\left[1+2{\cal N}_q\right] \label{tadpo}  
\ee 
which is reminiscent of the equilibrium tadpole contribution, but it
contains the out of equilibrium distribution functions ${\cal N}_q$. 

The  non-local contribution is given by
\bea
m^2_{bub} &=& \lim_{k\rightarrow 0} \lim_{t\rightarrow
\infty}\Gamma_k(t,t) \cr \cr
 &=& \lim_{k\rightarrow 0} \lim_{t\rightarrow
\infty}\frac{e^2}{4\pi^2} \int_0^{\infty} q^4 \; dq \int_{-1}^{1}dx \;
(1-x^2) \; \mbox{Im} 
\left[ \varphi_q(\tau) \; \varphi_{|\vec q + \vec k|}(\tau)
\int^{\tau}_0 d\tau'\varphi^*_q(\tau') \; \varphi^*_{|\vec q+\vec k|}(\tau')
\right]\; . \label{bubbly}
\eea
In the asymptotic time region, the mode functions are oscillatory and
the product 
$ \varphi_q(\tau) \; \varphi_{|\vec q + \vec k|}(\tau) $ oscillates
very fast for $ k \tau \gg 1 $. Hence, any contribution that does not
cancel the rapid time dependence of the phases will be averaged out.
The memory integral from $ 0 $ up to time $ \tau \rightarrow \infty $ can 
be split into an integral from $ \tau' = 0 $ up to a time $ \tau' = \tau_0 \geq
\tau_{NL} $ within which the mode functions 
are exponentially growing but with slow oscillations and from $ \tau' =\tau_0
$ up to  $ \tau' =\tau \rightarrow \infty $. In this second region
the mode functions have achieved their asymptotic 
forms (\ref{asimodoR}) and (\ref{asimodo}). The contribution from the
first domain cannot possibly cancel the fast 
oscillations from the mode functions at $ \tau $. Therefore this first 
contribution will vanish by the rapid oscillation of the 
mode functions at very large $ \tau $ provided $ k \tau \gg 1 $.
In the second region the integral can be performed using the asymptotic form 
of the mode functions and we find using eq.(\ref{occupation})
\bea
&&m^2_{bub}  \buildrel{\tau \gg
1, \; k\tau \gg 1}\over=   
-\frac{e^2}{4\pi^2}\lim_{k\rightarrow 0}\lim_{\tau\rightarrow \infty} 
\int_0^{\infty} \frac{q^4 dq}{\omega_q \; 
\omega_{|\vec q+\vec k|}} \int_{-1}^{1}dx \; (1-x^2)\cr \cr
&&\left[ \frac{1+{\cal N}_q+{\cal N}_{|\vec q+\vec
k|}}{\omega_q +\omega_{|\vec q+\vec k|}} 
 \left\{1-\cos\left[\left(\omega_q +\omega_{|\vec q+\vec
k|}\right)(\tau-\tau_0)\right] \right\} 
-\frac{{\cal N}_{|\vec q+\vec k|}-{\cal N}_q}{\omega_{|\vec q+\vec
k|}-\omega_q} \left\{1-\cos\left[\left(\omega_{|\vec q+\vec k|}-\omega_q
\right)(\tau-\tau_0)\right] \right\} \right]  \label{asymbub}
\eea 
This expression  is remarkable, the terms with the occupation numbers
are exactly of the same form as those obtained 
in an equilibrium description\cite{htl} and have a similar kinetic
interpretation, the first describes the production minus 
the annihilation of two scalars, and the second term is out of
equilibrium analogouos of Landau damping or bremsstrahlung (and its inverse) 
in the medium in terms of  the asymptotic
{\em non-thermal} occupation numbers $ {\cal N}_q(\infty) $ (\ref{nfuera}). 

In the $ \tau \rightarrow \infty $ limit, the terms with cosines inside
the integral (\ref{asymbub}) vanish. After taking the $ k\rightarrow 0 $
limit, the integral over $ x $ is immediate and the Landau damping
term leads to the derivative of the distribution function. 
Subtracting the vacuum contribution we
find
\be\label{burbren}
m^2_{bub,ren} = -\frac{e^2}{3\pi^2} \int_0^{\infty} \frac{q^4
dq}{\omega_q^2} \left[ { {\cal N}_q \over \omega_q}- {\omega_q \over
q} { d {\cal N}_q \over dq} \right] \; ,
\ee
thus keeping $ k $ fixed and taking $ \tau \rightarrow \infty $ we
recognize that the tadpole contribution and the one-loop bubble contribution 
have the same structure as the  equilibrium calculation 
but in terms 
of the out of equilibrium distribution functions $ {\cal N}_q $ .

Upon integration by parts in eq.(\ref{burbren}) and subtracting the
vacuum contribution in (\ref{tadpo})  we find the exact
cancellation between the bubble [eq.(\ref{burbren})] and tadpole
[eq.(\ref{tadpo})] contributions,  i.e.  
\be\label{nula}
m^2_{mag} =  m^2_{bub,ren}+ m^2_{tad,ren}= 0 \; ,
\ee
just as is the case in the equilibrium calculation\cite{htl,fradkin}.

\subsection{The effective magnetic mass $ m^2_{mag}(k,\tau) $ for $
k\tau \gg 1 $}

Having established that the magnetic mass vanishes
out-of-equilibrium, we can now study in detail the precise time
evolution of the effective magnetic mass $ m^2_{mag}(k,\tau) $ for
late times $ \tau \gg 1 $ and fixed but small $k$, so that $ k \tau \gg 1 $, by
analyzing the different contributions displayed in
eq.(\ref{asymbub}). It is at this point that we justify keeping the
oscillatory terms in (\ref{asymbub}) so as to highlight the different
time scales for the buildup of the different contributions. The two
terms have very different oscillatory behavior, whereas the term with
the {\em sum} of the frequencies maintains strong oscillations even if
$ k \ll 1 $, the second term proportional to the {\em difference} of
the frequencies evolves slower  in time for  small $ k $. This second term is  
recognized as the non-equilibrium analogous of Landau damping. In
order to extract the long time behavior we proceed as follows: i) take
$ k $ small and replace the difference in frequencies by a derivative
with respect to momentum, ii) neglect the strong oscillatory behavior
arising from the term $ \cos[2\omega_q\tau] $, to find that the
effective magnetic mass defined by eq.(\ref{magmassdef}) behaves as 
\bea
m^2_{mag}(k,\tau) \buildrel{k\tau \gg 1 , \tau \gg 1}\over=&&
-\frac{e^2}{4\pi^2}\int_0^{\infty} \frac{q^4 dq}{\omega_q^2} 
\int_{-1}^{+1}dx \; (1-x^2)\left\{ {{\cal N}_q \over \omega_q}
-{\omega_q \over q}{ d{\cal N}_q \over dq} \left[
1-\cos\left({q\,k \tau \over \omega_q}x \right)\right] \right\}\cr
&&+ \frac{e^2}{\pi^2} \int_0^{\infty} \frac{q^2
dq}{\omega_q} \; {\cal N}_q\label{asimag2}
\eea

\noindent where the first term inside the bracket is the contribution
of the two-particle cut after neglecting the strong 
oscillatory component,  the second term is the Landau damping term, the last term is the tadpole contribution, and we have
subtracted as usual the vacuum contribution which is renormalized in the absence of the medium. 

Upon integrating over $ x $ and integrating by parts the Landau
damping term, the time independent contributions cancel each other out
as discussed above and eq.(\ref{asimag2})  yields   
\be
m^2_{mag}(k,\tau) \buildrel{k \ll 1 , k \tau \gg 1}\over=
\frac{2 \, e^2}{\pi^2}\int_0^{\infty} q \, dq \; {\cal N}_q
\left\{ { 1 \over (k \, \tau )^3 }\left[ 
\sin\left({q \, k \, \tau\over \omega_q}\right) - {q \, k \, \tau\over
\omega_q} \cos\left({q \, k \, \tau\over \omega_q}\right) \right]
+ { m^2 \over k \, \tau \, \omega_q^2} \sin\left({q \, k \, \tau\over
\omega_q}\right) \right\}\label{asimag3}
\ee
 and clearly  in the long time limit,
$$
\lim_{\tau\rightarrow \infty} m^2_{mag}(k,\tau) = 0
$$
in agreement with eq.(\ref{nula}). However this analysis clearly reveals that  Landau damping  or in-medium bremsstrahlung
is the process with the {\em slowest time scale} in the long-wavelength limit.

The case of broken symmetry with $ \omega_q = q $ is particularly
clear. Eq.(\ref{asimag3}) then simplifies as
\be
m^2_{mag}(k,\tau) \buildrel{k\tau \gg 1 , \tau \gg 1}\over=
\frac{2\, e^2}{\pi^2}\left[ \int_0^{\infty} q \, dq\; {\cal
N}_q\right] { 1 \over (k \, \tau )^3 }\left[ 
\sin\left(k\,\tau\right)- k \, \tau \cos\left( k \, \tau \right)\right]
\label{asimagR}
\ee
We see that $ m^2_{mag}(k,\tau) $ oscillates around zero for large $
k\tau $ with an amplitude that decreases as $ {\cal O}\left({1 \over
(k \, \tau )^2 }\right) $ and period $ 2\pi/k $. 

\subsection{The effective magnetic mass $ m^2_{mag}(k,\tau) $ for $
k\tau \ll 1 $}

As we have noted above the asymptotic long time limit and the
long-wavelengh limit do not commute, this happens out-of-equilibrium
and  also in equilibrium where the zero frequency and the zero
momentum limit do not commute. 

However for {\em finite} time we can ask what is the behavior of the
effective mass in the long-wavelength limit.  This question 
{\em is relevant} for the evolution of the mean field in the
long-wavelength limit and for finite time. This corresponds to   
studying the effective magnetic mass (\ref{magmassdef}) in the
opposite limit  $ k\tau \ll 1 $ keeping $ \tau \gg 1 $.

For finite time the effective magnetic mass is a slowly varying function of 
$ k $ thus for  $ \tau \gg 1 $ but $k\tau \ll 1$, we shall simply set
$ k = 0 $ to explore the region $ k \ll 1 $.

In this case we find for the
effective magnetic mass (\ref{magmassdef}),
\begin{eqnarray}\label{burbasi2}
&&m^2_{mag}(k,\tau) \buildrel{k \ll 1 , k \tau \ll 1}\over= 
\frac{ e^2}{3 \pi^2} \int_0^\infty q^4 \, dq \int_0^{\tau} d\tau'\;
\mbox{Im}[\varphi_q^2(\tau)\; \varphi_q^2(\tau')^*] +{e^2 \over 2 \pi^2}
\int_0^\infty  q^2\; dq \; |\varphi_q(\tau)|^2 
\end{eqnarray} 
For the computation of the bubble contribution $ m^2_{bub}(k=0,\tau) $
here [the first term in eq.(\ref{burbasi2})] some remarks are in
order. Na\"{\i}vely, since $  m^2_{bub}(k=0,\tau) $ contains a product
of four mode functions, each of order $ g^{-1/2} $, one would expect
the result being of order $ g^{-2} $. However the following
interference argument reveals that $ m^2_{bub}(k=0,\tau) $ turns out
to be  of order $ 1/g $, just as the tadpole contribution [the second
term in eq.(\ref{burbasi2})]. 

Indeed, for $ 0<\tau'<\tau_{NL} $ the mode functions $ \varphi_q(\tau')
$ are given approximately by eq.(\ref{early.broken}). The exponentially
growing term dominates in eq.(\ref{early.broken}) while the
exponentially decreasing terms are of the order $ {\cal O}(g) $ [see
eqs.(\ref{tnolin}) and (\ref{tnolinp})]. The dominant term has a time
independent phase as can be seen read from eq.(\ref{early.broken}).   

Since the mode equations (\ref{modeeq}) have {\bf real} coefficients,
a solution with a constant phase during some time interval keeps such
phase constant for all times. Therefore, the phase of the modes is
time independent up to $ {\cal O}(g) $ corrections and the phases of $
\varphi_q^2(\tau) $ and $ \varphi_q^2(\tau')^* $ cancel up to $ {\cal
O}(g) $. Hence, Im$[\varphi_q^2(\tau)\; \varphi_q^2(\tau')^*]$ is a
factor $ g $ smaller than $ g^{-2} $. That is, it is of order $ g^{-1}
$ and not of order $ g^{-2} $. 

Due to this cancellation of the dominant  growing exponentials,
an analytical evaluation of $ m^2_{mag}(k,\tau) $ requires detailed  knowledge
of $ {\cal O}(g) $ corrections to the mode functions
(\ref{early.broken}), which is not available analytically. 

\begin{figure}[t]
  \begin{center}
 \centerline{\epsfig{file=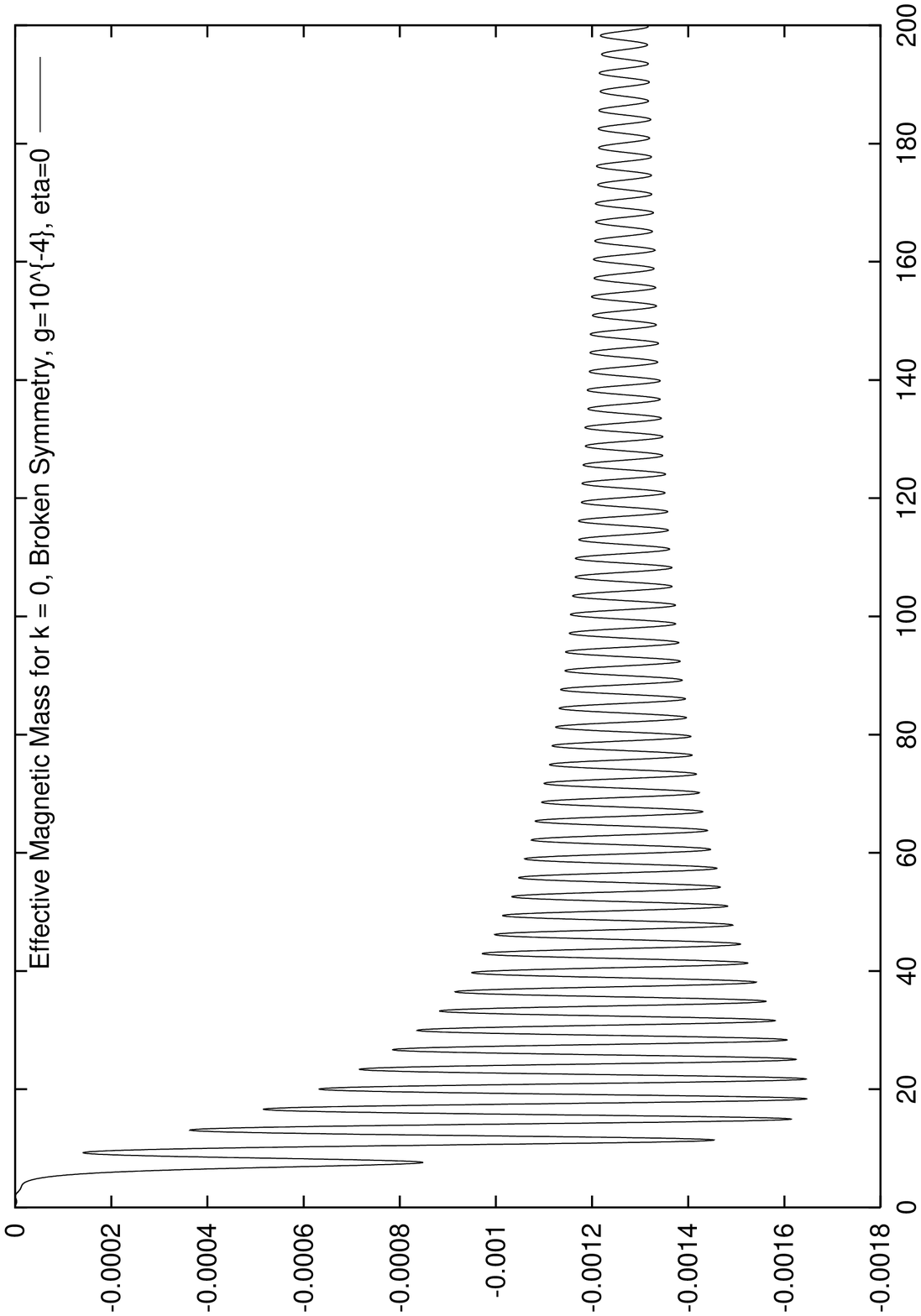,width=7in,height=8in}}
  \end{center}
\caption{Effective Magnetic Mass for $ k = 0 $ as a function of time,
Broken Symmetry, $ g=10^{-4}, \; \eta_0=0 $ \label{figmagR}}
\end{figure}

\begin{figure}[t]
  \begin{center}
  \centerline{\epsfig{file=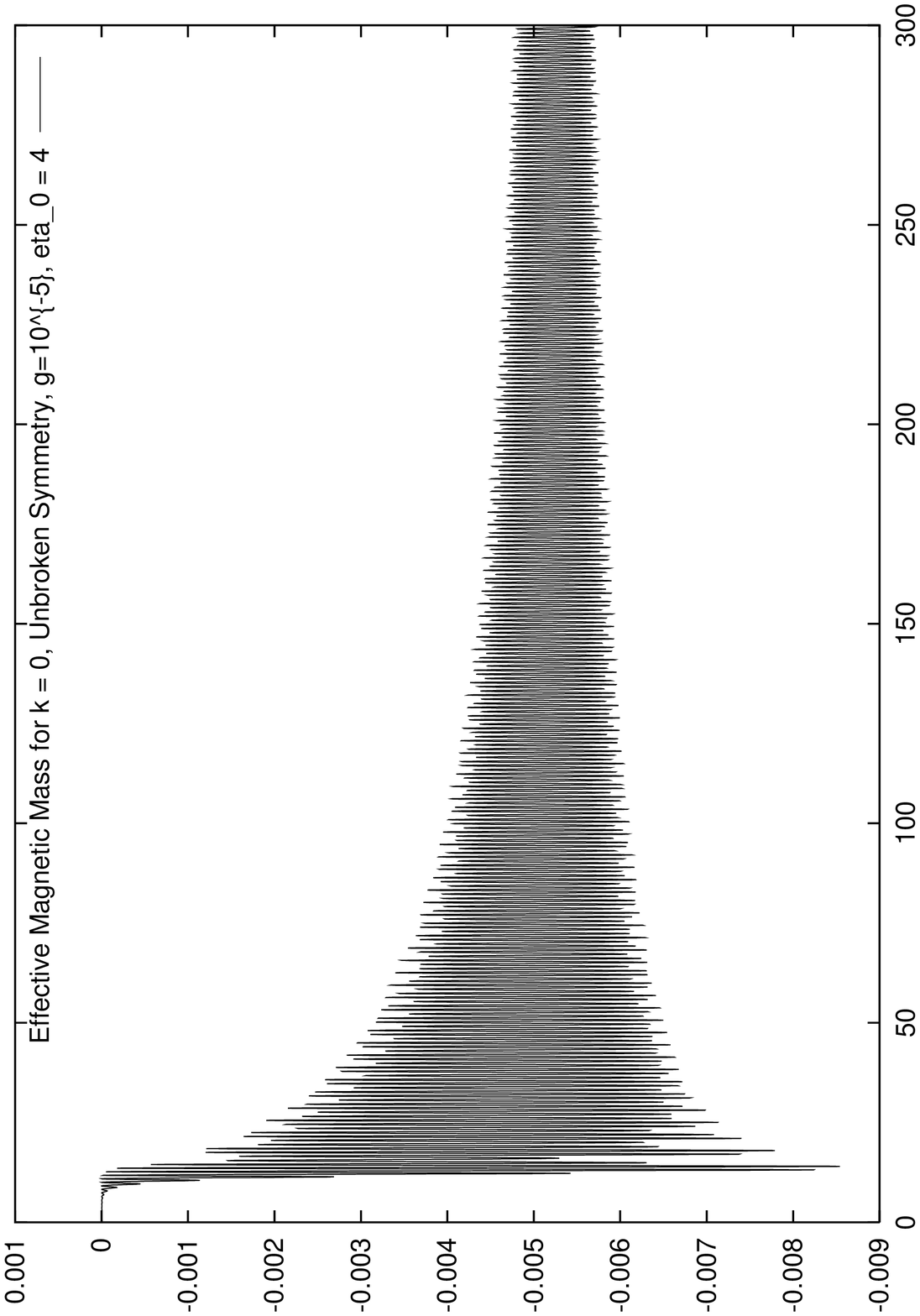,width=7in,height=8in}}
  \end{center}
\caption{Effective Magnetic Mass for $ k = 0 $ as a function of time,
Unbroken Symmetry, $ g=10^{-5}, \; \eta_0=4 $\label{figmag}} 
\end{figure}

Instead, we evaluated numerically the integrals in eq.(\ref{burbasi2}) using
the high precision modes obtained in refs.\cite{nuestros,destri}. The
results are displayed in figs. \ref{figmagR} and \ref{figmag}.

For late times,  $ m^2_{mag}(k=0,\tau) $ oscillates around the following constant values,
\begin{eqnarray}\label{magasi}
m^2 _{mag}(k=0,\tau \gg1) &=& -0.000513\ldots {e^2 \over g} \quad
\mbox{for ~Broken~symmetry,} \; \eta_0 = 0\; ,
\cr \cr
m^2 _{mag}(k=0,\tau \gg 1) &=& -0.00126\ldots {e^2 \over g}\quad
\mbox{for ~Unbroken~symmetry,} \;\; \eta_0 = 4 \; .
\end{eqnarray}

The coefficients of $ {e^2 \over g} $ are not very sensitive to the value 
of $g$ for small coupling $g \ll 1$, 
in the unbroken symmetry case the coefficient depends on the value of 
$\eta_0$. 
 
Such small numbers arise from a delicate
cancellation  of the negative contribution from the bubble diagram and
the positive contribution from the tadpole diagram. In the unbroken
case, the larger is $ \eta_0 $ the more negative is $
m^2_{mag}(k=0,\tau \gg 1) $. 

The negative sign of this effective squared mass
indicates the unexpected presence of a weak instability in the time
evolution of the mean field, which we conjecture to be 
linked to the strong photon production during this time scale. We
expect to report on a detailed study of these issues in a forthcoming
article.   

At this point it is important to remind the reader that had we studied a
situation in which the global gauge symmetry 
was spontaneously broken either by the initial state or by the
dynamics, there would have been a magnetic mass generated 
via the ordinary Higgs mechanism. Hence the vanishing of the properly
defined magnetic mass is in agreement with the 
fact that the gauge symmetry is not spontaneously broken by the dynamics. 

\section{Screening and  Debye mass generation out of equilibrium}

The Debye mass or inverse of the electric screening length, determines
the spatial 
extent over which electric charges are screened in the plasma. As in
the case of the 
magnetic mass, the Debye mass can be obtained from a linear response
problem. In this 
case the relevant linear response is that of the Lagrange multiplier
$ A_0 $ associated with
the Coulomb interaction, or longitudinal photon to an external charge
density ${\cal J}_0(\vec x,t)$.  
The Debye mass can thus be recognized from the equation of motion for
the expectation value of $A_0(\vec x,t)$ as 
a linear response to the external charge density. In terms of spatial
Fourier transforms and calling this expectation value  
${\cal A}_0(\vec{k},t)$ the equation of motion in linear response is
obtained by following the method described in detail 
in\cite{htl}. We obtain the following equation of motion for the
expectation value in linear response 
\begin{equation}
k^2{\cal A}_0(\vec{k},t)+\int_0^t dt'\; 
\Sigma^L_{k}(t,t')\; {\cal A}_{0}(\vec{k},t')={\cal J}_0(\vec k,t)\;,
\label{eq_long}
\end{equation}
where the longitudinal retarded self-energy $ \Sigma^L_{k}(t,t') $
is given to lowest order in $ e^2 $  by the following expression\cite{htl} 
\be
\Sigma^L_k(t,t')=-4e^2\int\frac{d^3
q}{(2\pi)^3}\;\mbox{Im}\left[\partial_{t'}G^>_q(t,t') \;
\partial_{t}G^>_{|\vec q+\vec k|}(t,t')-\partial_{t}
\partial_{t'}G^>_q(t,t') \;G^>_{|\vec q+\vec k|}(t,t')\right]\; ,
\label{longiselfenergy} 
\ee
and $ {\cal J}_0(\vec k,t) $ is the spatial Fourier transform of the
external source that generates the linear response. 

We remark that Schwinger terms arising from the time derivatives of
time ordered Green's functions had cancelled the 
tadpole contribution $ 2e^2 \langle \Phi^{\dagger}\Phi \rangle $ and
{\em after} this cancellation the remainder of 
the longitudinal photon polarization is given by (\ref{longiselfenergy}). The
reader is referred to \cite{htl} for further details of this
cancellation which is independent of whether the system is in or out
of equilibrium.   

Following the arguments presented previously in the case of the
magnetic mass above, we define the  Debye mass out of equilibrium as 
\begin{equation}
\label{Debye-mass}
m^2_{Deb} \equiv\lim_{k\to0}\lim_{\tau\to\infty}\left[\int_0^{\tau}
d\tau'\;\Sigma^L_k(\tau,\tau')\right]. 
\end{equation}
We emphasize again that the limits $ \tau\to\infty $ and $ k\to0 $
must be taken in the order specified above since they do not
commute. Taking the limits in the inverse order yields a vanishing
result.

Using the expressions for the Green's functions in terms of the mode
functions as given by eqs. (\ref{greater})-(\ref{lesser}) 
(in terms of the dimensionless mode functions $\varphi_q(\tau)$) we
finally obtain 
\begin{equation}\label{masaD}
m^2_{Deb}=\lim_{k\to0}\lim_{\tau\to\infty}
e^2\int\frac{d^3 q}{(2\pi)^3} \;\mbox{Im}\left\{
\left[\varphi_q(\tau)\dot \varphi_{|\vec q+ \vec k|}(\tau)-
\dot \varphi_q(\tau)\varphi_{|\vec q+ \vec k|}(\tau)\right]
\int_0^{\tau} d\tau' \;\dot \varphi_q(\tau')^* \varphi_{|\vec q+ \vec
k|}(\tau')^*\right\} 
\end{equation}
We  compute now  the Debye mass  generally in both cases under
consideration: broken and unbroken symmetry. The expression
(\ref{masaD}) displays a remarkable feature: unless the memory
integral develops singularities as $ k\rightarrow 0 $ the Debye mass will 
vanish identically in this limit. Separating the time integral
into a part between $ \tau' =0 $ and $ \tau' = \tau_{NL} $ and a second
part from $ \tau_{NL} $ to $ \tau $ we recognize that no singularities 
can arise from the first part. The contribution to the Debye mass from
the region of spinodal or parametric instabilities 
is regular in the limit $ k\rightarrow 0 $ and do not survive in the
 $ \tau \to \infty $ limit. Therefore, we conclude that despite the 
fact that there are strong non-equilibrium processes during the stages
of spinodal and parametric amplification, they 
are not directly associated with the generation of a Debye mass.
However, as it will become evident below, the late time {\em distribution} of
particles produced during these stages determines the Debye screening 
mass. In the second part of the integral the modes acquire their asymptotic
form. Just as in the discussion of the magnetic mass, only few of the
contributions survive the rapid dephasing in the limit $\tau
\rightarrow \infty$. 

Replacing the mode functions in eq.(\ref{masaD}) by their asymptotic
behavior, using the relation (\ref{occupation}) and neglecting the
oscillatory contributions in the limit of $ \tau \rightarrow \infty $
with $ k $ fixed, we obtain  
\bea
m^2_{Deb} = && \lim_{k\rightarrow 0}\lim_{\tau\rightarrow \infty}
\left\{ {e^2} \int   
\frac{ d^3q}{(2\pi)^3 \omega_{q}} \left[
\frac{\omega_{|\vec q+\vec k|}-\omega_q}{\omega_{|\vec q+\vec
k|}+\omega_q} 
 \left(1+{\cal N}_q+{\cal N}_{|\vec q+\vec k|}\right)
 \right.\right. \nonumber \\
&&\left. \left. -\frac{\omega_{|\vec q+\vec
k|}+\omega_q}{\omega_{|\vec q+\vec k|}-\omega_q} 
\left({\cal N}_{|\vec q+\vec k|}-{\cal N}_q\right)\right]
\right\}  \label{asymdeby}
\eea 
\noindent which is recognized as the longitudinal polarization
evaluated at zero frequency\cite{htl}. Again the 
different contributions have an obvious kinetic interpretation which
has been discussed in ref.\cite{htl}.  

Taking the zero momentum limit, we finally find
\be
m^2_{Deb} = -\frac{e^2}{\pi^2} \int_0^{\infty} dq~ q~ \omega_q~
\frac{d{\cal N}_q}{dq} \label{deby} 
\ee
This expression reveals at once the important feature that the Debye
mass is determined by the {\em derivative} of 
the distribution function of the charged fields with respect to
momentum. Although this happens in other contexts, it is 
seldom highlighted in the literature. 
Integrating  by parts at finite times the surface
term at  $ q = 0 $ vanishes since the distribution $ {\cal N}_q $ 
is regular at $ q = 0 $ for {\em finite} times. We obtain the final form
\be\label{debfin}
m^2_{Deb} = \frac{e^2}{\pi^2} \int_0^{\infty} {dq \over\omega_q} \;
\left[2\,q^2+{\cal M}^2(\infty)\right] {\cal N}_q \label{deby2} 
\ee
We now study each case separately.

\subsection{Broken symmetry}

\begin{figure}
\centerline{ \epsfig{file=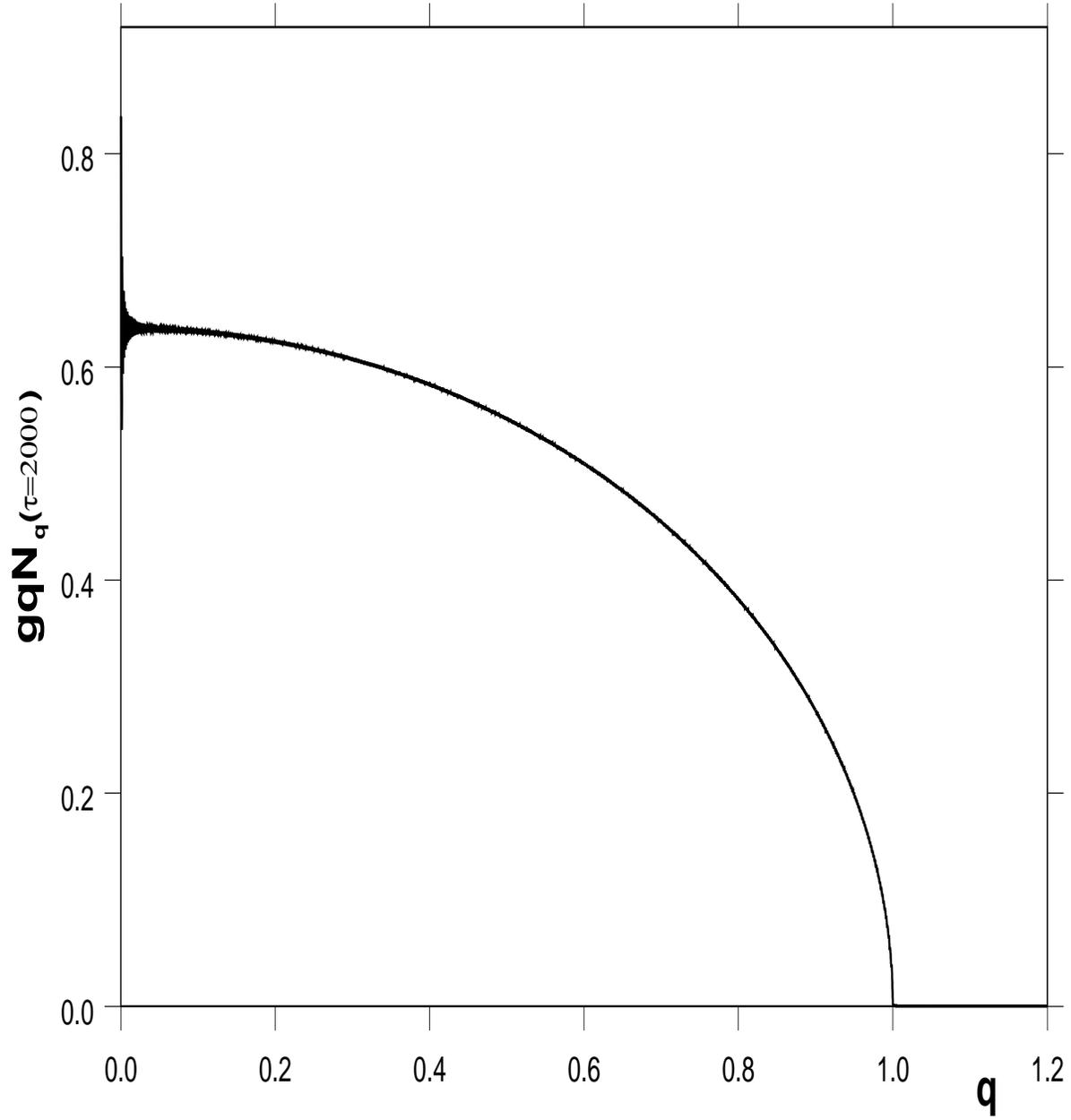,width=7in,height=8in}}
\caption{$ g\,q\,{\cal N}_q(\tau=2000)$ for the broken symmetry case as a
function of $ q $ for $ g=10^{-4}, \eta_0=0 $. The distribution 
saturates for most of the range at $\tau \approx \tau_{NL}$ but for
very small momentum.   \label{qNqbroken}} 
\end{figure}

In the broken symmetry case with $\omega_q=q ~;~ {\cal M}(\infty)=0$, 
the distribution
function $ {\cal N}_q $ is of $ {\cal O}(1/g) $ in the region $ 0\leq q
\leq 1 $ and near the origin 
behaves as $ {\cal N}_q(\tau) \sim 1/q $ for all times including $ \tau \to
\infty$. Fig. \ref{qNqbroken} shows 
$ g \; q \; {\cal N}_q(\tau) $ vs. $ q $.  Since $ \lim_{q\rightarrow
0} q^2 {\cal 
N}_q =0$ we are justified in neglecting the 
surface term in eq.(\ref{deby}) and eq.(\ref{deby2}) is valid even for $
\tau \to \infty $. Using equation (\ref{debfin}), the relation
(\ref{occupation}) and the sum rule (\ref{sumrule1}) we find for the
broken symmetry case 
\be
m^2_{D} = \frac{e^2}{g \; \pi^2}[1+{\cal O}(g)] \label{massdebyebroke}
\ee

\subsection{Unbroken symmetry}
\begin{figure}
\centerline{ \epsfig{file=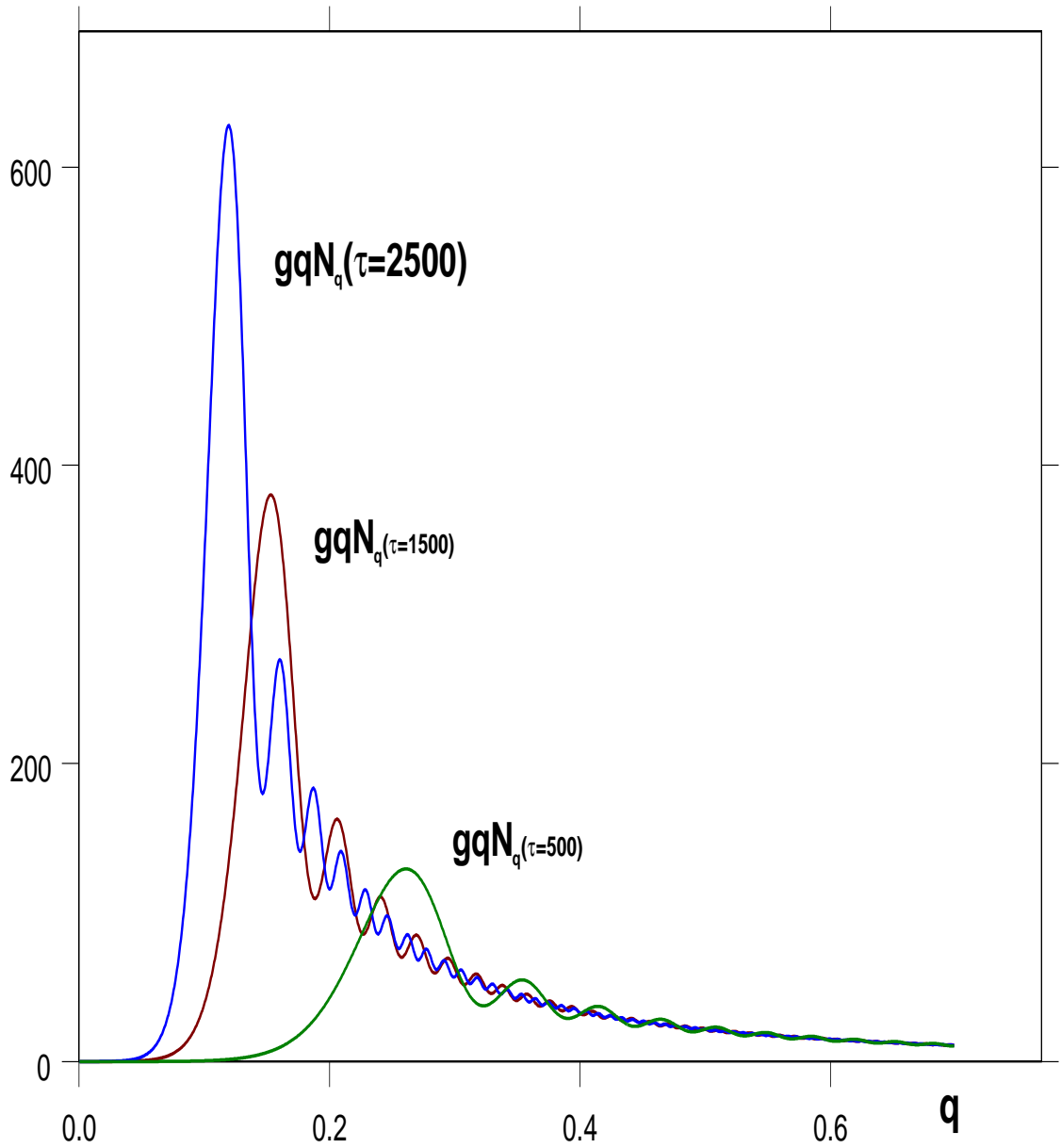,width=7in,height=8in}}
\caption{$g\, q\, {\cal N}_q(\tau=500,1500,2000)$ for the unbroken symmetry
case as a function of $q$ for $g=10^{-4}$. The distribution continues
to evolve as a function of $\tau$. The peak at low momentum is at
$q_0\approx \sqrt{\frac{K_1}{\tau}}$ and moves towards the 
origin while its magnitude increases.   \label{qNqunbroken}}
\end{figure}

As discussed in detail in section 3.2, in the unbroken symmetry case
the distribution function at times larger than $ \tau_{NL} $  
is dominated by the peak  of the non-linear resonances. The distribution
function {\em continues}  
to evolve at long times with two marked peaks in the region of
non-linear resonances (\ref{bandas}) inside which the amplitudes 
$ A_q(\tau)\; ; \; B_q(\tau) $ and consequently the distribution
function $ {\cal N}_q $ grows with a power law in time. The width of
these non-linear  
resonant bands diminishes in time, the resonance near $ q \approx
\eta_0/\sqrt{2} $ becomes subdominant and the resonance in the 
region $ 0<q< \sqrt{K_1/\tau} $ becomes the dominant one, the
peak growing in amplitude and the width of the resonance 
diminishing as time evolves. Fig. \ref{qNqunbroken} displays $ g\; q \;
{\cal N}_q(\tau) $ vs. $ q $ for different times. 

\begin{figure}[t]
  \begin{center}
  \mbox{\epsfig{file=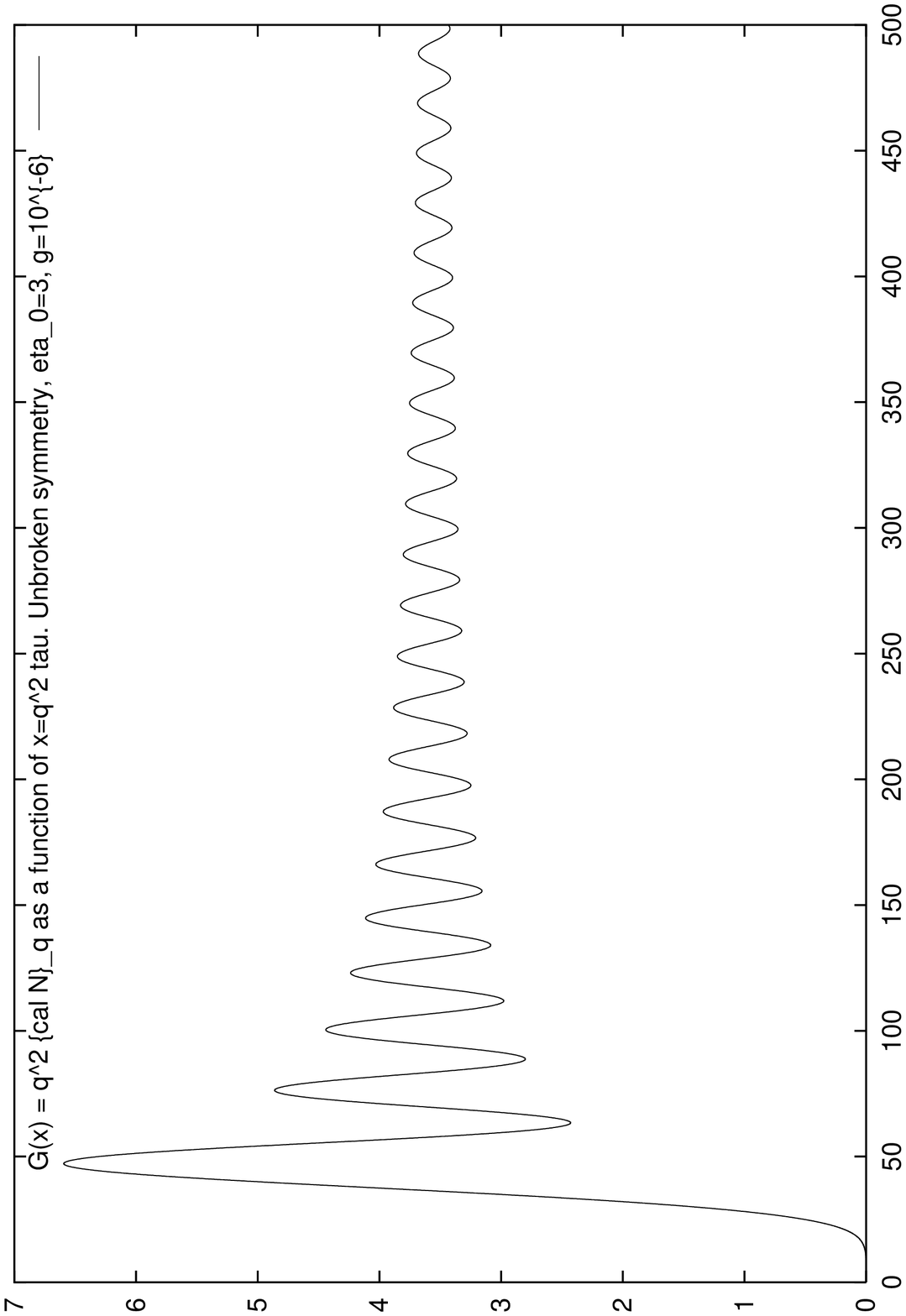,width=7in,height=8in}}
  \end{center}
\caption{ The function $ G(x) \equiv g\; q^2 {\cal N}_q(\tau) $ as a function
of $ x \equiv q^2 \tau $. Unbroken symmetry, $ \eta_0=3, \; g=10^{-6}
$\label{figG}}
\end{figure}

Our extensive numerical calculations shows that
for times $ \tau \gg \tau_{NL} $ the distribution function takes the
scaling form
\be\label{escalin}
{\cal N}_q(\tau) = { G(q^2 \, \tau) \over g \; q^2 }\quad \mbox{with} \quad
G(0) = 0 \; .
\ee
The function $ G $ only depends on time and $ q $ through the
combination $ x \equiv q^2 \tau $. We plot $ G(x) $ as a
function of $ x $ in fig. \ref{figG}. Notice that $ G(x) $ is of order
$ {\cal O}(g^0) $.

At finite times the distribution function $ {\cal N}_q $ is finite 
at $ q=0 $ and neglecting the surface terms in the integration
by parts leading to (\ref{deby2}) is justified.  

\begin{figure}
\centerline{ \epsfig{file=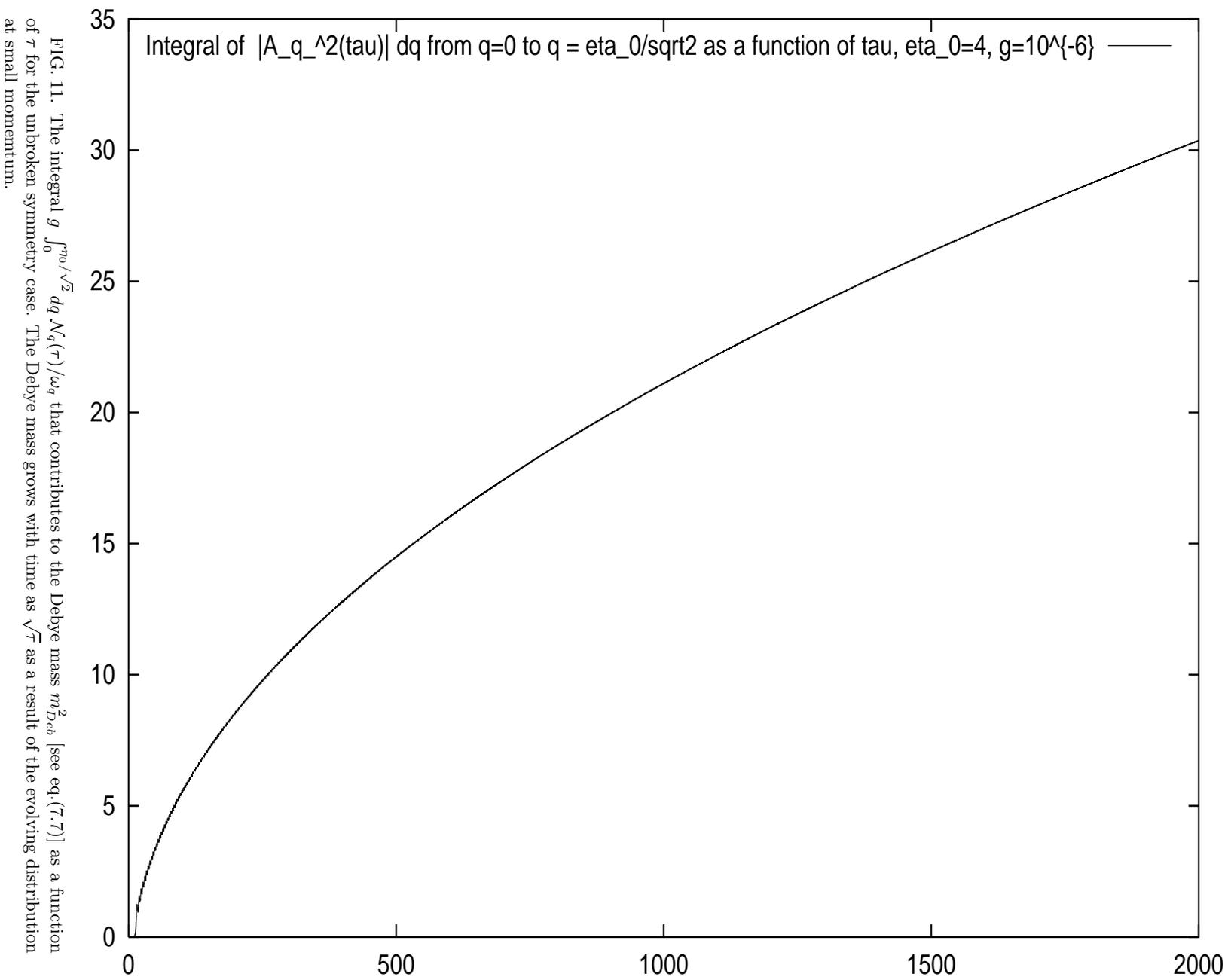,width=7in,height=8in}}
\caption{The integral $ g \; \int_0^{\eta_0/\sqrt2} dq \; {\cal N}_q(\tau)/
\omega_q $  
that contributes to the Debye mass $ m^2_{Deb} $ [see eq.(\ref{debfin})]
as a function of $ \tau $ for the unbroken symmetry case. The Debye mass
grows with time as $ \sqrt{\tau} $ as a result of the
evolving distribution at small momemtum.    \label{debyemasa}} 
\end{figure}

Since the derivative of the distribution
function is dominated by this peak, we find that even asymptotically
the Debye mass (\ref{deby}) {\bf continues to grow with time as $
\sqrt{\tau} $}.  Fig. \ref{debyemasa} displays the Debye mass as a function
of time, it is clear from this figure that the trend is that of monotonic 
increase as $ \sqrt{\tau} $ as one obtains inserting the scaling form
of the distribution function (\ref{escalin}) into eq.(\ref{debfin}). 
$$
\int_0^{\infty} {dq \over\omega_q} \; {\cal N}_q \buildrel{\tau \to
\infty }\over= \frac{\sqrt{\tau}}{2\,  g \; {\cal M}^2(\infty) } 
\int_0^{\eta_0^2 \,\tau/2  } {dx 
\over x^{3/2}} \; G(x) + {\cal O}(1)
$$
Notice that the integral converges for $ \tau = \infty $ since $
G(\infty) $ is finite [see fig. \ref{figG}].

The reason for this increase is that the distribution is 
dominated by the peak near $ q\approx 0 $ which continues to 
evolve as a consequence of the non-linear resonance with the width
ever decreasing in time and the peak continues to grow. 

In the infinite time limit the distribution will be peaked at zero momentum behaving as
$$ 
{\cal N}_q(\tau = \infty) = { G(\infty) \over g \;q^2  }
$$
leading to a {\em divergent} Debye mass because $ {\cal M}^2(\infty)
\neq 0 $ and thus the behaviour at $ q = 0 $ makes the integral in
eq.(\ref{debfin}) to diverge. 

This divergence suggests that higher order
contributions in the electromagnetic coupling must be taken into
account and perhaps a resummation of higher order terms can lead 
to a finite Debye mass, 
 but clearly this possibility
requires a more detailed study which is beyond the scope of this article.   

\section{Non-equilibrium transverse conductivity}

Consider applying an external transverse electric field $\vec{\cal
E}_{T,ext}(\vec x,t) = -\dot{\vec{\cal A}}_{T,ext}(\vec x,t)$ with 
$\vec{\cal A}_{T,ext}(\vec x,t)$ an external transverse vector
potential. The induced transverse current is obtained in linear 
response by coupling the external vector potential to the current in
the Lagrangian density ${\cal L}\rightarrow {\cal L}+ 
 \vec{J}_T\cdot{\cal A}_{T,ext}$. The transverse current induced by
the external vector potential is obtained in linear response 
in terms of spatial Fourier transforms as

\be
\langle J^i_T(\vec k,t) \rangle = i \int dt' \; \langle J^i_T(\vec
k,t)J^j_T(-\vec k,t')\rangle_{ret} \; {\cal A}^j_{T,ext}(\vec k,t')
\ee

\noindent where $\langle J^i_T(\vec k,t)J^j_T(-\vec
k,t')\rangle_{ret}$ is the retarded correlation function given by 

\be
\langle J^i_T(\vec k,t)J^j_T(-\vec k,t')\rangle_{ret}= \langle
J^{+,i}_T(\vec k,t)J^{+,j}_T( -\vec k,t')\rangle - 
\langle J^{+,i}_T(\vec k,t)J^{-,j}_T(-\vec k,t')\rangle
\ee
and the symbols $\pm$ refer to the time branches along the CTP
contour. Comparing with the expression for the photon polarization in 
the equation of motion (\ref{eq_photop}) we recognize that the
retarded current-current correlation function is given  
at lowest order (${\cal O}(\alpha)$) by

\be
i\langle J^i_T(\vec k,t)J^j_T(-\vec k,t')\rangle_{ret}=
-\Sigma^{bub}_k(t,t')\; {\cal P}^{ij}(\vec k) 
\ee
with $\Sigma^{bub}_k(t,t')$ given by (\ref{sigmabub}). 

Introducing  the non-equilibrium transverse conductivity  as follows
\be
\sigma^{ij}_{\vec k}(t,t')={\cal P}^{ij}(\vec k) \;\sigma_{k}(t,t')=
 - {\cal P}^{ij}(\vec k)
\int^{t'}_0 dt''\; \Sigma^{bub}_k(t,t'') ~~; ~~ t>t'\label{conductivity}
\ee

\noindent integrating by parts in eq.(\ref{conductivity}) 
and neglecting surface terms we obtain the linear response relation 
\be
\langle J^i_T(\vec k,t) \rangle = \int dt' \sigma^{ij}_k(t,t') \; {\cal
E}^j_T(\vec k,t') \label{linresp} 
\ee
Although the definition of the conductivity (\ref{conductivity}) may
not look familiar, it is straightforward to confirm that 
in the equilibrium case it leads to the usual relation between the
conductivity and the polarization in equilibrium.

In thermal equilibrium the polarization is a function of the time
difference and the system has been in equilibrium from $ t=-\infty $. Thus, 
extending the lower limit in (\ref{conductivity}) to $ t''=-\infty $ and
writing 
$$ 
\Sigma_{k,bub}^{equil}(t-t'')=\int_{-\infty}^{+\infty} d\omega \;
\tilde{\Sigma}_{k,bub}^{equil}(\omega) \; e^{i\omega(t-t'')} 
$$ 
it is straightforward to find the spatial and temporal Fourier 
transform of the conductivity to be given by
\be\label{condequil}
 \tilde{\sigma}_k^{equil}(\omega)={\tilde{\Sigma}_{k,bub}^{equil}(\omega)
 \over i\omega}
\ee
which is the usual relationship between the bubble polarization
and the equilibrium conductivity at lowest order in $\alpha$. 

\bigskip

Since in the out of  equilibrium case under consideration the initial state at time $t=0$
is the vacuum and the plasma is generated during the stage of strong
non-equilibrium evolution, the time integral in the conductivity kernel
(\ref{conductivity}) has the initial time ($ t=0 $) as the lower limit.  

The explicit expression for the conductivity at leading order in $ \alpha $
follows from eq.(\ref{sigmabub}) and is given by
\bea \label{condu}
\sigma_{k}(\tau,\tau')=-\frac{e^2}{4\pi^2} \int_0^{\infty} q^4 \; dq
\int_{-1}^{+1}dx \; (1-x^2) \; \mbox{Im}\left[ \varphi_q(\tau) \;
\varphi_{|\vec q + \vec k|}(\tau)\int^{\tau'}_0
d\tau''\varphi^*_q(\tau'') \; \varphi^*_{|\vec q+\vec k|}(\tau'') \right]
\eea

It is difficult
to compute explicitly the conductivity  in the full range of the two time 
variables, however we can provide explicit formulae in
the relevant regimes $ 1 < \tau'<\tau \leq \tau_{NL} $  
and when both time variables are in the asymptotic regime
$ \tau>\tau' > \tau_{NL} $ for fixed $ k $.  

\subsection{$ 1 <\tau'<\tau < \tau_{NL} $}

In this time regime there are no fast oscillatory solutions and we can
simply set $k=0$ to obtain an estimate for the 
long-wavelength limit of the conductivity.

{\bf Broken symmetry:} In this case the mode functions in this time
regime are given by (\ref{early.broken})-(\ref{alfas}). Furthermore
the modes that grow 
the most are those for $q\approx 0$ for which 
a non-relativistic approximation $q\ll1$ is reliable. The computation
of the conductivity proceeds in three steps, i) 
recognize the terms that contribute to the imaginary part in
eq.(\ref{sigmabub}), ii) carry out the integral in the variable $
\tau'' $, iii) perform the integral in the $q$ variable in the saddle point
approximation using the non-relativistic 
approximation for the mode functions. In the region $1<\tau'<\tau
\leq \tau_{NL}$ the dominant term of the conductivity 
in the long-wavelength limit is given by
\be
\sigma_{k\approx 0}(\tau,\tau') = \frac{e^2}{16}\left[
\frac{\tau'}{\pi^{3/2}\,\tau^{5/2}}\; e^{2\tau}- {e^{2\tau'}\over 15
\, \pi^2} \right][1+{\cal O}(k^2 \, \tau) ] \label{conducbroke} 
\ee
We remark that since the expression for the polarization
(\ref{sigmabub}) involves four mode functions one would 
naively conclude that the polarization and the conductivity would be
$ \sim e^{4\tau} $, however these terms are real and  
do not contribute to the imaginary part in eq.(\ref{sigmabub}). Therefore
at the end of the exponential growth of long-wavelength 
modes at $\tau \sim \tau' \sim \tau_{NL}$ the conductivity is of order
$ {\cal O}(e^2/g) $ and positive since the first term
dominates over the second one in eq.(\ref{conducbroke}) for $ \tau >
\tau' $. 

\vspace{3mm}

{\bf Unbroken symmetry:} For this case the mode functions are given by 
eqs.(\ref{earlyunbro})-(\ref{floquetindex}).
The calculation of the conductivity in this 
time regime follows the same steps as in the broken symmetry case, 
with the difference in the third step being 
that the saddle point in the $q$-integral is at the maximum of the
Floquet index $ q^* $  given by  eq.(\ref{maxfloquet}). After 
some tedious but straightforward calculation we find
\be
\sigma_{k\approx 0}(\tau,\tau') \buildrel{\tau_{NL} >\tau > \tau' \gg 1 }\over=
\frac{e^2\; \eta_0^5}{768 \pi^{3/2}}\;\sqrt{\frac{\hat q(\eta_0)}{\tau}}\;
{\tau' \left(\frac54 \eta_0^2 + 2 \right) \; e^{2 {\bar B(\eta_0)} \tau}
\over  \left(\eta_0^2
+1\right)^{1/4} \, \left(1 + \frac34 \eta_0^2\right)^{3/2} \sqrt{1 +
\frac54 \eta_0^2} }\left[ 1 + {\cal O}\left({\hat q(\eta_0)}, {1 \over
\tau}\right)\right] \label{conducunbroke}
\ee
where $ {\bar B(\eta_0)} = 4 {\hat q(\eta_0)} 
\sqrt{1+\eta_0^2}\left[ 1 - 4 {\hat q(\eta_0)} + {\cal
O}({\hat q^2(\eta_0)}) \right] $.

Hence, besides some quantitative differences, this result is
qualitatively similar to that in the broken symmetry 
case above with the same conclusion in the order of magnitude of the
transverse conductivity in the long-wavelength 
limit at the time scale $ \tau_{NL} $. 

\subsection{$\tau_{NL}<\tau'<\tau$, $k$ fixed}

In this regime we can use the asymptotic form of the mode functions
generally in both cases, broken and unbroken  
symmetry. When we studied the magnetic mass in the previous section,
we have noted that  the long-wavelength limit does not commute with
the long time limit. Thus, we will consider the long time limit but keeping 
$ k $ fixed. Furthermore, since the conductivity is a function of two
time variables, we will consider $ \tau \gtrsim \tau' $ 
and both arguments $ \tau $ and $ \tau' $ larger than $ \tau_{NL} $. 
In particular we will consider that phases involving the {\em sum} 
of the time arguments vary more rapidly and therefore dephase faster
 than those that depend on the {\em difference} 
of these time arguments, effectively deciding that $ \tau-\tau' $ is
{\em slower varying} than $ \tau+\tau' $ but both 
arguments are in their asymptotic regime. The calculation proceeds
along the same steps outlined in the case of the magnetic 
mass, the product $\varphi_q(\tau) \; \varphi_{\vec q+\vec k}(\tau)$ is
strongly oscillatory in the asymptotic regime.
Keeping only the oscillatory factors in the difference
of time arguments $ \tau - \tau'$, neglecting terms that oscillate much 
faster than these and using the relation (\ref{occupation}), we find  
\bea
\sigma_k(\tau,\tau')\buildrel{k\,\tau \gg 1 \ll k\,\tau' }\over=
&&\frac{e^2}{4\pi^2}\int \frac{q^4 \; dq}{\omega_q \;
\omega_{|\vec q + \vec k|}}\int^1_{-1}dx \; (1-x^2)\left\{ 
\left(1+{\cal N}_q+{\cal N}_{|\vec q+\vec k|} \right)
\frac{\cos\left[(\omega_q +\omega_{|\vec q + \vec k|})(\tau-\tau')
\right]}{(\omega_q +\omega_{|\vec q + \vec k|})} \right. \nonumber \\ 
&&\left. -({\cal N}_{|\vec q+\vec k|}-{\cal N}_q)
\frac{\cos\left[(\omega_{|\vec q + \vec k|}-\omega_q )(\tau-\tau')
\right]}{(\omega_{|\vec q + \vec k|}-\omega_q )}\right\} \label{sigmasi}
\eea 
This expression can be written in a more familiar form by
introducing the representation 
\be
\frac{\cos\left[(\omega_{|\vec q + \vec k|}\pm \omega_q)(\tau-\tau')
\right]}{(\omega_{|\vec q + \vec k|}\pm \omega_q )} =  
\int d\omega \; \frac{e^{i\omega(\tau-\tau')}}{2\omega}
\left[\delta(\omega-(\omega_{|\vec q + \vec k|}\pm
\omega_q))-\delta(\omega +(\omega_{|\vec q + \vec k|}\pm \omega_q) )
\right] 
\ee
\noindent which leads to the final relation
\be 
\sigma_k(\tau,\tau')= \int d\omega \; e^{i\omega(\tau-\tau')}\;
\frac{\tilde{\Sigma}_k^{bub}(\omega)}{i\omega} 
\ee
where $\tilde{\Sigma}_k(\omega)$ is the imaginary part of the Fourier
transform of the asymptotic   real-time,  retarded  
polarization out of equilibrium,
\bea
{\tilde{\Sigma}_k^{bub}(\omega)} &=& -\frac{i\,e^2}{8\pi^2}\int_0^\infty 
\frac{q^4 \; dq}{\omega_q  \;
\omega_{|\vec q + \vec k|}}\int^1_{-1}dx \; (1-x^2)\left\{ 
\left(1+{\cal N}_q+{\cal N}_{|\vec q+\vec k|} \right)\left[
\delta(\omega+\omega_q +\omega_{|\vec q + \vec k|})
-\delta(\omega-\omega_q -\omega_{|\vec q + \vec k|})\right]
\right. \nonumber \\ 
&-& \left.({\cal N}_{|\vec q+\vec k|}-{\cal N}_q)\left[
\delta(\omega-\omega_q +\omega_{|\vec q + \vec k|})
-\delta(\omega+\omega_q -\omega_{|\vec q + \vec k|})\right]\right\}
\label{sigtil} 
\eea 
Hence, we find the remarkable result that 
in the asymptotic limit the conductivity and the polarization  are
related in a manner akin to that in {\em equilibrium} (\ref{condequil}), i.e.,
\be\label{condfuequi}
\tilde{\sigma}_k(\omega) =
\frac{\tilde{\Sigma}_k^{bub}(\omega)}{i\omega} 
\ee  
\noindent we stress however,  that there is {\bf non-equilibrium}
information in this relationship because the distribution function of 
the produced particles is {\bf out of equilibrium}.

The expression for the conductivity (\ref{sigtil})-(\ref{condfuequi})
simplifies considerably for small $ k $ and 
in the static
limit $ \omega = 0 $. Similarly to the Debye mass, to lowest order in $\alpha$  
the conductivity is determinated by
the derivative of the distribution function,
\be
\tilde{\sigma}_k(0) \buildrel{k \to 0 }\over=  
-\frac{e^2}{4\pi^2\; k}\int_{k/2}^{\infty} q^2 \;
dq\; \frac{d{\cal N}_q}{dq}\; . \label{explicit.conduct}
\ee
In the broken symmetry case 
the $k\rightarrow 0$ limit of the integral can be computed by part
using the sum rule (\ref{sumrule1}) leading to
$$
\tilde{\sigma}_k(0) \buildrel{k \to 0 }\over= \frac{e^2}{4\pi^2\; g}
\frac{m_R^2}k
$$
This result can be compared with the equilibrium conductivity at
 $ \omega = 0 $ simply by replacing $ {\cal N}_q $ by its
thermal counterpart $ n_q = (e^{q/T} - 1 )^{-1} $ which yields
$$
\tilde{\sigma}_k^{equil}(0) \buildrel{k \to 0 }\over= {e^2 \; T^2 
\over 12 \; k} 
$$
This shows a {\em qualitative} comparison between the  high temperature limit
in thermal 
equilibrium and the small coupling limit out of equilibrium, which in
dimensionful units reads 
$$
g^{-1} \leftrightarrow {\pi^2 \over 3} \; \left( \frac{T}{|m_R|} \right)^2.
$$

We emphasize that this comparison is only {\em qualitative} and should
{\em not} be taken as a direct relation between the two cases since 
the non-equilibrium distributions are very far from thermal. 

Consider for instance the conductivity in the {\it unbroken} phase:
our analysis in the previous sections shows that the distribution
function ${\cal N}_q$ is strongly enhanced at small $q$ as a
consequence of the non-linear resonance at $ q=0 $, i.e, $ {\cal
N}_q\sim 1/q^2 $ for $ q\to0 $. This non-equilibrium effect leads to a
logarithmic enhancement for long-wavelenths 
$$
\tilde{\sigma}_k(0) \buildrel{k \to 0 }\over\sim \frac{e^2}{g\;
k}\;m_R^2\; \ln\left(\frac {m_R}{k}\right)\; ,
$$
which has no analog in the equilibrium counterpart.

We remark  that the non-equilibrium conductivity [eqs.(\ref{sigtil})-(\ref{condfuequi})] is {\bf finite} 
for all $ k $ including $ k = 0 $ at {\em finite time}. 

Only in the $ \omega = 0 $ limit, which corresponds to an integral up to infinite time, the conductivity
  has  a divergent
$ k \rightarrow 0 $ limit, such is the case for the equilibrium
static conductivity $ \tilde{\sigma}_k^{equil}(0) $.  Thus whereas the non-equilibrium conductivity is well behaved for long-wavelengths
at any {\em finite} time,  the long time limit will require a resummation of
diagrams that  must include the  the width of the charged scalar
particles. In the static limit the finite mean-free path of charged particles will provide a  cutoff for long-wavelength (long distance)
propagation and will lead to a finite long-wavelength static conductivity.   

Our analysis reveals i) the initial
stages of build up of the conductivity through the formation of the
non-equilibrium plasma, ii) an asymptotic description
at long  times in terms of the non-equilibrium distribution functions
which is akin to the equilibrium description. 

\section{Conclusions, discussion and further questions}

In this article we have studied the formation of a plasma of charged
particles from the strongly out 
of equilibrium processes of a) spinodal decomposition (or phase
separation) and b) parametric amplification. 
The model, scalar QED with $N$-charged scalar fields and a $U(1)$
photon, does not only provide an arena to 
study the questions of the formation of the plasma, electric and
magnetic screening, photon production and conductivity out 
of equilibrium, but also is phenomenologically relevant both in
cosmology and in heavy ion collisions as a description of the chiral
phase transition out of equilibrium. In cosmology an important
consequence of this study is the novel mechanism of generation of 
primordial magnetic fields at the time scale of the QCD phase
transition, whereas in heavy ion collisions the mechanisms studied 
here can lead to strong photon production with non-equilibrium
distributions that could be an important signature of non-equilibrium 
effects associated with the chiral phase transition. 

Spinodal decomposition describes the early stages during a quenched or
supercooled second order phase transition 
and the dynamics is determined by the exponential growth of the
fluctuations with wavevectors in the unstable 
band. Parametric amplification of quantum fluctuations occurs during
the stage when the order parameter is oscillating around 
the minimum of the potential with large amplitude. In this situation
there are resonances that amplify exponentially quantum 
fluctuations with wavevectors in the regions of  parametric
instability. In both cases, the explosive exponential growth of 
charged fluctuations lead to the formation of a non-equilibrium plasma
and to photon production and the generation of electric 
and magnetic fields. These unstabilities are shut-off by the
non-linear field interactions which are systematically and
consistently treated in the large $ N $ limit\cite{eri97,destri}. 
Thus we have combined the large $ N $ limit that allows  the non-perturbative aspects of the 
formation of the plasma and  a novel kinetic
description of photon production to study many relevant electromagnetic 
properties of the non-equilibrium plasma. Our conclusions and further questions can be summarized as follows:

\begin{itemize}

\item{ {\bf Photon production:} We have obtained a novel kinetic
equation to study photon production strongly 
out of equilibrium to lowest order in $\alpha$ and to leading order in
the large $ N $ expansion. We find that at the 
end of the linear stage dominated by the exponential growth of
instabilities in both cases, spinodal decomposition and 
parametric amplification, the photon distribution function is peaked
at low momentum with a typical photon density of 
${\cal O}(\alpha/\lambda^2)$ with $\lambda$ the scalar
self-coupling. In the case of a quenched phase transition we find 
that electric and magnetic fields  generated during the
non-equilibrium stage are correlated on distances given by a {\em
dynamical 
correlation length} $ \xi \approx \sqrt{t/|m_R|} $ for times 
$ \tau < \tau_{NL} $ with $ |m_R| $ the
(renormalized) mass scale of the scalar fields. These 
mechanisms of photon production {\em could be} an important source of
primordial magnetic fields in the early universe at a time 
scale of the chiral phase transition $ t  \approx 10^{-5} $ seconds after
the big bang and temperature scales $ {\cal O}(100)\mbox{Mev} $, 
however before coming to definite conclusions on the cosmological
implications, two important issues must be studied further: i) 
whether a strongly supercooled (quenched) chiral phase transition can
take place, given that the relaxation time scales for 
QCD are much shorter than the inverse of the expansion rate of the
Universe during the transition, and ii) the kinetic equation 
used to study photon production neglected an initial population and
therefore stimulated processes, these must be taken into 
account fully in the case of the cosmological phase transition since
the Universe is radiation dominated at that stage.  

Perhaps, phenomenologically more relevant is the case of the chiral
phase transition in Ultrarelativistic Heavy Ion Collisions, since 
it is quite possible that in this situation the phase transition
occurs out of equilibrium\cite{rajagopal}. In this case the photons 
produced through spinodal decomposition would have a non-equilibrium
spectrum that could be a potential experimental  
signature\cite{photop2}. In the case of broken symmetry we have found
that the presence of massless particles asymptotically, lead 
to collinear divergences in the bremsstrahlung contributions in the
medium. These infrared divergences result in a logarithmic growth 
of the photon density asymptotically. This growth is also present in
the equilibrium case and points out to a breakdown of the perturbative
kinetic equation. A thorough understanding of the photon distribution
function in this regime requires a consistent resummation  
using the dynamical renormalization group\cite{rgir}.   }

\item{ {\bf Magnetic screening mass:} We have introduced a definition
of the magnetic mass out of equilibrium which is the natural 
generalization of the equilibrium case. We find that the magnetic mass vanishes
through a cancellation mechanism similar to  that in the equilibrium case
despite the fact that the asymptotic
distribution functions are non-thermal.

To highlight the different processes that contribute to the magnetic
mass and their widely different time scales in the long-wavelength
limit we have introduced an {\em effective} magnetic mass that coincides
asymptotically with the proper definition of the magnetic mass. 

We find that the  non-equilibrium generalization of  Landau damping
begins to compete with the contributions from  two particle
excitations and mean-field on time scales that in the long wavelength
limit are far longer than those for these processes. This effective
magnetic mass  displays memory effects that correlates the spinodal or
parametric particle production at early times with the dynamics at late times. 
We also find some unexpected weak long-wavelength instability in the
time evolution of the mean transverse gauge field, which 
we conjecture to be related to the strong photoproduction during the
early stages of spinodal or parametric instabilities. } 

\item{ {\bf Electric (Debye) screening mass:} As in the case of the
magnetic screening mass, we define the 
electric (Debye) screening mass out of equilibrium as the
natural generalization of the equilibrium case. 
 
In the case of spinodal instabilities we find that the Debye mass is
given by $ m^2_{Deb} =  8\; |m_R|^2\;  e^2/\lambda + {\cal O}(\lambda^0) $.

In  the case of parametric amplification we find that the Debye mass {\em
diverges} asymptotically as $ \sqrt{\tau} $ with a coefficient of the
order $ {\cal O}(e^2\lambda^{-1}) $. This result is a consequence of  {\em massive} asymptotic states 
and the presence of non-linear resonances\cite{destri} that result in a  
peak in the distribution function of the charged particles
that moves towards zero momentum and whose width
vanishes at long times. Since 
the Debye mass is determined by the {\em derivative} of the
distribution function (\ref{deby}) 
in the case of massive particles a distribution
which is singular at small momentum such as the one resulting from the
non-linear resonances gives a divergent Debye mass. 

A divergent result to first order in $ \alpha $ suggests that a
resummation of electromagnetic corrections using for example the
dynamical renormalization group \cite{rgir} must be carried out. }

\item{ {\bf Transverse electric conductivity (Kubo):} The transverse
electric conductivity is an important transport 
coefficient which in the case of primordial magnetic fields limits the
propagation and correlation of these fields and in the QGP it 
enters in the calculation of Ohmic energy losses in the plasma. We
have obtained the non-equilibrium conductivity from 
Kubo's linear response out of equilibrium. The electric conductivity
is a complicated function of two time variables and the wavelength. We
solved in detail the {\em build-up} of conductivity during the early
stages of formation of the non-equilibrium plasma as well as the
asymptotically long time regime.  The long-wavelength conductivity
builds up exponentially because of the instabilities that lead to the
formation of the plasma, and at the end of the stage dominated by
linear instabilities it achieves a magnitude $ \sigma_{k\approx 0}
\approx |m_R| \; e^2/\lambda $. At asymptotically long times we find
that the conductivity has a similar structure to the equilibrium
conductivity but with non-equilibrium distribution functions replacing
the thermal ones. We find that the electric conductivity stays {\bf
finite} for all momenta including $ k = 0 $ at {\bf finite} times. } 

\end{itemize}
  
Our study offers a novel view of the
electromagnetic response of non-equilibrium plasmas  in a 
model that allows to extract quantitative and qualitative information
and that also bears  phenomenological interest from 
the point of view of generating seeds of primordial magnetic fields at
the chiral phase transition in the early universe and of 
describing non-equilibrium aspects of pion-photon dynamics in heavy
ion collisions. 

\appendix

\section{Kinetic equation for the photon distribution}

\begin{figure}
\centerline{ \epsfig{file=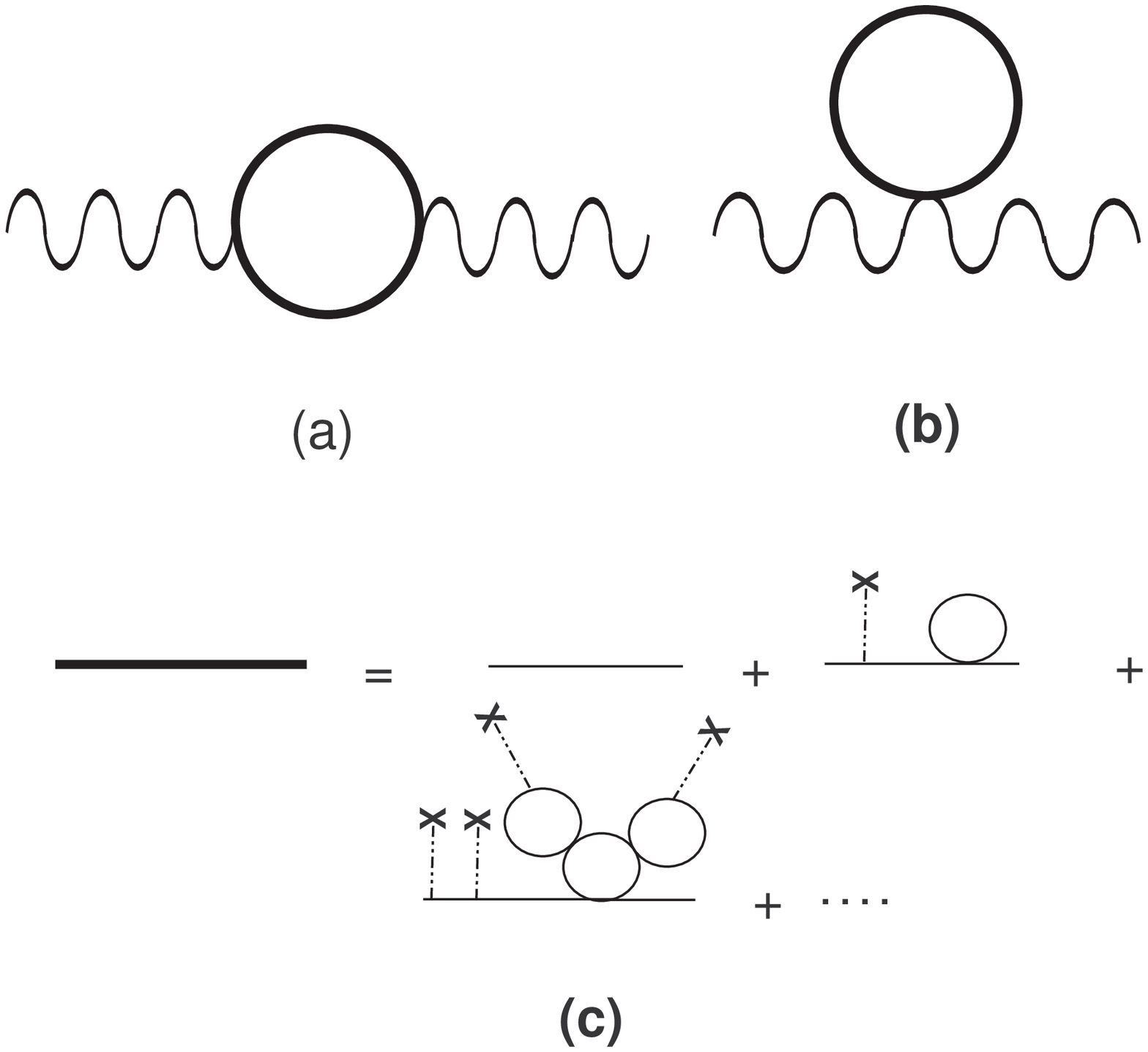,width=7in,height=8in}}
\caption{Photon polarization with full large $ N $ scalar propagators. The
dashed line with the 
cross at the end represents an insertion of the background
$\eta(\tau)$.  \label{loops}} 
\end{figure}

If the charged scalar fields were in  equilibrium the rate of photon
production would be determined by the 
imaginary part of the Fourier transform in frequency and momentum of
the polarization depicted in fig. \ref{loops}. The 
expression for photon production in equilibrium has been obtained
in\cite{toimela,rusk,kapustagale,lichard} for the 
case of the QGP. 

In the situation under consideration, strongly out of equilibrium, the
polarization is not time translational 
invariant and the frequency representation is not available. The time
evolution of the photon distribution function 
must be obtained from a kinetic equation. The validity of a kinetic
description requires a wide separation of time 
scales between the time scale over which the photon distribution
function changes and that of the phenomena that is 
strongly out of equilibrium. In the case under consideration the
non-equilibrium evolution of the scalar fields result 
from spinodal and parametric unstabilities and these occur on fast
time scales of order $ |m_R|^{-1} \ln(1/g) $, we 
expect that the change in the photon distribution will occur on time
scales that are {\em longer} by at least a factor 
$1/\alpha$. Hence under the assumption of weak electromagnetic
coupling, the photon distribution function will evolve 
much slower than the non-equilibrium dynamics of the charged scalar
field. Under these circumstances a kinetic description 
is valid. Furthermore since the charged scalar particles are far-off
shell, a simple Boltzmann equation for the photon 
distribution function will miss the important off-shell effects
associated with the non-equilibrium evolution of the 
scalar fields. This point becomes more important during the stage of
spinodal instabilities when there is no meaning to 
on shell particles.  

In this Appendix we obtain the kinetic equation for the 
photon number from first principles (see also \cite{photop2,htl}).

By using the simplest definition for the photon number or phase space
distribution (for homogeneous systems) 
\begin{equation}
\label{ph_number}
N_{ph}(k,t)=(2\pi)^3\frac{d^6N}{d^3xd^3k}= \sum_{\lambda=1,2}\langle
a_\lambda^\dagger(k)a_\lambda(k)\rangle =\frac1{2k} \langle\dot
A_T(-\vec k,t)\cdot\dot A_T(\vec k,t)+k^2 A_T(-\vec k,t)\cdot A_T(\vec
k,t)\rangle, 
\end{equation}
one extracts the time derivative of the distribution function as follows:
$$
\dot N_{ph}(k,t)=\frac1{2k}<\ddot A_T(t_1,-\vec k)\dot A_T(t_2,\vec k)+
\dot A_T(t_2,-\vec k)\ddot A_T(t_1,\vec k)$$
$$
\left.+k^2A_T(t_1,-\vec k)\dot A_T(t_2,\vec k)+k^2
\dot A_T(t_2,\vec k)A_T(t_1,\vec k)>\right|_{t_1=t_2=t}$$
$$
=\frac1{2k}
\frac{\partial}{\partial t_2}\left[\left(\frac{\partial^2}
{\partial t_1^2}+k^2\right)<A_T(t_1,-\vec k)\cdot A_T(t_2,\vec k)+
A_T(t_2,-\vec k)\cdot A_T(t_1,\vec k)>
\right]_{t_1=t_2=t}
$$
Therefore the photon density rate can be rewritten as
$$
{\dot N}_{ph}(k,t)=-\frac{i}{2k}
\frac{\partial}{\partial t_2}\left[\left(\frac{\partial^2}
{\partial t_1^2}+k^2\right)\left(\bar{\cal G}^>_{ii}(k;t_1,t_2)+
\bar{\cal G}^<_{ii}(k;t_1,t_2)\right)
\right]_{t_1=t_2=t}
$$
where $\bar {\cal G}_{ij}$ represent  the {\it exact} photon
propagator not to be confused with the free propagator
${\cal G}_{ij}$. 

In order to simplify this expression
we need the Schwinger-Dyson equations for ${\cal G}_>$
and  ${\cal G}_<$. Including a mean field contribution $\delta
\Omega^2(t)= 2e^2 \langle \Phi^{\dagger}(t)\Phi(t) \rangle $ in the
Hamiltonian we have
$$
\left(\frac{\partial^2}
{\partial t_1^2}+k^2\right)\bar{\cal G}^>_{ii}(k;t_1,t_2)= -\delta
\Omega^2(t)\bar{\cal G}^>_{ii}(k;t_1,t_2)+ 
\int[ 
\Pi_{im}^{++}(k;t_1,t)
\bar{\cal G}^<_{mi}(k;t,t_2)-\Pi^{+-}_{im}(k;t_1,t)\bar{\cal  G}^{++}_{mi}
(k;t,t_2)] dt
$$
and
$$
\left(\frac{\partial^2}
{\partial t_1^2}+k^2\right)\bar{\cal G}_{ii}^<(k;t_1,t_2)=-\delta
\Omega^2(t)\bar{\cal G}_{ii}^<(k;t_1,t_2)+ 
\int[ 
-\Pi^{--}_{im}(k;t_1,t)\bar{\cal G}^>_{mi}(k;t,t_2)+\Pi^{-+}_{im}(k;t_1,t)
\bar{\cal G}_{mi}^{--}(k;t,t_2)] dt.
$$
Using the definitions
$$
\Pi^{++}_{im}(t,t';k)=\Pi^>_{im}(t,t';k)\theta(t-t')+\Pi^<_{im}(t,t';k)
\theta(t'-t) 
$$
$$
\Pi^{--}_{im}(t,t';k)=\Pi^>_{im}(t,t';k)\theta(t'-t)+
\Pi^<_{im}(t,t';k)\theta(t-t') 
$$
$$
\Pi^{-+}_{im}(t,t';k)=\Pi^>_{im}(t,t';k),\quad\Pi^{+-}_{im}(t,t';k)=
\Pi^<_{im}(t,t';k)
$$
the photon production rate can be rewritten in the form
\begin{eqnarray}
\dot N(k,t)=&& \frac i{k}\delta \Omega^2(t) \frac{\partial}{\partial
t_1}\left[\bar{\cal G}^>_{ii}(k;t,t_1)+ 
\bar{\cal G}^<_{ii}(k;t,t_1)\right]_{t_1=t}\nonumber \\
&&-\frac i{k}
\int_{t_0}^t\left[\Pi^>_{im}(k;t,t')\; \partial_{t}\bar{\cal G}^<_{m i}(k;t',t)
-\Pi^<_{im}(k;t,t')\;\partial_{t}\bar{\cal G}^>_{mi}(k;t',t)\right]dt'.
\label{exactrate}
\end{eqnarray}

This expression is {\it exact} but  formal. To make progress, we
 consider the first order in 
 $\alpha$ by  replacing the full transverse 
photon propagator $\bar{\cal G}^>_{ij}$ with
 its free field form but with non-equilibrium distribution  
functions and neglecting the electromagnetic contribution to the
 Green's functions of the charged scalar field. 
If there is  an initial non-zero
photon distribution $N(k,t_0)$ the free Wightman functions read
$$
{\cal G}^>_{ij}(k;t',t)=
\frac i{2k} \; {\cal P}_{ij}(\vec k) \; \left\{e^{-ik(t'-t_)}[1+N(k,t_0)]+
e^{ik(t'-t)}N(k,t_0) \right\}\; ,
$$
$$
{\cal G}^<_{ij}(k;t',t)= \frac{i}{2k}\; {\cal P}_{ij}(\vec k)\; 
\left\{e^{ik(t'-t)}[1+N(k,t_0)]+ e^{-ik(t'-t)}N(k,t_0) \right\}\; .
$$
By inserting this propagator in the first term (contribution from the
mean field) in (\ref{exactrate}) we see that the mean 
field {\em does not} contribute to the photon production to ${\cal O}(\alpha)$. To  first order in $\alpha$
the production rate  obtains a rather simple form

\begin{equation}
\dot N(k,t)=\int_{t_0}^t\left[ \Gamma^{(1)}_+(k,t,t')\ (1+N(k,t_0))-
\Gamma_-^{(1)}(k,t,t')\ N(k,t_0)\ \right] dt'
\label{rateqn}
\end{equation}
with the time dependent rates
$$
\Gamma_+^{(1)}(k,t,t')=
-\frac i{2k}\left[{\cal P}^{ij}(\vec k) \; \Pi^{>(1)}_{ij}(k;t,t') \; 
e^{-ik(t-t')}+ {\cal P}^{ij}(\vec k) \;
\Pi^{<(1)}_{ij}(k;t,t') \; e^{ik(t-t')}
\right],
$$
$$
\Gamma_-^{(1)}(k,t,t')=
-\frac i{2k}\left[{\cal P}^{ij}(\vec k) \;\Pi^{>(1)}_{ij}(k;t,t') \;
e^{ik(t-t')}+{\cal P}^{ij}(\vec k) \;\Pi^{<(1)}_{ij}(k;t,t') \;
e^{-ik(t-t')}\right]. 
$$
The transverse self-energies are given to lowest order by
\bea
\Pi^>_{k,ij}(t,t') & = & -4ie^2 \int \frac{d^3q}{(2\pi)^3} \;
q_{Ti} \;q_{Tj} \; G^>_k(t,t') \;G^>_{\vec q+ \vec k}(t,t') \nonumber \\  
\Pi^<_{k,ij}(t,t') & = & -4ie^2 \int \frac{d^3q}{(2\pi)^3} \;
q_{Ti} \;q_{Tj} \; G^<_k(t,t') \;G^<_{\vec q+ \vec k}(t,t') \label{pis} 
\eea
where in the large $ N $ limit the Green's functions $ G^{<,>} $ are given
by (\ref{greater})-(\ref{lesser}) leading to the final 
form of the self-energies to be used to lowest order in $\alpha$ and
leading order in the large $ N $ limit,   
\bea
&&{\cal P}^{ij}(\vec k) \;\Pi_{ij}^{>(1)}(k;t,t')=ie^2 
\int \frac{d^3 q}{(2\pi)^3} \;q^2 \;(1-\cos^2\theta) \;f_q(t) \;f^*_q(t') \;
f_{|\vec q+\vec k|}(t) \;f^*_{|\vec q+\vec k|}(t') \nonumber \\
&&{\cal P}^{ij}(\vec k) \;\Pi_{ij}^{<(1)}(k;t,t')={\cal P}^{ij}(\vec
k) \;\Pi_{ij}^{>(1)}(k;t',t) \label{firstorderpis} 
\eea
In the case in which the initial state is the photon vacuum,
i.e. $N(k,t_0)=0$ the expression (\ref{rateqn}) 
simplifies considerably. Upon integrating eq.(\ref{rateqn}) in time
up to time $ t $, 
$$
N(k,t)=\int_{t_0}^t \dot{N}(k,t') \;dt'
$$ 
one obtains two terms each one with a nested double time integral
which can be written as a double 
integral up to the time $t$ by inserting a theta function. Upon
relabelling the time variables in one of the 
terms we  obtain the expression (\ref{ph.spectrum}), which is valid to
lowest order in $\alpha$ and for vanishing 
initial population of photons.  The photon production rate obtained in
equilibrium in\cite{toimela,rusk,kapustagale,lichard} 
also neglects the photon population in the initial state as well as
the stimulated emission and the loss process. In these 
references the photons are assumed to escape from the medium without
thermalizing and the photon production rate is valid to  
lowest order in $\alpha_{em}$ and to all orders in the strong coupling
constant. The kinetic equation (\ref{ph.spectrum}) 
is precisely the non-equilibrium counterpart of the rate obtained in
these references which is valid also to lowest order 
in the electromagnetic coupling and to leading order in the large $ N $.

 This equation is clearly only approximate since it neglects the buildup of
the population of photons. As the photon distribution increases in
time due to photon production there will be 
stimulated photon production from the Bose enhancement factor
resulting in enhanced photon production, but also 
processes in which a photon present in the plasma can decay into two
charged scalars as well as change in 
the population with momentum $k$ by bremsstrahlung or inverse
bremsstrahlung in the medium. These latter processes 
will result in a depletion of population of photons and must be
accounted for by a more complete kinetic description 
provided below which must eventually be studied numerically.
 However, if the initial state is the photon vacuum,  we expect
(\ref{ph.spectrum}) to be qualitatively correct for early 
and intermediate time scales, since Bose enhancement and the loss term
from photon annihilation and scattering will 
depend on the number of photons present in the appropriate region of
phase space. Hence first photons must be produced 
requiring an ${\cal O}(\alpha)$ and then the stimulated and loss
processes will take place requiring another 
power of $\alpha$. Thus we expect that the early stages of photon
production through spinodal decomposition or 
parametric amplification will be described reliably with the
simplified kinetic equation (\ref{ph.spectrum}), the late 
stages will require the full kinetic equation described below which
will involve a numerical study.  

The equation for the change of population (\ref{ph.spectrum}) as well
as the more general kinetic equation (\ref{rateqn})  
do not account for the change in the photon population, since $N(k,t_0)$
is the population at the initial time. Under the assumption of a wide
separation of time scales, which 
relies on the weak coupling expansion in $\alpha$ a dynamical
renormalization group analysis\cite{boltzshang} leads to 
a resummation of this kinetic equation by the replacement (to lowest
order) $ N(k,t_0) \rightarrow N(k,t) $, thus leading 
to the final form of the (lowest order) kinetic equation

\begin{equation} 
\label{kinetic}
  \dot N(k,t)=R_+^{(1)}(k,t)[1+N(k,t)]-R_-^{(1)}(k,t) \;N(k,t)
  \label{truekinetic} 
\end{equation}
with the {\em time dependent} forward and backward rates given by 
\be
R_\pm^{(1)}(k,t)=\int_{t_0}^tdt'\
\Gamma_\pm^{(1)}(k,t,t') \; . \label{forwbackrates} 
\ee
This resummation is akin to the Markovian approximation introduced
in\cite{htl,boylawrie} and is justified as a consistent 
expansion in the electromagnetic coupling. 

The resummation implied by this kinetic equation is  based on a
dynamical renormalization group analysis of
kinetics\cite{boltzshang,rgir} valid under the assumption of the
separation of time scales, between the time 
scale of non-equilibrium processes of the scalar fields and that of
the evolution of the photon distribution 
function, which is justified for small $ \alpha $. Physically the
process that gives rise to this kinetic equation 
is the following\cite{rgir,boylawrie,boltzshang}: evolve the system
from the initial time $ t_0 $ with (\ref{rateqn}) up to a time $ t_1 $  at
which the photon distribution has changed by a small amount of $ {\cal
O}(\alpha) $: 
\be
N(k,t_1) = N(k,t_0)+\int^{t_1}_{t_0} \dot{N}(k,t') \;dt' \label{phott1}
\ee
with $ \dot{N}(k,t') $ given by (\ref{rateqn}) in terms of the photon
distribution at $ t_0 $. At this time $ t_1 $ reset 
the occupation number of the photon states to (\ref{phott1}) and
evolve up to a further time $t _2 $ using (\ref{rateqn}) 
but now with the occupation number at time $t _1 $. The dynamical
renormalization group establishes the equation that performs this
operation infinitesimally and leads to
(\ref{truekinetic})\cite{boltzshang,rgir}. As explained  
in ref.\cite{boltzshang,boylawrie} the coarse graining 
results from neglecting off-diagonal correlations of the type
$ aa,a^{\dagger}a^{\dagger} $ in the time evolution of the density matrix. 

A similar resummation scheme is implied by the semiclassical Boltzmann
equation, in which if the occupation numbers are treated in lowest order, the
change is linear in time. Replacing the occupation numbers by the time
dependent ones in the Boltzmann equation leads to a resummation and
exponentiation of the time series\cite{boylawrie}. However, as discussed
in\cite{boylawrie} the Boltzmann equation assumes completed collisions that
result in a coarse graining in time and neglects all of the transient effects
and dynamics on short time scales.

In particular, for the case considered in the previous section with
vanishing photon occupation number in the initial state, the photon
distribution function at a given time $ t $ is given by 
\be
N(k,t)= \int_{t_0}^t dt_1\; 
R_+^{(1)}(k,t_1)\; e^{-\int_{t_1}^t\gamma(k,t_2)dt_2}
\label{totalphotnumb} 
\ee
with $\gamma(k,t)= R_-^{(1)}(k,t)-R_+^{(1)}(k,t)$ being the total
time dependent 
rate to ${\cal O}(\alpha)$. Clearly eq.(\ref{totalphotnumb}) provides a
resummation of the perturbative series as is 
generally the case in any kinetic description wherein the rates are
calculated perturbatively. The early time behavior 
of the growth of photon population is obtained from
(\ref{totalphotnumb}) by approximating $\gamma(k,t) \approx 0$,  
leading to the expression (\ref{ph.spectrum}).
A more detailed estimate of the photon population including the
reverse processes and depletion for a general range of 
momenta $k$ will undoubtedly require a numerical evaluation of the
memory kernels in eq.(\ref{totalphotnumb}), this is clearly a formidable task.

If the time evolution were slow for {\em all} fields, we could write
down a closed set of coupled kinetic equations 
for the distribution functions of photons and charged scalars. However, the
strongly out of equilibrium evolution of the scalar fields and fast
dynamics associated with the spinodal 
and parametric instabilities prevent such a kinetic description as
there is no natural separation of 
time scales for the evolution of the scalar fields. The evolution of
the scalar fields is therefore 
taken into account fully through the large $ N $ equations of motion and
enters in the Green's functions that 
define the forward and backward rates (\ref{forwbackrates}).

\vskip .5cm
{\large\bf Acknowledgements}\\
 
We thank D. Schiff for very useful discussions on photon production.
D. B. thanks the N.S.F. for partial support through grant awards:
PHY-9605186 and INT-9815064 and 
LPTHE (University of Paris VI and VII) for warm hospitality and
partial support, H. J. de Vega thanks the Dept. of Physics at the
Univ. of Pittsburgh for hospitality.  We thank the CNRS-NSF
cooperation programme  for partial support. M. S. thanks  the
Foundation `Aldo Gini' of Padova and INFN, Gruppo Collegato di Parma
for financial support during the early stages of this work.
This work was completed under support of Padova University.




\end{document}